\documentclass[apj,numberedappendix,letterpaper]{emulateapj}
\usepackage{epsfig}
\usepackage{amsmath}
\usepackage{enumerate}

\newcommand{\beq}{\begin{equation}}
\newcommand{\eeq}{\end{equation}}
\newcommand{\bea}{\begin{eqnarray}}
\newcommand{\eea}{\end{eqnarray}}
\newcommand{\non}{\nonumber \\}
\newcommand{\trm}[1]{\textrm{#1}}

\renewcommand{\vec}[1]{\mbox{\boldmath $\displaystyle #1$}}

\newcommand{\grad}{{\mbox{\boldmath $\nabla$}}}

\begin{document}

\shortauthors{WEINBERG, ARRAS, QUATAERT, \& BURKART}
\shorttitle{NONLINEAR TIDES IN CLOSE BINARY SYSTEMS}

\title{Nonlinear Tides in Close Binary Systems}
\author{Nevin N.~Weinberg,\altaffilmark{1} Phil~Arras,\altaffilmark{2} Eliot~Quataert,\altaffilmark{3} \& Josh Burkart\altaffilmark{3}}
\altaffiltext{1}{Department of Physics, and Kavli Institute for Astrophysics and Space Research, Massachusetts Institute of Technology, Cambridge, MA 02139, USA; nevin@mit.edu}
\altaffiltext{2}{Department of Astronomy, University of Virginia, P.O. Box 
400325, Charlottesville, VA 22904-4325}
\altaffiltext{3}{Astronomy Department \& Theoretical Astrophysics
  Center, 601 Campbell Hall, University of California Berkeley, CA
  94720, USA}
  
\begin{abstract}

  We study the excitation and damping of tides in close binary
  systems, accounting for the leading order nonlinear corrections to
  linear tidal theory.  These nonlinear corrections include two
  distinct physical effects: three-mode nonlinear interactions, i.e.,
  the redistribution of energy among stellar modes of oscillation, and
  nonlinear {\it excitation} of stellar normal modes by the
  time-varying gravitational potential of the companion.  This paper,
  the first in a series, presents the formalism for studying nonlinear
  tides and studies the nonlinear stability of the linear tidal flow.
  Although the formalism we present is applicable to binaries
  containing stars, planets, and/or compact objects, we focus on
  non-rotating solar type stars with stellar or planetary companions.
  Our primary results include the following: (1) The linear tidal solution almost
  universally used in studies of binary evolution is unstable over
  much of the parameter space in which it is employed.  More
  specifically, resonantly excited internal gravity waves in
  solar-type stars are nonlinearly unstable to parametric
  resonance for companion masses $M'\ga 10-100 M_\Earth$ at orbital
  periods $P\approx 1-10\trm{ days}$.  The nearly static
  ``equilibrium'' tidal distortion is, however, stable to parametric resonance
  except for solar binaries with $P\la 2-5 \trm{ days}$. (2) For
  companion masses larger than a few Jupiter masses, the dynamical
  tide causes short length scale waves to grow so rapidly that they
  must be treated as traveling waves, rather than standing waves.  (3) We show that the global three-wave treatment of parametric instability typically used in the astrophysics literature does not yield the fastest growing daughter modes or instability threshold in many cases.  
  We find a form of parametric instability in which a single
  parent wave excites a very large number of daughter waves ($N\approx
  10^3[P/\trm{ 10 days}]$ for a solar-type star) and drives them as a
  single coherent unit with growth rates that are a factor of $\approx
  N$ faster than the standard three-wave parametric instability. These are local instabilities viewed through the lens of global analysis; the coherent global growth rate follows local rates in the regions where the shear is strongest. In
  solar-type stars, the dynamical tide is unstable to this collective
  version of the parametric instability for even sub-Jupiter companion
  masses with $P \la$ a month. (4) Independent of the parametric
  instability, the dynamical and equilibrium tides 
  excite a wide range of stellar p-modes and g-modes 
  by nonlinear inhomogeneous forcing; this coupling
  appears particularly efficient at draining energy out of the
  dynamical tide and may be more important than either wave breaking 
  or parametric resonance at determining the nonlinear dissipation of the 
  dynamical tide.
 \end{abstract}


\section{Introduction}
\label{sec:intro} 

In binary systems at small separation, the differential gravity from
one body can induce a significant tide in the other. As long as the
tidal potential is time-varying, the tide and orbit exchange energy
and angular momentum. Dissipation damps away the variable component of
the tide, leading to an evolution of both the orbit and the spins of
the bodies. The final state of such dissipation, if it is allowed to
continue, is a circular orbit, with the spin and orbital angular
momentum aligned, and the bodies synchronized with the same side
always facing each other.

The rate at which the orbit and spins evolve depends on two quantities: the amplitude of the 
tide and how rapidly the energy in the tide is dissipated (``tidal friction"). By analogy 
with a damped, driven oscillator, energy dissipation at a rate $\dot{E}$ acting on a tide 
with frequency $\omega$ and energy $E$ induces a phase lag $\delta \sim \dot{E}/(\omega E)$ 
between the time of maximum tidal acceleration and high tide. This phase lag is often 
written in terms of the tidal quality factor $Q\simeq 1/\delta$, where large $Q$ implies 
little dissipation and vice versa. First modeled in the classic treatment of \citeauthor{Darwin:1879} (1879; 
see \citealt{Goldreich:66} for applications in the Solar System), tidal friction is 
fundamental to our understanding of the origin, evolution, and fate of a wide variety of 
binary systems, including planets and moons in the solar system, extrasolar planets, 
solar-type binaries, and compact object binaries.

What is the physical origin of the `friction' which circularizes the
orbits and synchronizes the spins of {\it completely fluid} bodies in
binary systems?  Most investigations treat the fluid motion as a
linear perturbation of the background state, and focus on linear
damping mechanisms such as radiative diffusion or viscosity from
eddies in convection zones.  However, these applications of linear
theory fail to explain many of the observed properties of binary
systems, as we describe below.

This paper goes beyond the standard linear theory treatment in order
to understand the effects of nonlinear interactions on tides in binary
systems. Although aspects of this problem have been studied previously
(e.g., \citealt{Press:75,Kumar:96,Goodman:98,Barker:10}), we present a
comprehensive formalism for understanding nonlinear tides in stars,
planets, and even compact objects.  The ideas and formalism we present
are general, but for concreteness all of our examples focus on tides
in slowly rotating solar type stars with either stellar or planetary
companions; in such stars internal gravity waves (g-modes) can be
nearly resonant with the tide.  Our goal is ultimately to quantify the
orbital and spin evolution of binaries accounting for nonlinear
damping and excitation of tides.  In this first paper, however, we
focus on determining the conditions under which standard linear theory
is invalid because nonlinear instabilities drain energy out of the
linear tide -- we leave calculations of the resulting Q-values and
orbital evolution to future work.

Our study of tides is motivated by a number of observational puzzles,
including:

1. {\it Orbital circularization of solar binaries}---Binaries
containing two solar-type stars, or a solar-type star and a gas giant
exoplanet, are observed to circularize much more rapidly than expected
from even the most detailed theoretical studies (e.g.,
\citealt{Ogilvie:04, Meibom:05}). The observed circularization rates
imply $Q\sim10^5-10^6$, a factor of $100-1000$ times smaller than
theoretical predictions. These observations test theories of tides in
solar-type stars and gas giant planets over a range of orbital periods
and amplitude of the tide.

2. {\it Orbital decay of hot Jupiters}---The tidal Q inferred from
observations of the circularization of solar-type binaries makes dire
predictions for the orbital decay of hot Jupiters (Jupiter-size
extrasolar planets with orbital periods $\la 1$ week).  Tides raised
in the star by the planet attempt to spin up the star, decreasing the
orbital semi-major axis to conserve total angular momentum.  If the
tidal Q for solar-type stars is really $10^5-10^6$, Darwin's theory
predicts that all 5 Gyr old hot Jupiters inside 4 day orbits should
have spiraled into their host (e.g., \citealt{Ogilvie:07}); by
contrast, there are $\sim 100$ such planets observed.  The observed
population of hot Jupiters could represent the tail end of a much
larger parent population, most of which have decayed into their host
stars \citep{Jackson:09}.  Alternatively, the (near universal) practice of calibrating Q
from one observation and applying it to another may be incorrect
(e.g., \citealt{Ogilvie:07}); it is certainly suspect without a deeper
understanding of the underlying physics.  For example, such a
calibration would be invalid if tidal excitation and dissipation
depends on orbital frequency and/or the amplitude of the tide (i.e.,
the mass of the secondary).  In this paper, we show that this is
indeed the case.

3. {\it Radii of hot Jupiters}---Tidal heating has been invoked to
explain the large observed radii of transiting hot Jupiters compared
to evolutionary models \citep{Gu:03, Arras:06, Arras:10}. 
Primordial eccentricity and asynchronous rotation
will give rise to tidal heating, the duration of which is set by
the value of $Q$. Steady state mechanisms, such as thermal tides
\citep{Arras:10}, may also be at work, for which the tidal
heating rate is also dependent on $Q$. A question which has received
less attention is the depth-dependence of tidal heating. In order
for tidal heating to effectively bloat radii, the heat must be
deposited in deep regions with long thermal times. Understanding the
depth dependence requires detailed calculations of wave excitation
and damping.  \vspace{0.2cm}

The observational puzzles above, and the physical motivation that
nonlinear effects can be important, led us to the present
investigation. At first glance, it may be surprising that nonlinear
effects are important, when basic estimates such as the height of the
tide relative to the stellar radius are small. However, the importance
of nonlinear effects is more subtle since wave amplitudes can become
large in highly localized regions.  Consider, e.g., surface waves on
the ocean which can travel great distances relatively undamped, but
are susceptible to wave breaking as the depth decreases near shore.

In the next subsection we review the results from linear tidal theory
in more detail; we also relate our nonlinear calculations to these
standard linear theory results.  In \S~\ref{sec:structure} we 
summarize the structure of the remainder of this paper.  Throughout we
presume a basic understanding of the physics of stellar g-modes and
p-modes (see, e.g., \citealt{Unno:89, Dalsgaard:03}).

\subsection{Linear Theory Predictions for Solar-Type Stars}

The tidal response of a fluid is typically decomposed into
``equilibrium'' and ``dynamical'' components.  The simplest
approximation for the tidal flow, the ``equilibrium tide," ignores
fluid inertia, allowing the fluid to follow gravitational
equipotentials. The equilibrium tide has large lengthscales, and hence
is weakly damped by diffusive effects (such as turbulent viscosity due
to convection), leading to circularization times much longer than are
observed (e.g., \citealt{Goodman:97}).  The ``dynamical tide" refers
to the resonant excitation of waves by the tidal acceleration.  This,
together with linear mechanisms of dissipation, is the key physics in
most {\it linear} predictions of tidal dissipation.

In this paper, we calculate the nonlinear stability of both the
equilibrium and dynamical tides.  The stability of the equilibrium
tide in particular has rarely been considered in the literature
(although see, e.g., \citealt{Press:75}, \citealt{Papaloizou:81}, and
\citealt{Kumar:98a} for some discussion along these lines).  It is,
however, an {\it a priori} promising source of enhanced orbital
evolution given that the equilibrium tide contains most of the energy
in linear theory.

Solar-type stars are composed of a stably stratified core and a neutrally stratified convective envelope.  Ignoring stellar rotation, tides in the convective envelope resemble the equilibrium tide. However, buoyancy in the core gives rise to short wavelength internal gravity waves that can have periods comparable to the orbital period of the binary; these waves have large amplitudes at the stellar center, but small amplitudes in the convection zone. Lastly, including both stellar rotation and fluid inertia, the core gravity waves are altered, and inertial waves restored by the Coriolis force are possible in both the radiative and convective zones.

The main linear dissipation mechanisms for solar-type stars are
radiative diffusion in the core and turbulent viscosity due to
convection in the envelope. \citet{Terquem:98} and \citet{Goodman:98}
carried out detailed calculations for nonrotating stars. Both studies
show that resonant gravity waves in the core are only weakly damped by
radiative diffusion, giving $Q$ values far too large to be of
interest. Goodman \& Dickson sharpened the question. They showed that even if
the train of gravity waves traveling inward from the core-envelope
boundary -- where they are excited by the tidal potential -- were
completely absorbed at the center the resulting dissipation would be
$Q \simeq 10^9\ [P/(10\ {\rm days})]^{8/3}$ (based on their eq. [13]), 
too large by 3 orders of magnitude. Goodman \& Dickson did not
present detailed calculations of the damping of the gravity waves at the
center, but argued that the waves would ``break,'' i.e., become highly
nonlinear, based on an estimate of the gravity wave amplitude required
for the waves to invert the stratification of the star.  This argument
is supported by the numerical simulations of \citet{Barker:10} and \citet{Barker:11}.

Drawing on Goodman \& Dickson's argument, \citet{Ogilvie:07} suggested that the dissipation
of gravity waves may be much less efficient for planetary companions
because the amplitude of the gravity waves are much smaller when they
are excited by a planet rather than a star (given the factor of $\sim
1000$ mass difference between the two); in particular, they argued
that the gravity waves would not overturn the stratification in the
core and the efficient dissipation invoked by \citet{Goodman:98} would
not apply in the case of hot Jupiter systems (see also \citealt{Barker:10}).  In this paper, we show
that nonlinear damping of the dynamical tide can be important even
when the stratification of the star is not overturned.  Linear theory
thus does not apply even for solar-type stars with planetary
companions.

Searching for a more efficient source of damping for tides in
solar-type stars, \citet{Witte:02} and \citet{Ogilvie:07} realized
that far shorter lengthscales, and hence enhanced damping, could be
achieved by including the Coriolis force. Ogilvie \& Lin showed that
$Q \sim 10^8$ can be achieved for solar-type stars based on the
excitation and damping of short wavelength inertial waves in the
convective envelope. This is, however, still not sufficient to explain
the observed circularization of solar binaries.

\subsection{Structure of this Paper}
\label{sec:structure}

The remainder of this paper is organized as follows.  In \S~\ref{sec:eom} we use perturbation theory to derive the nonlinear
equations of motion and the key dimensionless coupling coefficients
required to quantify the nonlinear couplings among stellar modes. Many
of the details of these calculations are given in Appendix
\ref{sec:app:coef_in_amp_eqn}.  The results in \S~\ref{sec:eom}
include two different ways of approaching the nonlinear problem.  In
the first (\S~\ref{sec:method1}), we expand quantities relative to the
unperturbed background state of the star.  In the second (\S~\ref{sec:method2} and Appendix \ref{sec:app:alternative_eom}), we
expand quantities relative to the linear tidal solution.  These two
approaches are formally equivalent, but we will show that having both
at our disposal is useful for understanding certain aspects of
nonlinear tides in stars.  In Appendix \ref{sec:properties_of_coefficient} we specialize to solar type stars and describe some of the key properties of the
nonlinear coupling coefficients derived in \S~\ref{sec:eom} and
Appendix \ref{sec:app:coef_in_amp_eqn}. 

Having derived the underlying equations of motion, we then review the
results of linear theory (\S~\ref{sec:linear}) and summarize the key
physical processes present in nonlinear tidal theory
(\S~\ref{sec:nonlinear_tide}).  In \S~\ref{sec:stability_analysis} we
present the nonlinear stability analysis that is at the heart of this
paper: this allows us to determine the conditions under which the
linear tidal flow is nonlinearly unstable to the parametric
instability. We also find a
`collective' parametric instability involving many modes that grows
significantly faster than the standard three-mode parametric
instability.  Section \ref{sec:nonlinear}
presents examples of the types of nonlinear instabilities that afflict
stellar tides, including both resonant nonlinear coupling (parametric
instability) and nonlinear inhomogeneous driving.   In \S\S~\ref{sec:dyntide} and \ref{sec:eqtide} we apply
the results of \S~\ref{sec:stability_analysis} to solar type stars and
assess the parametric stability of the dynamical and equilibrium
tides, respectively.  In \S~\ref{sec:inhomog_driving} we study
off-resonant non-linear excitation of both stellar g-modes and p-modes
by the equilibrium and dynamical tides.  In
\S~\ref{sec:global_vs_local} we assess when the nonlinear interactions
are sufficiently strong to invalidate the assumption that tides excite
global normal modes.  Finally, in \S~\ref{sec:conclusions} we
summarize our results, discuss their implications for solar binaries
and solar-type stars with hot Jupiter companions, and discuss key
directions for future research.

The rigorous study of the synchronization and circularization of
binary systems requires that one include the rotation of the star
being tidally forced.  However, including even the simplest case of
rigid rotation greatly complicates the analysis. While the centrifugal
force may be ignored to a good approximation, inclusion of the
Coriolis force changes the properties of waves with mode frequency
smaller than rotation frequency, and new wave families
appear. Specifically, eigenfunctions must be expanded in a sum over
spherical harmonics, instead of a single harmonic for nonrotating
stars. This greatly complicates the calculations and for simplicity we neglect Coriolis forces in this paper.   

\section{Statement of the Problem}
\label{sec:eom}

We consider a primary star of mass $M$ and radius $R$ subject to a tidal acceleration from the secondary of mass $M'$. In a spherical coordinate system $(r,\theta,\phi)$ centered on the primary, we take the orbit of the secondary to be $(D(t),\pi/2,\Phi(t))$, where $D(t)$ is the separation (neglecting backreaction on the orbit) and $\Phi(t)$ is the true anomaly corresponding to a Keplerian orbit with semi-major axis $a$, eccentricity $e$, and orbital period $P_{\rm orb}=2\pi [a^3/G(M+M')]^{1/2}$. Defining the dynamical time of the primary $P_0=2\pi/\omega_0=2\pi(R^3/GM)^{1/2}$, the dimensionless strength of the tidal acceleration relative to internal gravity can be parameterized by 
\bea
\label{eq:epsilon}
\varepsilon & = & \frac{M'}{M}\left( \frac{R}{a} \right)^3 \non
&\simeq& 10^{-4}\left(\frac{M'}{M+M'}\right)\left( \frac{P_0}{2.8 \trm{ hr}}\right)^2\left(\frac{P_{\rm orb}}{10 \trm{ days}}\right)^{-2}.
\eea
We treat the tidal acceleration $\grad U \propto \varepsilon (GM/R^2)$ as a small quantity compared to the internal gravity $GM/R^2$ due to either the small mass or long orbital period of the secondary. Linear theory computes the response of the star to the external tidal forcing to ${\cal O}(\varepsilon)$. In this work, we include the lowest order nonlinear terms, and hence consider effects at both ${\cal O}(\varepsilon)$ and ${\cal O}(\varepsilon^2)$. As we will describe, the latter includes nonlinear driving of high-order modes by the dynamical tide and the equilibrium tide. 

\subsection{Second order equations of motion}
\label{sec:second_order_eom}
To derive the second order equations of motion, let $\vec{x}'$ be the position of a fluid element in the perturbed star, $\vec{x}$ the position of the same fluid element in the background state, and $\vec{\xi}$ the Lagrangian displacement vector. They are related by $\vec{x}'=\vec{x}+\vec{\xi}(\vec{x},t)$. Likewise, $\rho'$ is the true density and $\rho$ is the background density. We write the \emph{internal} forces due to pressure, buoyancy, and perturbed gravity that are linear in $\vec{\xi}$ as $\vec{f}_1[\vec{\xi}]$ and those due to leading-order nonlinear  interactions as $\vec{f}_2[\vec{\xi}, \vec{\xi}]$. Explicit expressions for $\vec{f}_1$ and $\vec{f}_2$ are given in Schenk et al. (2002; see their eq. [I37]).

The \emph {external} forcing terms due to the companion can be derived from the interaction term of the fluid Hamiltonian \citep{Newcomb:62}. To second order, the fluid interaction Hamiltonian is 
\bea
H_{\rm int} & = & \int d^3x' \rho'(\vec{x}') U(\vec{x}')
= \int d^3x \rho(\vec{x}) U(\vec{x}+\vec{\xi})
\nonumber \\ & \simeq &
\int d^3x \rho(\vec{x}) \left[U(\vec{x})
+ \vec{\xi} \cdot \grad U +
\frac{1}{2}\xi^i\xi^j\nabla_i \nabla_j U\right], 
\hspace{0.3cm}
\label{eq:Hint}
\eea
where $d^3x' \rho'(\vec{x}')= d^3x \rho(\vec{x})$ since mass is conserved. The first term in equation (\ref{eq:Hint}) can be ignored since it contains no dependence on $\xi$. Taking a functional derivative $\delta H_{\rm int}/\delta \vec{\xi}$ of this expression with respect to $\vec{\xi}$ leads to the tidal acceleration
\bea
\vec{a}_{\rm tide} & = & - \frac{1}{\rho} \frac{\delta H_{\rm int}}{\delta \vec{\xi}}
= - \grad U - (\vec{\xi} \cdot \grad) \grad U.
\label{eq:atide}
\eea
The first term in equation (\ref{eq:atide}) is the standard linear, time-dependent, inhomogeneous tidal forcing which acts to excite oscillation modes. The second term is the nonlinear tidal forcing. Since it is linear in mode amplitude, it may lead to an exponential growth or damping of waves, as in the Mathieu equation (see also \citealt{Papaloizou:81}).

Gathering terms, the second-order equation of motion including linear forces, tidal forcing, and 3-wave nonlinear interactions, is then
\bea
\rho \ddot{\vec{\xi}} & = & \vec{f}_{\rm 1}[\vec{\xi}]
+ \vec{f}_{\rm 2}[\vec{\xi},\vec{\xi}]
+ \rho \vec{a}_{\rm tide}.
\label{eq:xieqn}
\eea
We consider two approaches to solving equation (\ref{eq:xieqn}). Both approaches involve expanding the spatial dependence of all quantities in terms of the linear adiabatic eigenmodes of the star. In the first method, we expand quantities relative to the star's unperturbed background state. In the second method, we expand quantities relative to the star's linearly perturbed state. As we will describe, each approach has its own conceptual and practical advantages and we will make use of both throughout the paper.

\subsubsection{Method 1}
\label{sec:method1}

Following \citet{Schenk:02}, expand the six-dimensional phase space vector as
\begin{eqnarray}
\label{eq:xiexpansion}
 \left[ \begin{array}{c}
        \vec{\xi}(\vec{x},t)\\
        \dot{\vec{\xi}}(\vec{x},t)
        \end{array} \right] =
\sum_a q_a(t) \left[ \begin{array}{c}
        \vec{\xi}_a(\vec{x}) \\
        - i \omega_a \vec{\xi}_a(\vec{x})
        \end{array} \right].
\end{eqnarray}
The eigenmode labeled ``$a$" is specified by its frequency $\omega_a$, eigenfunction $\vec{\xi}_a(\vec{x})$, and total amplitude $q_a$. The sum over $a$ runs over all mode quantum numbers, mode families, and frequency signs to allow both a mode and its complex conjugate. Plugging this expansion into equation (\ref{eq:xieqn}) and using the orthogonality of eigenmodes leads to a set of coupled oscillator equations for each mode (see Appendix \ref{sec:app:coef_in_amp_eqn})
\bea
\dot{q}_a + i\omega_a q_a & =&  - \gamma_a q_a + i\omega_a U_a(t)
\nonumber \\ && + i\omega_a \sum_b U_{ab}^*(t) q_b^*
+ i \omega_a \sum_{bc} \kappa_{abc}^* q_b^* q_c^*.
\label{eq:modeampeqn}
\eea
The left hand side of equation (\ref{eq:modeampeqn}) describes an uncoupled oscillator. The terms on the right hand side represent linear damping ($\gamma_a$), the linear ($U_a$) and nonlinear ($U_{ab}$) tidal force, and three-wave coupling ($\kappa_{abc}$). The latter two represent effects that come in at $\mathcal{O}(\varepsilon^2)$. We choose a normalization in which, at unit amplitude, each mode has energy (see eq. [\ref{eq:normintegral}])
\bea
\label{eq:Enorm}
E_0=2 \omega_a^2\int d^3x \rho \vec{\xi}_a^\ast \cdot \vec{\xi}_a=\frac{GM^2}{R}.
\eea 
In terms of this normalization, the coefficients in equation (\ref{eq:modeampeqn}) become \citep{Schenk:02}
\bea
\label{eq:Ua}
U_a(t) & = & - \frac{1}{E_0} \int d^3x \rho 
\vec{\xi}_a^* \cdot \grad U,\\ 
\label{eq:Uab}
U_{ab}(t) & = & 
 -\frac{1}{E_0}\int d^3x \rho \vec{\xi}_a\cdot\left(\vec{\xi}_b\cdot\grad\right)\grad U,\\
\label{eq:kappa}
\kappa_{abc} & = &  \frac{1}{E_0} \int d^3x \,
\vec{\xi}_a \cdot \vec{f}_{\rm 2}[\vec{\xi}_b,\vec{\xi}_c].
\eea

The tidal potential $U$ is expressed in terms of spherical harmonics as
\bea
U(\vec{x},t) & = & - \sum_{\ell\geq 2,m} \frac{GM'W_{\ell m} r^\ell}{D^{\ell+1}(t)}
Y_{\ell m}(\theta,\phi) e^{-im\Phi(t)},
\label{eq:U}
\eea
where $W_{lm}=4\pi(2\ell+1)^{-1}Y_{\ell m}(\pi/2,0)$. The $W_{\ell m}$ are nonzero only if $\ell-m$ is even; for the $\ell=2$ harmonic, which dominates for small $R/D$, $W_{20}=-\sqrt{\pi/5}$ and $W_{2\pm2}=\sqrt{3\pi/10}$. Corrections to the tidal potential due to the extended mass distribution of the secondary occur at quadrupole-quadrupole order ($\mathcal{O}(\varepsilon^4)$ for a twin binary) and can therefore be neglected at $\mathcal{O}(\varepsilon^2)$.

Plugging equation (\ref{eq:U}) into  equations (\ref{eq:Ua}) and (\ref{eq:Uab}) leads to the following dimensionless overlap integrals:
\bea
I_{a\ell m} & = & \frac{1}{MR^\ell} \int d^3x \rho
\vec{\xi}_a^* \cdot \grad \left( r^\ell Y_{\ell m} \right)
\label{eq:Ialm}
\eea
and
\bea
J_{ab\ell m} & = &
 \frac{1}{M R^\ell} \int d^3x \rho \vec{\xi}_a \cdot\left(\vec{\xi}_b\cdot\grad\right)\grad\left(r^\ell Y_{\ell m}\right).
\label{eq:Jablm}
\eea
In terms of these overlap integrals, the time-dependent coefficients in equation  (\ref{eq:modeampeqn}) can be written solely as functions of the orbit
\bea
\label{eq:Ua_via_Ia}
U_a(t) & = &  \frac{M'}{M} \sum_{\ell m} W_{\ell m} I_{a\ell m} \left( \frac{R}{D(t)} \right)^{\ell+1}  e^{-im\Phi(t)}
\\
U_{ab}(t) & = & \frac{M'}{M} \sum_{\ell m}
W_{\ell m} J_{ab\ell m} \left( \frac{R}{D(t)} \right)^{\ell+1}  e^{-im\Phi(t)}.
\eea

For analytic work we find it convenient to expand the time-dependence of the Keplerian orbit as a sum of harmonic terms using the Hansen coefficients
\bea
\left( \frac{a}{D} \right)^{\ell+1} e^{-i m \Phi} & = &
\sum_{k=-\infty}^\infty
X^{\ell m}_k(e) e^{-ik\Omega t},
\label{eq:hansenexpansion}
\eea
where
\bea
X^{\ell m}_k(e) & = & \frac{\Omega}{2\pi} \int_0^{2\pi/\Omega} dt \, e^{ik\Omega
t-im\Phi} \left( \frac{a}{D} \right)^{\ell+1}
\nonumber \\ &=& \delta_{k,m}
+ \frac{e}{2}\Big[(\ell+1-2m)\delta_{k,m-1}
\non & & +(\ell+1+2m)\delta_{k,m+1} \Big]
+ {\cal O}(e^2).
\label{eq:hansen}
\eea 
The linear tidal potential can
be written in terms of the Hansen coefficients as $U_a(t)=\sum_k
U_a^{(k)} e^{-ik\Omega t}$, where \bea
\label{eq:Uak}
U_a^{(k)}&=&\frac{M'}{M} \sum_{\ell m} W_{\ell m} I_{a\ell m} X_k^{\ell m}\left(\frac{R}{a}\right)^{\ell+1},
\eea
and similarly for the nonlinear tidal coefficient $U_{ab}$.

The coupled mode amplitude equations can be integrated once the values of $\omega_a, \gamma_a, I_{a\ell m}, J_{ab\ell m}$ and $\kappa_{abc}$ are known. In Appendix \ref{sec:app:coef_in_amp_eqn} we describe our calculation of each of these coefficients.

\subsubsection{Method 2}
\label{sec:method2}

In method 1 we expressed the amplitude equation  (eq. [\ref{eq:modeampeqn}]) satisfied by each mode $a$  in terms of the mode's total amplitude $q_a$, i.e., the amplitude relative to the background state of an unperturbed star. In method 2 we instead express the amplitude equation in terms of the nonlinear amplitude $r_a \equiv q_a - q_{a, \rm lin}$, where $q_{a, \rm lin}$ is the linear amplitude found by ignoring nonlinear coupling and the nonlinear tide (i.e., setting $\kappa_{abc}=U_{ab}=0$ in eq. [\ref{eq:modeampeqn}]; see \S~\ref{sec:linear} and eq. [\ref{eq:qa_lin}]). Although the form of the amplitude equation for $r_a$ is somewhat less compact than that for $q_a$, we find that this approach offers some  advantages, particularly when studying the stability of the linear solution (see \S~\ref{sec:driving_rate}). 

Starting from the second-order equation of motion (eq. [\ref{eq:xieqn}]), write the total displacement as a sum of the linear and nonlinear displacement $\vec{\xi}=\vec{\xi}_{\rm lin} + \vec{\xi}_{\rm nl}$. The linear displacement $\vec{\xi}_{\rm lin}$ is found by solving the linear inhomogeneous equations, as described in \S~\ref{sec:app:xi_lin_calc}. Noting that $\rho \ddot{\vec{\xi}}_{\rm lin} = \vec{f}_1[\vec{\xi}_{\rm lin}]-\rho\grad U$, we then have
\bea
\rho \ddot{\vec{\xi}}_{\rm nl}&=&\vec{f}_1[\vec{\xi}_{\rm nl}]
+\vec{f}_2[\vec{\xi}_{\rm lin},\vec{\xi}_{\rm lin}]+2\vec{f}_2[\vec{\xi}_{\rm lin},\vec{\xi}_{\rm nl}]
\non &&
+\vec{f}_2[\vec{\xi}_{\rm nl},\vec{\xi}_{\rm nl}]
-\rho \left[\left(\vec{\xi}_{\rm lin}+\vec{\xi}_{\rm nl}\right)\cdot\grad\right]\grad U.
\eea
Expanding $\vec{\xi}_{\rm nl}$ using equation (\ref{eq:xiexpansion}) with $(\vec{\xi},\dot{\vec{\xi}},q_a) \rightarrow (\vec{\xi}_{\rm nl},\dot{\vec{\xi}}_{\rm nl},r_a)$ leads to an equation for each mode's nonlinear amplitude  
\bea
\dot{r}_a+(i\omega_a +\gamma_a) r_a &=& 
i\omega_a\left(V_a^\ast+K_a^\ast\right)
\non && 
+ i\omega_a\sum_b \left(U_{ab}^\ast+2K_{ab}^\ast\right) r_b^\ast
\non &&
 +i\omega_a\sum_{bc} \kappa_{abc}^\ast r_b^\ast r_c^\ast,
\label{eq:ra_amp_eqn}
\eea
where
\bea
\label{eq:Ua_nl}
V_a(t)&\equiv& -\frac{1}{E_0}\int d^3x \, \rho \vec{\xi}_a\cdot\left(\vec{\xi}_{\rm lin}\cdot\grad\right)\grad U,\\
\label{eq:kap_a}
K_a(t) &\equiv&\frac{1}{E_0}\int d^3x\,\vec{\xi}_a \cdot \vec{f}_2\left[\vec{\xi}_{\rm lin},\vec{\xi}_{\rm lin}\right],\\
\label{eq:kap_ab}
K_{ab}(t) &\equiv&\frac{1}{E_0}\int d^3x\,\vec{\xi}_a \cdot \vec{f}_2\left[\vec{\xi}_{\rm lin},\vec{\xi}_b\right],
\eea
and $U_{ab}(t)$ and $\kappa_{abc}$ are given by equations (\ref{eq:Uab}) and (\ref{eq:kappa}). We describe our calculation of the coefficients $V_a$, $K_a$ and $K_{ab}$ in Appendix \ref{sec:app:alternative_eom}.

\subsection{Additional points}
Given the amplitude $q_a$ of each mode, we can calculate the orbital evolution of the system (e.g., $\dot{a}$ and $\dot{e}$). In Appendix \ref{sec:app:Interaction_energy} we derive the expressions describing the orbital evolution assuming the excited modes are all standing waves.

So far we have assumed that the tidally forced body is non-rotating.
This may be a good approximation for tidal forcing of slowly rotating
solar-type stars by close-in planets, however it is inadequate when studying
circularization, where the body is pseudo-synchronized with the orbit.
Within the approximations in this paper, where the Coriolis force is ignored,
we can still use the correct rotating frame forcing frequencies by 
replacing the stellar azimuth $\Phi$ in the inertial frame with the
rotating frame value $\Phi_{\rm rot}=\Phi-m\Omega_s$, where $\Omega_s$ 
is the rotation rate. This would alter the phase of the tidal potential
to be $\exp[ -i(k\Omega-m\Omega_s)t]$, where $k\Omega
- m\Omega_s$ is the co-rotating frame frequency. Note too that equation (\ref{eq:Enorm}) is the mode energy for a non-rotating star, and it assumes that the potential and kinetic energies are equal. In a rotating star, equation (\ref{eq:Enorm}) can still be used for normalization purposes, but to calculate the true energy the full kinetic and potential terms must be summed. Furthermore, one must choose a frame (co-rotating or inertial) to evaluate the wave amplitude ($q$), orbital phase ($\Phi$), eigenfunctions ($\vec{\xi}$), frequency ($\omega$), and energy.

Our treatment to this point has been for orbits with arbitrary
eccentricity. Since the binary systems that are the focus of this
paper typically have $e \ll 1$  and $R \ll a$, for analytic work
we use the dominant $\ell=2$ component of the potential and harmonics up to linear order in $e$.
 For circularization the four harmonics are $(m=0,k=\pm 1)$, $(m=\pm 2, k=m)$,
$(m=\pm 2, k=m-1)$ and $(m=\pm 2, k=m+1)$.
For synchronous rotation $\Omega_s = \Omega$, the forcing frequency
in the rotating frame is $(k-m)\Omega$ and the 
$k=m$ term can be ignored since it has zero frequency. 

\section{Linear Tide}
\label{sec:linear}

\begin{figure}
\epsscale{1.1}
\plotone{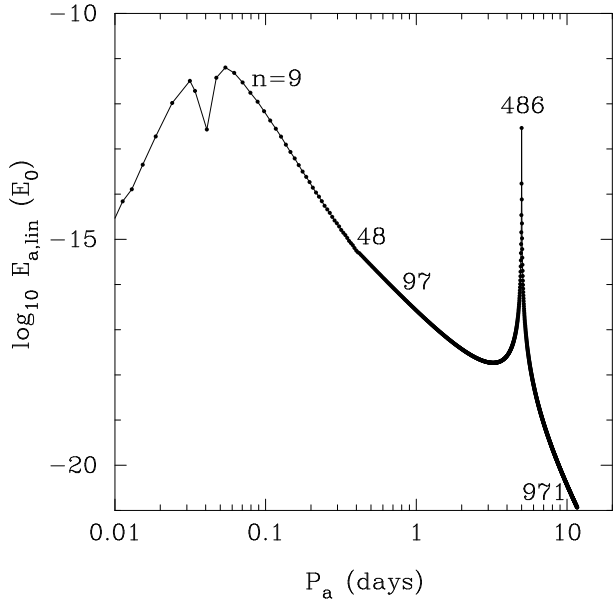}
\caption{Linear energy $E_{a, {\rm lin}}$ of individual modes (in units of $E_0$; eq. [\ref{eq:Enorm}]) as a function of mode period $P_a$ for the $\ell=k=m=2$ harmonic of the tide in a non-rotating solar-type star with $M'=M=M_\odot$ and $P_{\rm orb}=10\trm{ days}$. A line connecting the individual modes is drawn for clarity. The radial order $n$ of the g-modes with period $P\simeq 0.1, 0.5, 1, 5,$ and $10\trm{ days}$ are labeled. The broad peak centered on $P\simeq0.05\trm{ days}$ is the equilibrium tide and the narrow peak at $P\simeq5\trm{ days}$ is the dynamical tide.}
\label{fig:ElinP}
\end{figure}

The linear response of the star can be found by ignoring the nonlinear coupling ($\kappa_{abc}=0$) and the nonlinear tide ($U_{ab}=0$) in equation (\ref{eq:modeampeqn}). Each mode then rings as an independent oscillator according to the equation
\bea
\label{eq:linear_amp_eqn}
\dot{q}_a+i\omega_a q_a = -\gamma_a q_a+ i\omega_a U_a(t)
\eea 
whose steady-state solution is
\bea
\label{eq:qa_lin}
q_{a, \rm lin}(t) = \sum_{k=-\infty}^\infty \frac{\omega_a U_a^{(k)}}{\omega_a -k\Omega - i\gamma_a} e^{-ik\Omega t}.
\eea
The linear response can be broken up into a zero frequency equilibrium tide and a dynamical tide, $q_{a, \rm lin}(t) = q_{a, \rm eq}(t) + q_{a, \rm dyn}(t)$, where
\bea
\label{eq:qa_eq}
q_{a, \rm eq}(t) &\equiv& \sum_{k=-\infty}^\infty U_a^{(k)} e^{-ik\Omega t}
\eea
and
\bea
\label{eq:qa_dyn}
q_{a, \rm dyn}(t)
&\equiv& q_{a, \rm lin}(t)   - q_{a, \rm eq}(t)
\\ &=& \sum_{k=-\infty}^\infty \left( \frac{k \Omega + i \gamma_a}{\omega_a - k\Omega -
 i\gamma_a} \right) U_a^{(k)} e^{-ik\Omega t}.
\eea
The linear energy $E_{a, {\rm lin}}=|q_{a, \rm lin}|^2$ in an $\ell_a=2$ mode for the $\ell=k=m=2$ harmonic of the tide in a non-rotating solar-type star is 
\bea
\label{eq:Ealin}
E_{a, {\rm lin}} &=& U_a^2\left[\left(\frac{\Delta_a}{\omega_a}\right)^2 + \left(\frac{\gamma_a}{\omega_a}\right)^2\right]^{-1}
\non &\approx& 10^{-16}\left(\frac{M'}{M+M'}\right)^2 \left(\frac{P_{\rm orb}}{10\trm{ days}}\right)^{-4}  \left(\frac{P_a}{1 \trm{ day}}\right)^{-11/3}
\non &&\times\left[\left(\frac{\Delta_a}{\omega_a}\right)^2 + \left(\frac{\gamma_a}{\omega_a}\right)^2\right]^{-1}E_0
\eea
where $\Delta_a \equiv \omega_a - k\Omega$  is the detuning and the second expression is valid for $P_a \ga 1 \trm{ day}$ (i.e., when the linear overlap varies as $I_{a\ell m}\propto P_a^{-11/6}$; see eq. [\ref{eq:linear_overlap_aprox}]). In Figure \ref{fig:ElinP} we show $E_{a, {\rm lin}}$ for a solar binary with $P_{\rm orb}=10\trm{ days}$. 

As we describe in \S~\ref{sec:app:linear_coef}, the linear overlap integral $I_{a\ell m}$, and thus $U_a$, is largest
for low order $\ell=2$ modes with frequency $\omega_a \approx
\omega_0$. This is because the structure of these modes most closely
resembles the shape of the tidal potential. For the orbits of
interest, $\omega_0 \gg \Omega$ and these low order modes are
off-resonant with the tide; they are forced to oscillate at a frequency
much lower than their natural frequency.  These modes, seen as the
broad hump in Figure \ref{fig:ElinP} given by $E_{a,{\rm lin}}\simeq E_{a, \rm eq}=
U_a^2$, comprise the equilibrium tide and represent the nearly
hydrostatic response of the star. Their net dissipation (by radiative
damping in the stellar interior and/or turbulent damping in the
convective zone) yields solar binary circularization times much longer
than observed \citep{Goodman:97}.

The dynamical tide is comprised of resonant high-order g-modes with $\omega_a \simeq k\Omega$ (the peak at $P_{\rm orb}/2$ in Figure \ref{fig:ElinP}).  The dispersion relation of high order g-modes is $\omega_a \simeq \alpha_0 (l_a/n_a)$ where $\alpha_0\simeq 4\times10^{-3} \trm{ rad s}^{-1}$ for a solar model. The frequency spacing at fixed $\ell_a$ is therefore $\delta \omega_a
\approx \omega_a^2/(\alpha_0 \ell_a)$ and statistically the detuning
of the most resonant mode with the tide is $|\Delta_a/\omega_a|\approx
10^{-3} P_{10}^{-1}$ for $\ell_a=2$. We plot the linear energy of the most resonant mode for the $\ell=k=m=2$ harmonic of the tide in Figure \ref{fig:ElinEth}. Numerically we find an average energy in the dynamical tide of
\bea
E_{\rm dyn} \simeq 2\times10^{-13}\left(\frac{M'}{M+M'}\right)^2 \left(\frac{P_{\rm orb}}{10 \trm{ days}}\right)^{-17/3}E_0\non
\label{eq:Ealindyn}
\eea
(roughly the lower envelope of $E_{a,{\rm lin}}$ in Fig. \ref{fig:ElinEth}). For the periods of
interest, this is $\sim100$ times smaller than the linear energy of the low order modes with frequency $\simeq
\omega_0$. However, despite their smaller energy,
the dynamical tide in solar binaries is more dissipative
than the equilibrium tide for two reasons: (i) the resonant modes have
a higher radiative damping rate owing to their short wavelengths and,
more importantly, (ii) the resonant modes appear to undergo such
strong nonlinear damping in the core that within one travel time
across the core they deposit nearly all their energy
(\citealt{Goodman:98}; see also \S~\ref{sec:global_vs_local}).
Nonetheless, the combined dissipation from the linear dynamical and
equilibrium tides is $\sim3$ orders of magnitude too small to explain
the observed circularization times of solar binaries.

\section{Nonlinear Tide}
\label{sec:nonlinear_tide}

In linear tidal theory, the dynamics are dominated by $\ell=2$ modes since, by angular momentum conservation, only they can couple to the dominant $\ell=2$ harmonic of the tide. In nonlinear theory, by contrast, modes with $\ell\neq2$ can couple to the $\ell=2$ harmonic of the tide. Angular momentum conservation requires only that the coupling coefficient $\kappa_{abc}$ satisfy the selection rules $|\ell_b-\ell_c| \le \ell_a \le \ell_b + \ell_c$ with $\ell_a+\ell_b+\ell_c$ even and $m_a +m_b+m_c=0$.  The nonlinear overlap integral $J_{bc\ell m}$ is subject to the same selection rules but with $\ell_a$ and $m_a$ replaced by the $\ell$ and $m$ of the tidal potential. 
This freedom in the $\ell$ of the modes opens up a large region of the $(\ell, n)$ parameter space that is inaccessible in linear theory. In particular, since the g-mode dispersion relation is $\omega \propto \ell / n$ at the frequencies of interest, high $\ell$ and $n$ modes can be resonant with the tide and thus dynamically important. 

Nonlinear theory not only opens up the parameter space of modes accessible to the tide, it also modifies how modes interact with the orbit. This is because in nonlinear theory, the orbital evolution depends not only on the linear overlap $I_{a\ell m}$ but also on the nonlinear overlap $J_{ab\ell m}$ (see Appendix \ref{sec:app:Interaction_energy}). For two modes $a$ and $b$ of similar (short) wavelength, $J_{ab\ell m}\gg I_{a\ell m}$ so that nonlinear effects can be particularly important. Physically, this is because the coupled modes have an effective wavelength that is long and thus they induce a large density perturbation per unit mode energy. In this paper we focus on determining the stability of the linear solution and do not attempt to calculate the rate of orbital evolution. Determining the full implications of the above effect is therefore deferred to a future paper. 
 
There are three types of nonlinear terms at $\mathcal{O}(\varepsilon^2)$, each the result of different forms of three-wave coupling. This is most easily seen in the amplitude equation for the nonlinear amplitude $r_a$ (eq. [\ref{eq:ra_amp_eqn}]); the three terms on the right hand side each represent a different form of three-wave coupling.\footnote{These three terms can also be seen in the $q_a$ amplitude equation (eq. [\ref{eq:modeampeqn}]) by substituting $q_{b}=q_{b, \rm lin}+r_b$ and $q_{c}=q_{c, \rm lin}+r_c$ into the nonlinear terms $\sum_b[U_{ab}+\sum_c \kappa_{abc} q_c]q_b$.} In order of their appearance in equation (\ref{eq:ra_amp_eqn}), the three types of three-wave coupling are: (1) ``linear-linear coupling" (LLC)---the coupling of a nonlinear wave (or more precisely, the nonlinear correction to a wave) to a pair of linear waves (equilibrium tide plus dynamical tide), (2) ``linear-nonlinear coupling" (LNC)---the coupling of a nonlinear wave to a linear wave and another nonlinear wave, and (3) ``nonlinear-nonlinear coupling" (NNC)--- the coupling of a nonlinear wave to a pair of nonlinear waves. Conceptually, the terminology LLC and LNC is  most useful when the linearly excited modes are near their linear energy. If nonlinear interactions drive these modes far from their linear energy, it no longer makes sense to call them linear waves and there is no longer a clear dichotomy between LLC, LNC, and NNC. We discuss this issue further in \S\S~\ref{sec:simple_inhomog} and \ref{sec:inhomog_driving}.

Like the linear inhomogeneous driving term $U_a$, the LLC term $V_a + K_a$ in equation (\ref{eq:ra_amp_eqn}) acts as a (nonlinear) inhomogeneous driving term.  Thus, just as all modes that satisfy momentum conservation in linear theory have some linear amplitude $q_{a, \rm lin}$ due to the driving term $U_a$, all modes that satisfy momentum conservation in nonlinear theory have some nonlinear amplitude $r_a$ due to the driving term $V_a+K_a$. Insofar as the $\ell=2$ tide dominates, momentum conservation implies that only modes with $\ell_a=0,2,$ or 4 are driven significantly by LLC.

The LNC term $(U_{ab}+2K_{ab})r_b$ has the form of a network of coupled Mathieu equations; it can thus lead to parametric driving. As with all parametrically driven systems, only modes that are sufficiently resonant with the driving frequency and have sufficiently small linear damping are unstable to LNC (see eq. [\ref{eq:stability_criteria}]). This is in contrast to LLC, where all modes with nonzero couplings are driven regardless of their frequencies and damping rates.

Finally, the NNC term $\kappa_{abc}r_b r_c$ has a quadratic dependence on the nonlinear mode amplitudes. Whereas the stability of LNC is independent of the amplitude of the nonlinear modes (it depends only on their frequency and damping rate), only modes with sufficiently large nonlinear amplitudes are affected by NNC. NNC describes how the energy in nonlinear waves gets redistributed amongst other nonlinear waves and is thus related to the saturation of nonlinear instabilities. 

In this paper we focus on calculating the stability of the linear state. When the star is exactly in the linear state (i.e., the nonlinear amplitudes $r_a$ are vanishingly small), the only nonlinear interaction is the LLC term $V_a + K_a$. With just an infinitesimal perturbation from the linear state, the LNC term can become important, which represents parametric driving of `daughter' waves by the `parent' linear tide. In the next section, we carry out a stability analysis of LNC. Since our present focus is the stability of the linear state and not the saturation of possible nonlinear instabilities, we do not study NNC in this paper.

\section{Parametric stability analysis}
\label{sec:stability_analysis}

In this section we derive the stability condition for parametric driving by the linear tide (the LNC term $[U_{ab}+2K_{ab}]r_b$ of eq. [\ref{eq:ra_amp_eqn}]). The stability can be determined in the usual manner: perturb the linear solution and see if the perturbations grow exponentially due to nonlinear forces faster than they damp. If they do then the linear state is unstable to parametric driving.  In \S~\ref{sec:instability_criteria} we derive the criteria that determines the stability of the linear solution and in  \S~\ref{sec:driving_rate} we determine the parametric driving rate $\Gamma_{bc}$ on which the stability depends.  We show that $\Gamma_{bc}$ can be decomposed into two distinct types of driving: driving by the equilibrium tide and driving by the dynamical tide (\S\S~\ref{sec:eq_tide_driving_rate} and \ref{sec:dyn_tide_driving_rate}, respectively; the stability of the dynamical tide and equilibrium tide are discussed in more detail in \S\S~\ref{sec:dyntide} and \ref{sec:eqtide}). We also describe a collective form of parametric instability; in \S~\ref{sec:dyntide_collective} we show that as a result of this instability, the dynamical tide may drive a very large number of modes to large amplitude in a very short time ($\sim P_{\rm orb}$).

\subsection{Instability Criteria}
\label{sec:instability_criteria}

Consider the parametric (LNC) driving of  a set of daughter modes $\{b,c\}$ with $\omega_{b,c}\approx \omega/2\equiv k\Omega/2$, where $\omega$ is the tidal driving frequency. For such daughters, parametric driving will typically dominate over  inhomogeneous (LLC) driving, especially if $\ell_{b,c} \neq 0,2,$ or 4. To a good approximation we can therefore neglect the daughters inhomogeneous driving term $V_b+K_b$ and, by equation (\ref{eq:ra_amp_eqn}), the amplitude equation for each daughter is
\bea
\label{eq:daughter_amp_eqn}
\dot{r}_b+(i\omega_b + \gamma_b)r_b = i\omega_b\sum_c\left[U_{bc}^\ast(t)+2K_{bc}^\ast(t)\right]r_c^\ast. 
\eea
Consider a particular harmonic $\omega$ of the tide such that $U_{bc}(t)=U_{bc}e^{-i\omega t}$ and similarly for $K_{bc}(t)$. If we plug in $r_b(t)= Q_b(t) e^{i\omega t/2}$ and similarly for $r_c$, then the harmonic time dependences cancel, yielding
\bea
\dot{Q}_b + \left(i\Delta_b + \gamma_b\right)Q_b&=&i \sum_c \left(\frac{\omega_b}{\omega_c}\right)^{1/2} \Gamma_{bc}^\ast Q_c^\ast,
\eea
where we defined the daughter pair ``driving rate"
\bea
\label{eq:Gamma_bc}
\Gamma_{bc}\equiv \sqrt{\omega_b\omega_c} \left(U_{bc} + 2K_{bc}\right).
\eea 
Writing these equations as $\dot{\mathbf Q}=H{\mathbf Q}$, the solutions are ${\mathbf Q}\propto \exp(st)$ where the eigenvalues $s$ are the solutions of the characteristic equation $\det(H-sI)=0$. The system is unstable if there is an $s$ such that $Re(s)>0$. 
 For $N$ coupled daughter modes, $H$ is a non-symmetric $2N\times2N$ matrix with $2\times2$ block components $H_{jk}=(\omega_j / \omega_k)^{1/2}A_{jk}+\delta_{jk}B_k$, where the indices $j,k$ run over the daughter modes, and
\bea
\hspace{-0.3cm}
A_{jk}=\begin{bmatrix}
\textrm{Im}(\Gamma_{jk}) & \hspace{0.0cm}& \textrm{Re}(\Gamma_{jk})\\
&&&\\
\textrm{Re}(\Gamma_{jk}) & \hspace{0.0cm}& -\textrm{Im}(\Gamma_{jk})
\end{bmatrix},
\hspace{0.1cm}
B_k=\begin{bmatrix}
 -\gamma_k & \hspace{0.0cm} & \Delta_k\\
 &&&\\
 -\Delta_k & \hspace{0.0cm}  & -\gamma_k 
\end{bmatrix}.
\eea
Here $\Delta_k=\omega_k+\omega/2$ is the daughter detuning rather than the previously defined parent detuning $\Delta_a=\omega_a-\omega$. 

While in general the eigenvalues of the characteristic equation must be solved for numerically, we give examples below where the eigenvalue expressions are simple and offer insight into how the stability depends on $N$, $\omega_b$, $\gamma_b$, $\Delta_b$, and $\Gamma_{bc}$. From these examples and numerical experiments we deduce an approximate stability criteria: a set of $N$ daughters with $\omega_b \omega_c >0$ is unstable if each pair $(b,c)$ in the set satisfies
\bea
\label{eq:stability_criteria}
 N|\Gamma_{bc}| \ga \sqrt{ \gamma_b \gamma_c+\Delta_{bc}^2},
\eea
where $\Delta_{bc}=\Delta_b+\Delta_c=\omega+\omega_b+\omega_c$. When sufficiently far from the stability boundary, the growth rate of an unstable collection of modes is $\sim N|\Gamma_{bc}|$. If $\omega_b \omega_c <0$, the system can only be unstable if $\Gamma_{bc}$ is asymmetric in $b\leftrightarrow c$. However, since $U_{bc}$ and $K_{bc}$ are symmetric in $b\leftrightarrow c$, so is $\Gamma_{bc}$. High-frequency modes with $\omega_b \omega_c <0$ and small $\Delta_{bc}$ (i.e., a pair with a beat frequency $|\omega_b+\omega_c| \simeq |\omega|$) are therefore stable. 

We emphasize that to be collectively unstable, and thus have a growth rate $\sim N |\Gamma_{bc}|$, {\em each pair} in the set of $N$ daughters must satisfy the inequality (\ref{eq:stability_criteria}). The inequality's dependence on $N$ shows that the daughters can be collectively unstable even if each daughter pair would be stable on its own (i.e., even if $|\Gamma_{bc}| < \sqrt{\Delta_{bc}^2 + \gamma_b \gamma_c}$ for every pair). Conversely, a subset of pairs in the set can be unstable even if the set as a whole is collectively stable.  This implies that even if the tide excites $N\gg1$ daughters, these daughters are not necessarily undergoing a collectively instability. Rather, they may each be undergoing the standard parametric instability with a growth rate  $\ll N |\Gamma_{bc}|$ (if the daughters are coupled to each other in a chain, they may all grow at a single, coherent rate $\ll N |\Gamma_{bc}|$ corresponding to one of the eigenvalues $Re(s)$).

The case $N=2$ is the one most often found in the literature (``three-wave coupling"). As we show in \S\S~\ref{sec:nonlinear} and \ref{sec:dyntide}, the collectively unstable case with $N\gg2$ results in mode dynamics that are very different from the $N=2$ case; in particular, the driving rate of the $N$ daughters is much more rapid (by a factor of $\approx N/2$).  In \S~\ref{sec:dyn_tide_driving_rate} we show that due to collective driving, the global growth rate of daughter modes approaches their maximum local growth (in the case of dynamical tide driving, the latter is the local growth rate within the small driving region near the center of the star). 

Physically, collective driving occurs because the $N$ unstable daughters each have comparable group velocities and thus their superposition within the growth region remains coherent over a growth time. They therefore act as a single ``mode" that is much more strongly peaked than the individual eigenmodes. This results in an effective coupling coefficient $\sim N$ times larger than the three-mode coupling coefficient $\kappa_{abc}$. 

The astrophysics literature has traditionally focused on three-wave coupling using global normal modes (however see \citealt{Ryu:96}). What we call collective driving is not necessarily captured by such an analysis, even though it is simply a consequence of spatially localized coupling of many modes.  In particular, the rapid `collective' growth at nearly the maximum parent shear rate (see \S~\ref{sec:dyn_tide_driving_rate} below) is {\em not} captured in standard calculations of three mode coupling coefficients and growth rates (e.g., \citealt{Kumar:96, Wu:01, Weinberg:08}).  The reason is fundamentally that the standard basis set of global stellar normal modes does not necessarily capture the most rapidly growing daughters in the system, which are in fact a superposition of standard global stellar normal modes (those that are strongly peaked near where the parent's shear peaks). 
In future work, it will be important to revisit previous astrophysical applications of parametric resonance given this limitation of previous work.

The stability analysis of equation (\ref{eq:daughter_amp_eqn}) applies only if the nonlinear interactions are sufficiently weak that the tide excites standing waves (i.e., global normal modes). As we describe in \S~\ref{sec:global_vs_local}, if the nonlinear interactions are so strong that the daughters' local growth rate (eq. [\ref{eq:Gammabb}]) is larger than the inverse of their local group travel time (eq. [\ref{eq:tgroup_local}]), then the driving must be treated as a local interaction in which the tide excites traveling waves. We note, however, that the instability criteria (eq. [\ref{eq:stability_criteria}]) is independent of the uncertainty of traveling versus standing waves. This is because the daughter growth rates are infinitesimal at threshold; the growth times are therefore always much longer than the group travel time very near threshold (a system is near threshold for certain values of the companion mass and orbital separation; see Fig. \ref{fig:massperiod}). Thus, while collective driving may lead to growth rates that are much faster than that of standard three-mode coupling, a collectively unstable system is not necessarily in the regime in which the excitation of the daughters must be treated using traveling waves rather than global normal modes. 

\subsubsection{Examples}

\begin{enumerate}[(i)]
\item  

\underline{$N$ daughters, $\gamma_b=\gamma, \Delta_b=\Delta,  \Gamma_{bc}=\Gamma$}:
Two of the eigenvalues are $s=-\gamma\pm \sqrt{|N\Gamma|^2-\Delta^2}$ and the other $2(N-1)$ are $s=-\gamma+i\Delta$. The system is therefore unstable if $N|\Gamma| > \sqrt{\gamma^2+\Delta^2}$. We obtain the same result if instead $\Gamma_{jk}=(-1)^{j+k}\Gamma$;  the sign of $\kappa_{ajk}$ varies in such a manner for daughters that are $\Delta n=1$ neighbors.
\item  

\underline{$N=2$ daughters,  $\Gamma_{bc}=\Gamma$, $\Gamma_{bb}=\Gamma_{cc}=0$}:
This system is often considered in the literature (e.g., \citealt{Wu:01, Kumar:96}). The four eigenvalues are 
\bea
s&=&\frac{1}{2}\bigg\{-(\gamma_b+\gamma_c)\pm i(\Delta_b-\Delta_c)\pm\Big[4|\Gamma|^2-\Delta_{bc}^2\non&&+(\gamma_b-\gamma_c)^2\pm2i\Delta_{bc}(\gamma_c-\gamma_b)\Big]^{1/2}\bigg\},
\eea
with the first and third ``$\pm$" linked. The system is unstable if 
\bea
\label{eq:stability_3wave}
|\Gamma| >
\sqrt{\gamma_b \gamma_c} \left[1+\frac{\Delta_{bc}^2}{(\gamma_b +\gamma_c)^2}\right]^{1/2}.
\eea
Since this case does not include self-coupling, we can have $\Delta_{bc}\ll \Delta_b, \Delta_c$. Thus, even if the daughters' frequencies are not individually close to $-\omega/2$, their sum can be close to $-\omega$ and the pair can have a low parametric threshold.  
\end{enumerate}

\subsection{Parametric Driving Rate}
\label{sec:driving_rate}

We now discuss the parametric driving rate of a daughter pair $(b,c)$. In this case $N=2$ and the driving rate $N|\Gamma_{bc}| \approx |\Gamma_{bc}|$ (eq. [\ref{eq:Gamma_bc}]). Expanding the coefficients in terms of the tidal harmonics, the driving by the $k$'th harmonic is given by
\bea
\label{eq:Gamma_kbc}
\Gamma^{(k)}_{bc}&=&\sqrt{\omega_b\omega_c}\left[U_{bc}^{(k)}+2K_{bc}^{(k)}\right]
\non&=&
\frac{M'}{M}\sum_{\ell m} W_{\ell m} X_{k}^{\ell m} \left(\frac{R}{a}\right)^{\ell+1}\left[\overline{\Gamma}_{bc}^{({\rm eq})}+\overline{\Gamma}_{bc}^{({\rm dyn})}\right],
\hspace{0.5cm}
\eea   
where we define the equilibrium tide and dynamical tide driving rates
\bea
\overline{\Gamma}_{bc}^{({\rm eq})}\equiv \kappa_{bc}^{({\rm eq})} \sqrt{\omega_b \omega_c},
\hspace{0.3cm}
\overline{\Gamma}_{bc}^{({\rm dyn})}\equiv \kappa^{({\rm dyn})}_{bc}\sqrt{\omega_b \omega_c}.
\eea 
and their dimensionless coupling coefficients
\bea
\label{eq:kappa_eq}
\kappa_{bc}^{({\rm eq})}&\equiv& J_{bc\ell m}+ 2\kappa^{(I)}_{bc} + 2\kappa^{(H, {\rm eq})}_{bc},\\
\label{eq:kappa_dyn}
\kappa^{({\rm dyn})}_{bc}&\equiv& 2\kappa^{(H, {\rm dyn})}_{bc}.
\eea
 In Appendix \ref{sec:properties_of_coefficient} we describe the properties of the coefficients $\kappa_{bc}^{({\rm eq})}$ and $\kappa_{bc}^{({\rm dyn})}$ for the case of a solar-type star.
Here $\kappa^{(H, {\rm eq})}_{bc}$ and $\kappa^{(H, {\rm dyn})}_{bc}$ are the homogeneous parts of the linear tide coupling coefficient and $\kappa^{(I)}_{bc}$ is the inhomogeneous part. As we explain in \S~\ref{sec:app:kapU},  the homogeneous parts are found by directly replacing mode $a$ in our final expression for $\kappa_{abc}$ (lines \ref{eq:e411}-\ref{eq:e419}) with $\vec{\xi}_{\rm eq}$ and $\vec{\xi}_{\rm dyn}$, respectively (this expression for $\kappa_{abc}$ is similar to that typically used for three-mode coupling in the literature, with a few corrections). The inhomogeneous part $\kappa^{(I)}_{bc}$ corresponds to terms in the coupling coefficient that are not present in the existing treatments of three-mode coupling in the literature. They arise because we are considering non-linear coupling among {\em tidally forced} modes, rather than freely oscillating modes. More specifically, the inhomogeneous terms arise because in deriving our final expression for $\kappa_{abc}$, we assume all three modes satisfy the homogeneous equations of motion (eqs. [\ref{eq:eom1}, \ref{eq:eom2}]). Here, however, mode $a$ is the linear tide $\vec{\xi}_{\rm lin}$ which instead satisfies the inhomogeneous equations of motion (eqs. [\ref{eq:eom1}, \ref{eq:eom2}] with $\delta\phi \rightarrow\delta\phi +U$); 
substituting the equations of motion which include the inhomogeneous term $U$
 introduces additional terms which we call $\kappa^{(I)}_{bc}$. Because  $\kappa^{(I)}_{bc}$ does not contain an explicit dependence on $\vec{\xi}_{\rm lin}$, it does not separate into an equilibrium tide piece and a dynamical tide piece.

We find that for a solar-type star $|\Gamma_{bc}^{(\rm eq)}| \ll |\Gamma_{bc}^{(\rm dyn)}|$. However, the driving rates are not the only diagnostic of an instability's importance.  In particular, since the majority of the linear tidal (interaction) energy is stored in the equilibrium tide (\S~\ref{sec:linear}), an instability of the equilibrium tide could in principle lead to faster orbital evolution than an instability of the dynamical tide. Equilibrium tide driving thus offers a potentially important source of energy loss that is distinct from the dynamical tide driving considered by \citet{Goodman:98}.

\subsubsection{Relation between global and local driving rates}

While the exact expressions for the coupling coefficients are complicated, intuitively one expects the \emph{local} driving rate of short wavelength daughter waves to approximately equal the driving frequency $\omega$ times the local `shear' of the parent wave $\sim d\xi_r/dr$.  Indeed, in \S~\ref{sec:app:kappa_aprox} we show that the dominant terms in the full expressions for $\kappa_{abb}$ are approximately equal to
\bea
\label{eq:dkappa_dlnr}
\frac{d\kappa_{abb}}{d\ln r}\approx T \frac{dE_b}{d\ln r} \frac{d\xi_{r,a}}{dr}
\eea  
where $T$ is an angular integral ($|T|\sim 1$) and $dE_b/dr\simeq\rho r^2 N^2 \xi_{r,b}^2$ is the radial energy density of the daughter waves. Assuming high-order daughters (eqs. [\ref{eq:xirWKB}-\ref{eq:normWKB}]), this implies that the global driving rate of a self-coupled daughter $b$ by a parent $a$ is approximately given by
\bea
\label{eq:Gamma_global}
\overline{\Gamma}_{bb} \equiv \omega \kappa_{abb}\approx\frac{\int v_{r,b}^{-1} \overline{\Gamma}_{bb}^{(\rm local)} dr}{\int v_{r,b}^{-1} dr}\approx\frac{\int N \overline{\Gamma}_{bb}^{(\rm local)} d\ln r}{\int N d\ln r}
\eea
where $\overline{\Gamma}_{bb}^{(\rm local)}(r)=\omega T d\xi_{r,a}/dr$ is the local daughter driving rate, $v_{r,b}\simeq \omega_b / k_{r,b}$ is the daughter group velocity, and the limits of integration are the inner ($r_1$) and outer ($r_2$) turning points of the daughter (note that $N$ here is the Brunt-V\"ais\"al\"a frequency, not the number of modes). The global driving rate is therefore the average of the local driving rate weighted by the time the daughters spend at each radius (or, equivalently, weighted by $N/r$).

\subsection{Equilibrium tide driving rate}
\label{sec:eq_tide_driving_rate}
Naively, the nonlinear tide $U_{bc}$ and the equilibrium tide part of the three wave coupling $\sum_{a}\kappa_{abc}q_{a, \rm eq}$ have a completely different origin.  However, in Appendix \ref{sec:app:kapU} we show that the leading order terms cancel in the integrand of $J_{bc\ell m}+ 2\kappa^{(I)}_{bc}$ (in eq. [\ref{eq:kappa_eq}] for $\kappa_{bc}^{(\rm eq)}$). Specifically, the fractional difference between  $J_{bc\ell m}$ and $2\kappa^{(I)}_{bc}$ is $\sim (\omega / N)^2\sim10^{-5} P_{10}^{-2}$ and there is thus a large cancellation between these terms in the growth rate of daughter pairs due to the equilibrium tide, $\overline{\Gamma}_{bc}^{({\rm eq})}$. Physically, this is because  nonlinear tidal driving ($U_{bc}$) and internal nonlinear driving by the equilibrium tide ($\sum_{a}\kappa_{abc}q_{a, \rm eq}$) are fundamentally part of the same process; together they describe the nonlinear driving of daughter modes by the nearly hydrostatic response of the star to its companion. In their paper on the parametric excitation of modes in close binary systems, \citet{Papaloizou:81} considered driving by the nonlinear tide $U_{bc}$. However, they neglected three wave coupling  $\sum_{a}\kappa_{abc}q_a$. Their analysis therefore overestimates the  equilibrium tide driving rate by a factor of $\sim 10^5$ for solar-type stars. 

In Appendix \ref{sec:app:kappa_aprox_low_order_parent} we show that $d\kappa_{bc}^{(H, \rm eq)}/d\ln r$ is approximately given by equation (\ref{eq:dkappa_dlnr}), where now mode $a$ represents the equilibrium tide and $d\xi_{r,a}/dr\approx\Lambda (r/R)^{\ell+1}$ at $r\sim R$. The magnitude of $d(J_{bc\ell m}+2\kappa^{(I)}_{bc})/d\ln r$ is easily shown to be of the same order (see eq. [\ref{eq:Jplus2kapI}]). Thus, the \emph{local} daughter driving rate by the equilibrium tide is $\overline{\Gamma}_{bb}^{(\rm local)}\sim \omega (r/R)^{\ell+1}$. Although the local driving rate is $\sim \omega$ at $r\sim R$, the global driving rate $\overline{\Gamma}^{(\rm eq)}_{bb}$ (eq. [\ref{eq:Gamma_global}]) is considerably smaller than this local value. This is because the global driving rate is weighted by the time the daughters spend at each radius and the daughters propagate more slowly at small radii ($v_{r,b}\propto r / N$), where the shear is small (equivalently, the energy in the daughter modes peaks at small radii, where the shear is small).  For a solar binary, the local driving rate by the equilibrium tide is everywhere much smaller than the inverse of the daughters' group travel time (see the next paragraph). It is therefore appropriate to treat the daughter excitation using global normal modes and their growth rate is given by the global driving rate.

We can obtain an approximate expression for  $\overline{\Gamma}^{(\rm eq)}_{bb}$ by noting that for a solar model $N\approx 100 \omega_0 (r/R)$ for $r\la0.05R$ and $N\approx 3\omega_0$ (to within a factor of two) between $r\approx 0.05 R$ and the convection zone $r_c\simeq0.7 R$. With these approximations we find $\overline{\Gamma}^{(\rm eq)}_{bb}\approx 0.2T(\ell+1)^{-1} (r_c/R)^{\ell+1}\omega \approx 0.01 \omega$, in good agreement with the full integration of $\kappa^{(\rm eq)}_{bb}$ given in \S~\ref{sec:Jeq}.\footnote{The integral in the numerator of equation (\ref{eq:Gamma_global}) is dominated by the contribution at $r\sim R$, where to a good approximation $g\propto r^{-2}$ and thus $\xi_{r, \rm eq}\simeq -U/g\propto r^{\ell+2}$ (see eq. [\ref{eq:xir_eq}]). The core, where $g\propto r$ and $\xi_{r, \rm eq} \propto r^{\ell-1}$, contributes only a few percent to the integral.} The global driving rate by the equilibrium tide is therefore a factor of $\sim 100$ slower than the local driving rate at $r\sim R$.  For the dominant harmonic of a synchronized binary (see eq. [\ref{eq:Gamma_kbc}]), 
\bea
\label{eq:eqtide_growth_rate_estimate}
\Gamma_{bb}^{(\rm eq)}\sim \varepsilon e \overline{\Gamma}^{(\rm eq)}_{bb} \sim 10^{-4} \left(\frac{eM'}{M+M'}\right) P_{10}^{-3} \trm{ yr}^{-1},
\eea
where $\varepsilon = (M'/M)(R/a)^{\ell+1}$ (eq. [\ref{eq:epsilon}]).

We can use this estimate of $\Gamma_{bb}^{(\rm eq)}$ to evaluate the stability of the equilibrium tide. The linear damping rate of a high order daughter is $\gamma_b \sim n_b^2/t_{\rm KH}$, where for a solar model $t_{\rm KH}\approx3\times10^7\trm{ yr}$ and $n_b\simeq 500 \ell_b P_{10,b}$.  \emph{If we ignore the daughter detuning} we find that the equilibrium tide is unstable ($\Gamma^{(\rm eq)}_{bb}>\gamma_b $) for orbital periods 
\bea
P \la 4 \ell_b^{-2/5} \left(\frac{eM'}{M+M'}\right)^{1/5} \trm{ days}.  
\eea
While there may be brief intervals during which the detuning $\Delta_{bc}\la \gamma_b$ and the above limit applies, the detuning, on average, sets a more stringent constraint. This is because the equilibrium tide driving rate peaks strongly for self-coupled pairs ($\Gamma^{(\rm eq)}_{bc}\ll \Gamma^{(\rm eq)}_{bb}$ for $|n_b-n_c|\ga1$; see \S~\ref{sec:Jeq}); thus, although there may exist daughter pairs with very small detuning $\Delta_{bc}$, in general such pairs have $|n_b-n_c|\gg 1$ and they couple very weakly to the equilibrium tide. Using the relation for $\gamma_b$ and self-coupled detuning $\Delta_{bb}\approx \omega_b/2n_b\approx 0.2 \ell_b^{-1} P_{10,b}^{-2} \trm{ yr}^{-1}$ and minimizing the right hand side of equation (\ref{eq:stability_criteria}) with respect to $n_b$, we find that, on average, the equilibrium tide is only unstable to parametric resonance for
\bea
P \la 0.5 \left(\frac{eM'}{M+M'}\right)^{3/7} \trm{ days}.
\eea
These estimates agree well with the more exact treatment given in \S~\ref{sec:eqtide}.

Although we defined the parametric growth rate $\Gamma_{bc}$ in terms of the method 2 coupling coefficient $K_{ab}$ (\S~\ref{sec:method2}), we could have equivalently defined it in terms of the method 1 coefficient $\kappa_{abc}$ (\S~\ref{sec:method1}). To do so, write the coupling coefficient in functional form $\kappa_{abc}=\kappa[\vec{\xi}_a, \vec{\xi}_b, \vec{\xi}_c]$. Then
\bea
K_{bc}&=&\kappa\left[\vec{\xi}_{\rm lin}, \vec{\xi}_b, \vec{\xi}_c\right]
=\kappa\left[\sum_a q_{a, {\rm lin}} \vec{\xi}_a, \vec{\xi}_b, \vec{\xi}_c\right]
\non &=& 
\sum_a \kappa\left[\vec{\xi}_a, \vec{\xi}_b, \vec{\xi}_c\right]q_{a, {\rm lin}}
=\sum_a \kappa_{abc}q_{a, \rm lin}. 
\eea
 Our definition of $\Gamma_{bc}$ (eq. [\ref{eq:Gamma_bc}]) would thus involve a sum over all parents, including the low-order modes that comprise the equilibrium tide and the high-order modes that comprise the dynamical tide (see \S~\ref{sec:linear}). In principle, we could calculate $\Gamma_{bc}$ by numerically summing the product of each parent's $\kappa_{abc}$ and $q_{a, {\rm lin}}$ (eqs. [\ref{eq:kappa}, \ref{eq:qa_lin}]). However, in practice our calculations of individual $\kappa_{abc}$ are only accurate to a part in $\sim 10^3$ (even after considerable effort to minimize the numerical error; see \S~\ref{sec:app:kappa}). Since $U_{bc}$ and the equilibrium tide part of $2K_{bc}$ ($=2\sum_a \kappa_{abc} q_{a, \rm eq}$; see eq. [\ref{eq:qa_eq}]) cancel to a part in $\sim10^5$, we cannot accurately calculate the equilibrium tide part of $\Gamma_{bc}$ by a term-by-term sum over modes.\footnote{We have verified that they cancel to the precision of our sum over modes.} We find that the only way to accurately calculate this part of $\Gamma_{bc}$ is to use our method 2 formulation, in which we use the full solution to the inhomogeneous linear equations $\vec{\xi}_{\rm lin}$ rather than its mode decomposition $\vec{\xi}_{\rm lin}=\sum q_{a, {\rm lin}} \vec{\xi}_a$.
 
\subsection{Dynamical tide driving rate}
\label{sec:dyn_tide_driving_rate}
Unlike the shear of the equilibrium tide, which peaks at $r\sim R$, the shear of dynamical tide peaks in the core (since $d\xi_r/dr\simeq k_r \xi_r \propto r^{-2}$). 
The driving of daughters by the dynamical tide therefore occurs primarily near the dynamical tide's inner turning point $r_1$. If we define the nonlinearity parameter of the dynamical tide $A\equiv \max(k_r \xi_r)\propto r_1^{-2}$ (see \citealt{Goodman:98, Ogilvie:07}), then by equation (\ref{eq:Gamma_global}), the global driving rate of daughters is $\Gamma^{(\rm dyn)}_{bb}\simeq \omega T A r_1 / r_2$, where $r_2\approx r_c\simeq 0.7 R$. The inner turning point is located where $\omega \approx N\simeq 100 \omega_0 r/R$ and thus $r_1/r_2\approx 0.2/n\approx 2\times10^{-4} P_{10}^{-1}$, where $n$ is the radial order of the dynamical tide (this implies $\Gamma^{(\rm dyn)}_{bb} \propto \omega A/n$, a scaling also noted by \citealt{Barker:11b}). The global driving rate of the dynamical tide is therefore
\bea
\label{eq:dyntide_growth_rate_estimate}
\Gamma^{(\rm dyn)}_{bb}\approx 0.01 \left(\frac{M'/M}{10^{-3}}\right) P_{10}^{-11/6} \trm{ yr}^{-1},
\eea
where we took $A\simeq 470(M'/M)P_{10}^{1/6}$ corresponding to a tide raised in a slowly rotating solar-type star by a planet of mass $M'$ \citep{Ogilvie:07}. The ratio $r_1/r_2$ is approximately the ratio of the global driving rate to the maximum local driving rate (for comparison, the global driving rate by the equilibrium tide is $\sim 1\%$ of the maximum local driving rate). The above estimate of  $\Gamma^{(\rm dyn)}_{bb}$ agrees well with the more exact value found by multiplying our numerical integration of $\kappa_{abc}$ with $E_{\rm dyn}^{1/2}$ (see also eqs. [\ref{eq:kappa_analytic}] and [\ref{eq:Ealindyn}]); it also agrees with the rate found by \citet{Barker:11b}. At $P=10\trm{ days}$, the driving rate by the dynamical tide is $\sim 10^5$ times faster than the driving rate by the equilibrium tide (eq. [\ref{eq:eqtide_growth_rate_estimate}]). 

The above estimate assumes that only a single daughter pair is excited and thus ignores the possibility of collective driving. The collective driving rate due to the dynamical tide is $\approx N\Gamma_{bb}^{(\rm dyn)}$. In Appendix \ref{sec:kappa_dyn} we show that $N\approx n$ (see also \S~\ref{sec:dyntide_collective}). Thus, $N\Gamma_{bb}^{(\rm dyn)}\approx 0.2\omega T A$, i.e., the collective global driving rate is approximately equal to the maximum local driving rate of individual daughter pairs ${\Gamma}_{bb}^{(\rm local, max)}=\omega T A$.

\section{Illustration of nonlinear instabilities}
\label{sec:nonlinear}

In this section we consider simple coupled networks in order to illustrate how nonlinear interactions can redistribute the energy of excited modes.  In the first three subsections we give examples of parametrically unstable systems (i.e., LNC; see \S~\ref{sec:nonlinear_tide}) and in the last subsection we give an example of inhomogeneous driving (LLC). More specifically, in \S~\ref{sec:simple_3wave} we consider a parametrically unstable three-wave system consisting of a linearly resonant parent $a$ coupled to a pair of daughter modes $b,c$ with frequency $\simeq \omega_a/2$. We assume that this is the only source of daughter driving (i.e., we set $U_{bc}=0$). Because the dynamical tide is comprised of linearly resonant parents (\S~\ref{sec:linear}), this example helps illustrate the parametric instability of the dynamical tide discussed in \S~\ref{sec:dyntide}. In \S~\ref{sec:simple_nltide} we consider the same three-wave system as in \S~\ref{sec:simple_3wave} but include nonlinear tidal driving of the daughters ($U_{bc}\neq0$). Because the equilibrium tide drives daughters at a rate that depends on an effective $U_{bc}$ (\S~\ref{sec:driving_rate}), this example helps illustrate the parametric instability of the equilibrium tide discussed in \S~\ref{sec:eqtide}. In \S~\ref{sec:collective_instability} we show an example of a collectively unstable system in which a single linearly resonant parent is coupled to $N$ daughters all with frequency $\simeq \omega_a/2$. We show that the collective instability can lead to the very rapid growth of a large number of daughters and thereby dominate the mode dynamics. Finally, in \S~\ref{sec:simple_inhomog}, we consider a linearly resonant parent $a$ coupled to itself and a pair of daughter waves with frequency $\gg\omega_a$. This example helps illustrate nonlinear inhomogeneous driving by the linear tide discussed in \S\S~\ref{sec:nonlinear_tide} and \ref{sec:inhomog_driving}.

\subsection{Three-wave parametric instability}
\label{sec:simple_3wave}

\begin{figure}
\epsscale{1.1}
\plotone{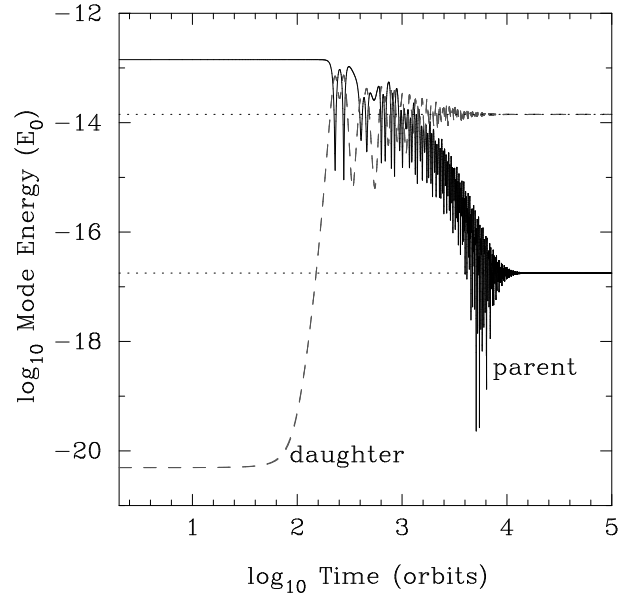}
\caption{Effect of three-wave coupling on a linearly resonant g-mode
  {\em without} nonlinear tidal driving ($U_{ab}=0$). The parent
  (\emph{solid black line}) and daughter modes (\emph{dashed gray
    line}; since they are similar, only one of the daughters is shown) are initially at the linear tide solution
  (eq. [\ref{eq:Ealin}]). For the parameters of the system we use $\ell_a=\ell_b=\ell_c=2$, $\Gamma_{bc}=\kappa_{abc} E_{a, {\rm lin}}^{1/2}\Omega=10^{-2}\Omega$ and $\gamma_a=\gamma_b=\gamma_c=\Delta_{bc}=10^{-4}\Omega$.  These parameters are similar to those of a solar binary. The linear solution is unstable and the daughters initially grow on a timescale of $\sim 1/ \Gamma_{bc}\sim 100$ orbits. After
  $\sim10^4$ orbits the system reaches a new steady state nonlinear
  equilibrium (\emph{dotted lines}; see Appendix \ref{sec:app:stability_3mode}). }
\label{fig:simple_3wave}
\end{figure}

The three-wave coupling considered here involves a linearly resonant parent mode $a$ coupled to two linearly driven daughter modes $b,c$ of approximately half the parent's frequency (this is sometimes referred to as the parametric subharmonic instability; see \citealt{Muller:86}). We artificially set the nonlinear tide coefficient to zero for all three modes ($U_{ij}=0$ for $\{i,j\}=\{a,b,c\}$). The amplitude equation for the parent is thus
\bea
\dot{q}_a +(i\omega_a +\gamma_a) q_a &=&  i\omega_a\left[U_a(t)+2\kappa_{abc}q_b^\ast q_c^\ast\right] 
\eea
and similarly for the daughters. We see by equations (\ref{eq:Gamma_bc}) and (\ref{eq:stability_3wave}) that the daughters are unstable if the parent's linear energy $E_{a, \rm lin}=|Q_{a, \rm lin}|^2$ is above the parametric threshold
\bea
\label{eq:Eth}
E_{\rm th}&\simeq&\frac{\gamma_b \gamma_c}{4\kappa_{abc}^2 \omega_b \omega_c}\left[1+\frac{\Delta_{bc}^2}{(\gamma_b+\gamma_c)^2}\right] 
\non &\approx& 10^{-20}\left(\frac{\ell_b \ell_c P_{\rm orb}}{10\trm{ days}}\right)^2
 \left[1+\frac{\Delta_{bc}^2}{(\gamma_b+\gamma_c)^2}\right]E_0
\eea
where in the second line we assumed $P_b\simeq P_c\simeq P_{\rm orb}$ and plugged in the analytic expressions (\ref{eq:linear_gamma_aprox}) and
(\ref{eq:kappa_analytic}) for $\gamma_{b,c}$ and $\kappa_{abc}$. 

In Figure \ref{fig:simple_3wave} we show the evolution of such a parametrically unstable three mode system. The modes are initialized at their linear energy (eq. [\ref{eq:Ealin}]). The linear solution is unstable and the daughters grow at a rate $\Gamma_{bc}\approx \kappa_{abc} E_{a, {\rm lin}}^{1/2}\Omega$. As we show in Appendix \ref{sec:app:stability_3mode}, this system eventually settles into a stable equilibrium with energies approximately given by $E_a\simeq E_{\rm th}$, $E_b\simeq (\gamma_c \omega_b/\gamma_b \omega_c)^{1/2}|U_a/2\kappa_{abc}|$, and $E_c = (\gamma_b\omega_c/\gamma_c\omega_b)E_b$.

\subsection{Nonlinear tidal driving}
\label{sec:simple_nltide}

\begin{figure}
\epsscale{1.1}
\plotone{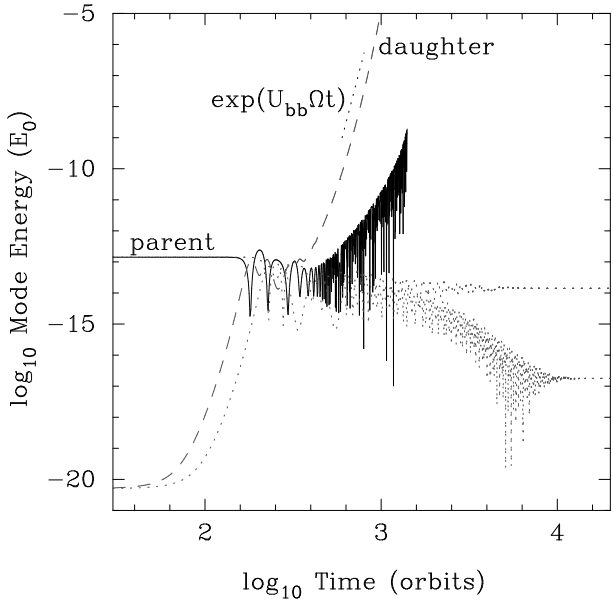}
\caption{Same three mode system as Figure \ref{fig:simple_3wave} but now including self-coupled nonlinear tidal driving of the daughters. We use $U_{bb}=U_{cc}=\kappa_{abc} E_{a,{\rm lin}}^{1/2}/4=2.5\times10^{-3}$, an artificially large value for a solar binary. For comparison, the thin dotted lines show the parent and daughter of Figure \ref{fig:simple_3wave} in which $U_{bb}=U_{cc}=U_{bc}=0$. The system is unstable and parametric coupling to the parent cannot saturate the exponential growth. In reality, the equilibrium tide is only unstable to parametric resonance in solar binaries with $P_{\rm orb}\la 2-5\trm{ days}$ (\S~\ref{sec:eqtide}).}
\label{fig:simple_nltide}
\end{figure}

Just as modes with frequencies $\simeq \omega$ are {\em linearly} resonant with the tide through the $U_a$ term in  equation (\ref{eq:modeampeqn}), modes with frequencies $\simeq \omega/2$ are {\em nonlinearly} resonant with the tide through the $U_{ab}$ term. The daughters in
the previous subsection are therefore subject to nonlinear tidal driving in
addition to three-wave coupling with the linearly resonant parents.

In Figure \ref{fig:simple_nltide} we show the evolution of the same three mode system as Figure \ref{fig:simple_3wave} but now including nonlinear tidal driving of the daughters. Such driving is analogous to driving by the equilibrium tide if it were to be parametrically unstable. In order to contrast the driving by the nonlinear tide with that by three wave coupling, we chose the ratio of their driving rates to be $1/4$, setting $U_{bc}=U_{bb}=U_{cc}=\kappa_{abc} E_{a,{\rm lin}}^{1/2}/4=2.5\times10^{-3}$ in the daughters' amplitude equation. Our choice of $J_{bc\ell m}$ in Figure \ref{fig:simple_nltide} is a factor of $\sim 10^3$ larger than the true effective value of the equilibrium tide coupling coefficient $\kappa_{bc}^{(\rm eq)}$  for solar binaries  (see \S~\ref{sec:driving_rate} and eq. [\ref{eq:kappa_eq}]). We use an artificially large value in order to illustrate the nature of the nonlinear instability. 

 Since $U_{bc}<\kappa_{abc} E_{a,{\rm lin}}^{1/2}$, the daughters' driving is initially dominated by three-wave coupling to the parent. There occurs a short-lived nonlinear equilibrium with energies similar to the equilibrium of Figure \ref{fig:simple_3wave}. However, this equilibrium is destabilized by $U_{bc}$ and after $\approx 300$ orbits the nonlinear driving by $U_{bc}$ begins to take over and the daughters grow exponentially at a rate $\Gamma_{bc}\simeq U_{bb} \Omega$. Three-wave coupling to the parent does not saturate the daughters; instead the daughters drag the parent with them and the parent also grows exponentially.

\subsection{Collective instability}
\label{sec:collective_instability}

\begin{figure}
\epsscale{1.1}
\plotone{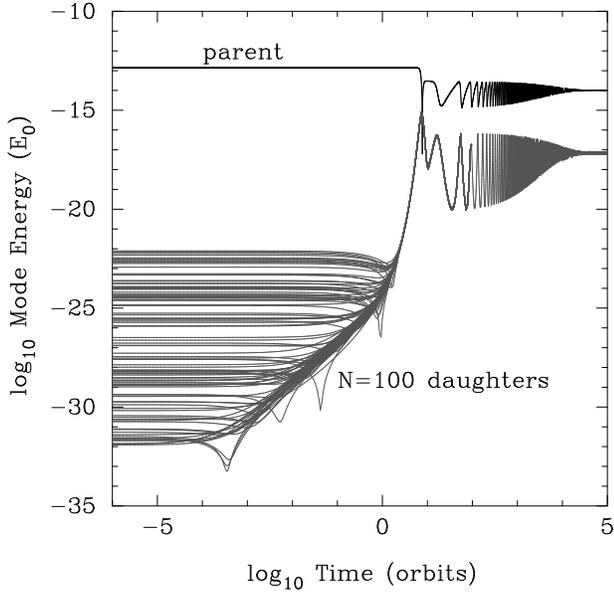}
\caption{Collective parametric driving of a set of $N=100$ coupled daughters. The parent is the same as in Figure \ref{fig:simple_3wave} with $\Gamma_{bc}=0.01\Omega$ while each daughter's damping rate and detuning is increased by a factor of $\simeq 10^3$ to $\gamma_b\simeq\gamma_c\simeq \Delta_{bc}\simeq 0.1\Omega$. The system is collectively unstable even though each daughter pair is stable according to the three-wave instability criteria (eq. [\ref{eq:stability_3wave}]).}
\label{fig:simple_3wave_collective}
\end{figure}

In Figure \ref{fig:simple_3wave_collective} we show a system of $N=100$ daughters that are each coupled to each other and the same linearly driven parent as in Figure \ref{fig:simple_3wave} (a total of $10^4$ couplings). The magnitude of the daughter damping rates and detunings are $\gamma,\Delta_{bc}\simeq 0.1 \Omega$, corresponding to high $\ell$, somewhat resonant daughters.We find that the stability of the daughters is not sensitive to their initial conditions. In order to illustrate this, we assign each daughter a random initial phase and amplitude such that their initial energies vary over 10 orders of magnitude. 

Since $\Gamma_{bc}=\kappa_{abc}E_{a, {\rm lin}}^{1/2}\Omega=0.01\Omega$, on their own each daughter pair would be stable according to the three-wave criteria (eq. [\ref{eq:stability_3wave}]). However, since the set of $N$ daughters satisfies the instability criteria of equation (\ref{eq:stability_criteria}), they are collectively unstable. The daughters with the lowest initial energy grow fastest and within a few orbits the entire set is growing coherently at a rate $N|\Gamma_{bc}|\approx \Omega$. This is a factor of  $\simeq N$ times faster than that of the standard three-wave parametric system considered in \S~\ref{sec:simple_3wave}. Once coherent, the daughters oscillate in unison at a frequency $\omega/2$ and eventually settle into a stable equilibrium with energies that are all equal to within $\approx 30\%$. 

The rapid daughter growth rate implies that the linearly driven parent loses energy at a much faster rate than the parent in the three-wave system of \S~\ref{sec:simple_3wave}. This, in turn, implies that a collectively unstable system can potentially drain energy out of the orbit much faster. As we discuss in \S~\ref{sec:global_vs_local}, this may imply that the driving is in the traveling wave (rather than standing wave) limit for much lower mass companions than indicated by ordinary three-wave coupling. In \S~\ref{sec:dyntide_collective} we show that the dynamical tide due to Jupiter-mass planets in few day orbits is collectively unstable, driving $N\approx10^3 P_{10}$ daughters with $\ell\approx 30 P_{10}^{-1}$ to significant amplitudes within a few orbits.

\subsection{Three-wave inhomogeneous driving}
\label{sec:simple_inhomog}

\begin{figure}
\epsscale{1.1}
\plotone{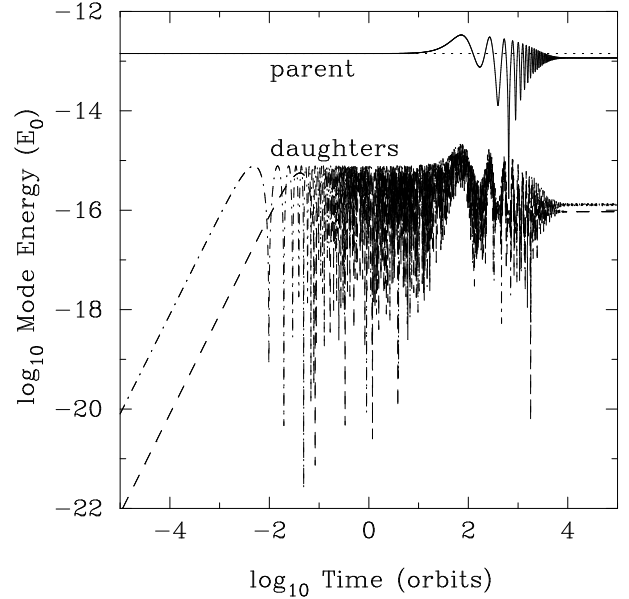}
\caption{Nonlinear inhomogeneous driving of two high-frequency daughter waves by the same linearly resonant parent as in Figure \ref{fig:simple_3wave}. The daughters have frequencies $\omega_b=10\omega_a$ and $\omega_c=100\omega_a$ (\emph{dashed} and \emph{dashed-dotted lines}, respectively) and  $\gamma_a=\gamma_b=\gamma_c=10^{-4}\Omega$. The daughters couple to the self-coupled parent (\emph{solid line}) with coupling coefficients $\kappa_{aab}=\kappa_{aac}=10^5$. The linear tide solution (eq. [\ref{eq:Ealin}]; \emph{dotted line} shows $E_{a, \rm lin}$) is used as the initial condition. The system settles into a nonlinear equilibrium in which the daughter energies are much larger than their linear values while the parent's energy is only slightly smaller than its linear value.}
\label{fig:simple_3wave_inhomog}
\end{figure}

The examples in the last three subsections all illustrate the excitation of modes through the parametric instability (LNC). As described in \S~\ref{sec:nonlinear_tide}, the linear tide can also excite modes through nonlinear inhomogeneous driving (LLC). In terms of the $r_b$ version of the amplitude equation (eq. [\ref{eq:ra_amp_eqn}]), nonlinear inhomogeneous driving of a mode $b$ arises from the coupling coefficient $V_b+K_b$. This coefficient accounts for the full linear tide and thus inherently includes a sum over all linear parent modes. Following the simple examples of the previous  subsections, here we instead consider inhomogeneous driving by just a single linearly resonant parent mode $a$. We couple this parent to itself and two high-frequency, non-resonant, daughter modes $b,c$. In terms of the $q_a$ version of the the amplitude equation (eq. [\ref{eq:modeampeqn}]), we have
\bea
\dot{q}_a+(i\omega_a + \gamma_a)q_a &=& i\omega_a \left[U_a(t)+2 \left(\kappa_{aab}q_b^\ast+\kappa_{aac} q_c^\ast\right)q_a^\ast \right]\non
\dot{q}_b+(i\omega_b + \gamma_b)q_b &=& i\omega_b \left[U_b(t)+\kappa_{aab}(q_a^\ast)^2\right]\non
\dot{q}_c+(i\omega_c + \gamma_c)q_c &=& i\omega_c \left[U_c(t)+\kappa_{aac}(q_a^\ast)^2\right].
\eea
We show an example of such a system in Figure \ref{fig:simple_3wave_inhomog}. Unlike the parametric instability considered in \S~\ref{sec:simple_3wave}, which is only unstable if $E_{a, \rm lin} > E_{\rm th}$ (eq. [\ref{eq:Eth}]), the linear solution of this system is always invalid because the parent appears as an inhomogeneous driving term in the daughter amplitude equations. Thus, just as the steady state solution to the linear equation for high-frequency modes driven at a frequency $\omega\ll\omega_a$ is $q_{b, \rm lin}\approx U_b e^{-i\omega t}$ (eq.  [\ref{eq:qa_lin}] assuming $\gamma_b\ll \omega_b$), the daughter steady state solution here is $q_b \approx  \kappa_{aab} E_a e^{2i\omega t}$ (assuming $ \kappa_{aab} E_a \gg U_b$).  For Figure \ref{fig:simple_3wave_inhomog} we chose coupling coefficients $\kappa_{aab}=\kappa_{aac} \ll E_{a, \rm lin}^{-1/2}$; the parent's steady state energy is therefore only slightly smaller than $E_{a, \rm lin}$. Depending on the tidal factor $\varepsilon$, the actual inhomogeneous driving coefficient $V_b+K_b$ can be so large that the parent's steady state energy is very different from $E_{a, \rm lin}$ (see \S~\ref{sec:inhomog_driving}). In that case, the simple view of ``linear-like" driving no longer holds and the other forms of three-wave coupling (LNC and NNC) can become important.

\section{Parametric Instability of the Dynamical Tide}
\label{sec:dyntide}

As discussed in \S~\ref{sec:linear}, the linear tide can be decomposed
into a dynamical tide (high-order modes resonant with the tide) and an
equilibrium tide (low-order, off-resonant modes with large overlap
integrals $I_{a\ell m}$). In this section we show that for the close
binary systems discussed in \S~\ref{sec:intro}, the dynamical tide is parametrically
unstable to nonlinear three-wave interactions. We first  
(\S~\ref{sec:dyntide_criteria}) assume that each daughter is coupled to the dynamical tide and only one other daughter ($N\le2$) and determine the parameter space over which the daughters are unstable.  We then (\S~\ref{sec:dyntide_collective}) allow each daughter to couple to the dynamical tide and many other daughters ($N\gg2$). We show that these collections of modes are collectively unstable and have growth rates that can be orders of magnitude larger than the $N\le2$ systems.

Since the dynamical tide is comprised of several parents on either side of the linear resonance, in principle, one should allow each daughter pair to couple to more than one parent. However, we find that the most linearly resonant parent dominates the dynamics and for simplicity we only show results for driving by this single parent. Furthermore, although we present results for standing waves, we will show in \S~\ref{sec:global_vs_local} that the local nonlinear interaction rates
are so fast in both solar-type binaries and hot Jupiter systems that the global standing wave
assumption may not be valid; one may then have to work instead in the traveling
wave limit.

\begin{figure}
\epsscale{1.1}
\plotone{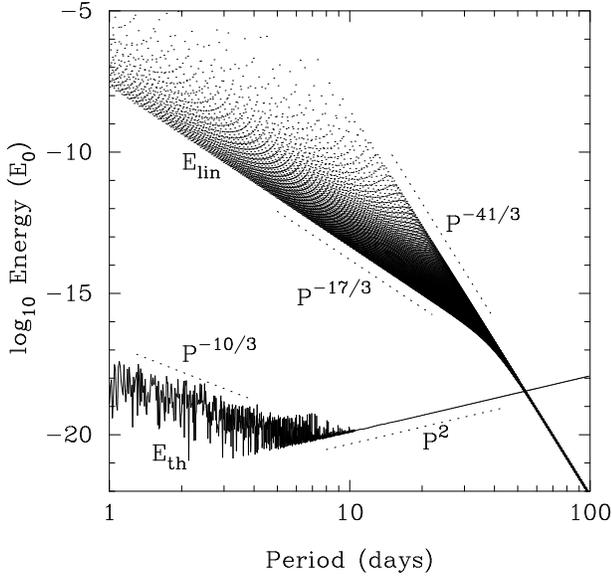}
\caption{Linear energy $E_{\rm lin}$ in the modes most resonant with the $\ell=k=m=2$ harmonic of the dynamical tide and those modes' minimum nonlinear energy threshold $E_{\rm th}$ for three-mode coupling ($N\le2$). The results are shown as a function of orbital period $P_{\rm orb}$ for $M'=M=M_\odot$. The $E_{\rm lin}$ points are spaced uniformly in $P_{\rm orb}$; the density of points indicates the likelihood distribution of $E_{\rm lin}$ near a given $P_{\rm orb}$. The dotted lines indicate how the various quantities scale with $P_{\rm orb}$. }
\label{fig:ElinEth}
\end{figure}

\subsection{Dynamical tide instability for $N\le2$ daughters}
\label{sec:dyntide_criteria}

For a three-mode system, the dynamical tide is unstable if the linear energy of the resonant modes $E_{\rm lin}$ (eq. [\ref{eq:Ealin}] for modes with small $\Delta_a$) exceeds the parametric threshold $E_{\rm th}$ (eq. [\ref{eq:Eth}]) for three-mode coupling to a daughter pair. In Figure \ref{fig:ElinEth} we show $E_{\rm lin}$ and the minimum $E_{\rm th}$ as a function of $P_{\rm orb}$ for $M'=M=M_\odot$. To find the minimum $E_{\rm th}$, we use the analytical expressions for $\kappa$, $\gamma$, etc. derived in Appendix \ref{sec:app:coef_in_amp_eqn} and at each $P_{\rm orb}$ search  the $(\ell, n)$ parameter space of daughter pairs that satisfy momentum conservation and the angular selection rules described in Appendix \ref{sec:properties_of_coefficient}. We find that the dynamical tide in solar binaries is unstable for $P_{\rm orb} \la 40\trm{ days}$ and that $E_{\rm lin} \gg E_{\rm th}$ over much of that range.

\begin{figure}
\epsscale{1.1}
\plotone{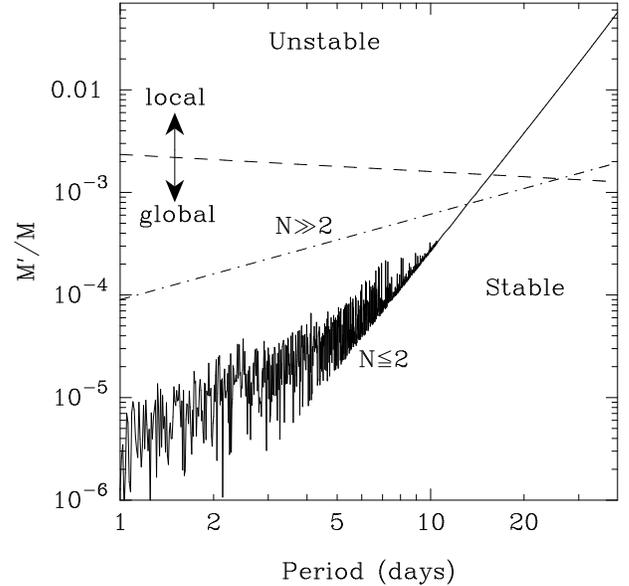}
\caption{Minimum mass of the companion $M'$ for which the linear
  dynamical tide is unstable in solar-type stars $M=M_\odot$ as a
  function of orbital period. We show results for the $\ell=k=m=2$ harmonic assuming circular orbits and a non-rotating primary ($X_{k}^{\ell m}=1$).  The solid line is the minimum companion mass for $N\le2$ while 
  the dashed-dotted line is the minimum companion mass  for the collective instability $N\gg2$ (eq. [\ref{eq:mass_threshold_collective}]). Above the dashed line, the local nonlinear
  interactions are so strong in the stellar core that the daughter standing
  wave approximation is no longer valid (eq. [\ref{eq:local_planet}] for $\ell_b=2$). The collective instability ($N\gg2$) may invalidate the standing wave approximation for even lower companion masses (\S~\ref{sec:global_vs_local}).}
\label{fig:massperiod}
\end{figure}

The linear energy scales with the mass of the companion $M'$ as
$E_{\rm lin}\propto U_a^2\propto [M'/(M+M')]^2$ while $E_{\rm th}$ is
independent of $M'$. The ratio $E_{\rm lin}/E_{\rm th}$ calculated for
solar binaries in Figure \ref{fig:ElinEth} can therefore be used to
solve for the minimum $M'$ for which the dynamical tide is unstable in
solar-type stars. We show this minimum companion mass $M'$ as a function of orbital
period in Figure \ref{fig:massperiod}. We find that for $N\le2$ the dynamical tide is unstable for Jupiter
mass planets $M'= M_J$ out to $P_{\rm orb}\simeq10\trm{ days}$ and for $\sim10$ Earth mass planets out to $P_{\rm orb} \simeq \trm{few days}$. 

The analytic scalings shown as dotted lines in Figure \ref{fig:ElinEth} can be derived using the analytic approximations to $\gamma, \Delta, U_a$ and $\kappa$ given in Appendix \ref{sec:app:coef_in_amp_eqn}. The rapid variation in $E_{\rm th}$ and the minimum $M'$ for $P_{\rm orb}\la 10\trm{ days}$ are due to particular modes coming in and out of resonance. This is especially pronounced at short orbital periods since the frequency spacing of resonant modes is larger at smaller $P_{\rm orb}$. 

We note that to calculate the minimum $M'$ we used the value of $E_{\rm
  lin}$ given by the lower envelope in Figure  \ref{fig:ElinEth} (eq. [\ref{eq:Ealindyn}]). Since this ignores the coincidental possibility of strong resonances, the minimum $M'$ shown is somewhat
conservative.  Furthermore, by approximating the dynamical tide with just the most resonant parent, we have ignored the driving of daughters by the slightly less resonant neighboring parents; we find that including these parents decreases the minimum $M'$ by a factor of order unity.

\subsection{Dynamical tide instability for $N\gg2$ daughters}
\label{sec:dyntide_collective}

In the previous section we assumed that each daughter couples to only one other daughter, in addition to the parent (either $N=1$ daughters for a self-coupled daughter pair or $N=2$). However, since $|\kappa_{abc}|$ is nearly constant for all daughter pairs with $|n_b - n_c| \la n_a$ (see Appendix \ref{sec:properties_of_coefficient}), each daughter in fact couples to $N\approx n_a\simeq 1000 (\ell_a/2) P_{a, 10}$ daughters.  All the modes are unstable if each daughter pair $(b,c)\in N$ satisfies (eq. [\ref{eq:stability_criteria}])
\bea
2N\kappa_{abc}E_{a}^{1/2} \ga \sqrt{\frac{\gamma_b \gamma_c+\Delta_{bc}^2}{\omega_b \omega_c}}.
\label{eq:dyn_tide_collective}
\eea
In order to derive the threshold energy  $E_{\rm th}$ of the system, let the radial order $n_0$ characterize the most nonlinearly resonant daughter at a given $\ell$ and consider the $N\approx n_a$ daughters on either side of $n_0$ that couple equally well to the parent.  Of these $N$ modes, the daughter pair that maximizes the right hand side of equation (\ref{eq:dyn_tide_collective}), and thereby determines the stability of the system, will be the pair that has the largest detuning and damping rate. This is the self-coupled pair $b=c$ with $n_b\simeq n_0+n_a/2$; it has a detuning $|\Delta_{bb}/\omega_b|\simeq n_a/n_0$ and a damping rate $|\gamma_b / \omega_b| \simeq B(n_0+n_a)^2\simeq Bn_0^2$, where $B\simeq 4\times10^{-11} P_{b, 10}$ (see eq. [\ref{eq:linear_gamma_aprox}]) and the second equality is appropriate if $\ell_0\gg \ell_a=2$. Solving for the $n_0$ that minimizes the right hand side of equation (\ref{eq:dyn_tide_collective}), we find
\bea
n_{0, \ast}\simeq\left(\frac{n_a}{B\sqrt{2}}\right)^{1/3}\simeq 1.6\times10^4\left(\frac{\ell_a}{2}\right)^{1/3},
\eea
independent of period. This corresponds to a daughter with $\ell=\ell_{0, \ast}\simeq (n_{0, \ast} / 500) P_{10, b}^{-1}\simeq32 P_{10,b}^{-1}$.  Since $\ell_{0, \ast}\gg\ell_a=2$, daughters with $(\ell,n)=(\ell_{0, \ast}\pm2, n_{0, \ast}\pm n_a)$  have nearly the same detuning  and $\kappa_{abc}$ as those with $(\ell,n)=(\ell_{0, \ast}, n_{0, \ast}\pm n_a)$. Thus a more accurate estimate is $N\approx 3 n_a$ rather than $N\approx n_a$. The minimum threshold energy is then $E_{\rm th}=(\kappa_{abc}n_{0, \ast})^{-2}/24$. For a resonant parent at the linear energy $E_{a, {\rm lin}}$ of the $\ell=k=m=2$ harmonic of the tide, this corresponds to a threshold mass ratio
\bea
\frac{M'}{M}\ga 9\times10^{-5} \left(\frac{P_{\rm orb}}{1 \trm{ day}}\right)^{5/6},
\label{eq:mass_threshold_collective}
\eea
where we used equation (\ref{eq:kappa_analytic}) and assumed the primary spins slowly relative to the orbit. This minimum mass ratio for collective parametric driving of daughters is shown in Figure \ref{fig:massperiod}. Note that this threshold is somewhat conservative in that we again used equation (\ref{eq:Ealindyn}) for the linear energy and thus ignore the coincidental possibility of strong resonances. 

At mass ratios just slightly above threshold (by a factor of two, say), collectively unstable systems have growth rates $\approx 2N\kappa_{abc}E_a^{1/2}\Omega$. Their growth rates are therefore $N\approx10^3 P_{10}$ times larger than $N\le2$ systems. Thus, even though collective systems have a somewhat higher threshold (see Fig. \ref{fig:massperiod}), they probably dominate the mode dynamics whenever they are unstable.

\begin{figure}
\epsscale{1.1}
\plotone{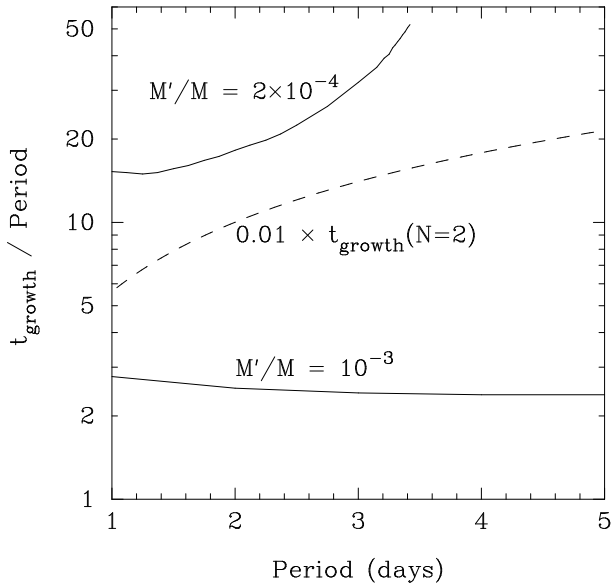}
\caption{Growth time $t_{\rm growth}$, in units of orbital period, of daughters coupled to the linearly resonant parent. The solid lines show $t_{\rm growth}$ for collectively unstable systems $N\gg2$ with $M'/M=2\times10^{-4}$ and $10^{-3}$. The dashed line shows the maximum $t_{\rm growth}$ (multiplied by $0.01$) for an $N\le2$ system with $M'/M=10^{-3}$.}
\label{fig:growth_collective}
\end{figure}

In Figure \ref{fig:growth_collective} we show the growth times $t_{\rm growth}$ of the fastest growing collectively unstable systems for two companion masses; for systems just slightly above threshold, the daughters all grow on timescales of $\la10$ orbits. We determine $t_{\rm growth}$ by numerically solving for the eigenvalues of large networks of daughters (the eigenvalues of matrix $H$ in \S~\ref{sec:stability_analysis}) and using our analytic expressions for $\gamma_a$, $I_{a\ell m}$, and $\kappa_{abc}$ given in Appendix \ref{sec:app:coef_in_amp_eqn}. To find the daughter set with the smallest $t_{\rm growth}$ we search over daughter $\ell$ and vary $N$. For a given $\ell$, we couple the $N$ consecutive daughters on either side of the most resonant daughter to each other and the most resonant parent. We include daughters with $\ell\pm2$ in the set since their coupling also satisfies momentum conservation for $\ell_a=\ell_{\rm tide}=2$.  We find that $t_{\rm growth}$ decreases with increasing $N$ until $N \approx 2 n_a\approx 2\times 10^3 (\ell_a/2) P_{a, 10}$. For $N\ga 2n_a$ a large fraction of the least resonant daughters do not couple to each other since they have $|\Delta n| > n_a$ and thus $\kappa_{abc}\simeq 0$. We find that these additional daughters are stable. 

Estimating $t_{\rm growth}$ using equations (\ref{eq:Ealindyn}) and (\ref{eq:kappa_analytic}), we find $t_{\rm growth} / P_{\rm orb} \approx 3\times10^{-3}(M/M')P_{10}^{-1/6}$. This is in good agreement with the numerical results for mass ratios sufficiently above threshold.

\section{Parametric Instability of the Equilibrium Tide}
\label{sec:eqtide}

In this section we consider the stability of the equilibrium tide to parametric driving. Like the dynamical tide, the equilibrium tide oscillates at the driving frequency $\omega$ and is thus nonlinearly resonant with short wavelength daughters with natural frequencies near $\omega/2$.  Unlike the dynamical tide, however, the amplitude of the equilibrium tide is smallest in the core; at small radii $\xi_{r, {\rm dyn}}\propto r^{-2}$ whereas $\xi_{r, {\rm eq}}\propto r^{\ell-1}$. Since the daughters' eigenfunctions peak in the core ($\xi_r\propto r^{-2}$), they couple much more weakly to the equilibrium tide than the dynamical tide. We will show that the coupling is in fact so weak that the equilibrium tide is only unstable to parametric resonance in solar binaries with $P_{\rm orb}\la 2-5 \trm{ days}$.   

A resonant daughter pair $(b,c)$ is unstable to driving by the $k$-th harmonic of the equilibrium tide if (eq. [\ref{eq:stability_criteria}])
\bea
N\Gamma^{(k, {\rm eq})}_{bc} \ga \sqrt{\gamma_b \gamma_c + \Delta_{bc}^2},
\label{eq:driving_rate_eqtide}
\eea
where (\S~\ref{sec:driving_rate}) 
\bea
\Gamma^{(k, {\rm eq})}_{bc}&=& \sqrt{\omega_b \omega_c} \frac{M'}{M}\sum_{\ell m} W_{\ell m} X_{k}^{\ell m} \left(\frac{R}{a}\right)^{\ell+1}\kappa_{bc}^{(\rm eq)}.
\eea 
In Appendix \ref{sec:properties_of_coefficient} we show that the equilibrium tide coupling coefficient $\kappa_{bc}^{(\rm eq)}$:  (1) is a weak function of daughter $\ell$ and period, (2) has a magnitude of $\approx 0.01-0.05$ for self-coupled modes $b=c$, and (3) decreases rapidly with increasing $|\Delta n|=|n_b-n_c|$.

We now estimate the period out to which the equilibrium tide is unstable to parametric resonance. Since the coupling coefficient peaks strongly for self-coupled modes $b=c$, the daughters are not collectively driven and $N=1$. For self-coupled pairs the detuning is $|\Delta_{bb}/\omega_b|\approx A n_b^{-1}$ and the damping rate is $\gamma_b/\omega_b\simeq Bn_b^2$, where numerically we find an average detuning $A\simeq0.5$ and $B\simeq4\times10^{-11}P_{10,b}$ (eq. [\ref{eq:linear_gamma_aprox}]). Since the coupling coefficient is a weak function of $\ell_b$ and $P_b$, we can solve for the stability threshold by minimizing the right hand side of equation (\ref{eq:driving_rate_eqtide}) with respect to $n_b$. For the representative case of the $\ell=m=k-1=2$ harmonic in a synchronized solar binary ($\Omega\simeq 2\omega_b$), we find that the equilibrium tide is unstable for orbital periods
\bea
P \la 1.2\left(\frac{e\kappa^{(\rm eq)}}{0.01}\right)^{3/7}\trm{ days}.
\eea
There can be brief epochs during which the detuning happens to be much smaller than average, i.e. $A\ll 0.5$. In the zero detuning limit, the stability criteria is determined by the damping rate of the most resonant $\ell_b=1$ daughter and we find
\bea
\label{eq:eqtide_stability_zero_detuning}
P(\Delta_{bb}=0) \la 5.1\left(\frac{e\kappa^{(\rm eq)}}{0.01}\right)^{1/5}\trm{ days}.
\eea
We have assumed in these estimates that the condition $|\Delta n|=0$ implies self-coupling and thus $\ell_b=\ell_c$. In fact, pairs with $\ell_b=\ell_c\pm 2$ and $|\Delta n|=0$ also have $\kappa_{bc}^{(\rm eq)}\approx 0.01-0.05$ for the $\ell=2$ harmonic of the tide. In a numerical search over potential daughter pairs, we find that relaxing this assumption increases the period out to which the equilibrium tide is unstable by $\la 30\%$.

\section{Nonlinear Inhomogeneous Driving}
\label{sec:inhomog_driving}

There are two mechanisms by which the linear tide can drive other modes (see \S~\ref{sec:nonlinear_tide}): parametric driving, which we considered in the previous two sections, and nonlinear inhomogeneous driving, which we consider here. Both forms of driving drain energy from the linear tide and can thus act as sources of enhanced dissipation relative to linear theory. In order to determine the extent to which they  influence the orbital and rotational evolution, we need to solve for the nonlinear equilibrium (i.e., saturation). In general, the nonlinear equilibrium depends on a balance between all three forms of three-wave interactions (LLC, LNC, and NNC in \S~\ref{sec:nonlinear_tide}). While we defer such a calculation to a future paper, in this section we solve for the nonlinear equilibrium in the absence of parametric driving and NNC. Such a simplification is justified only if the linearly excited modes remain near their linear energies despite their nonlinear coupling to other modes.  Although we show that this is not always the case, this calculation nevertheless demonstrates that inhomogeneous driving may be a significant source of nonlinear dissipation.

The steady-state solution to the nonlinear amplitude equation in the absence of parametric driving and NNC (eq. [\ref{eq:ra_amp_eqn}] but ignoring the last two terms on the right hand side) is that of a driven oscillator:
\bea
\label{eq:inhomog_soln}
r_a(t)&=&\sum_{kk'} \frac{\omega_a\left[V_a+K_a\right]^{(k+k')}}{\omega_a-(k+k')\Omega-i\gamma_a}e^{-i(k+k')\Omega t},
\eea
where
\bea
\left[V_a+K_a\right]^{(k+k')}&=&\left(\frac{M'}{M}\right)^2 \sum_{\ell m}\sum_{\ell' m'}W_{\ell m}W_{\ell' m'} X_{k}^{\ell m} X_{k'}^{\ell' m'}
\non &&\hspace{-2.5cm} \times \left( \frac{R}{a} \right)^{\ell+\ell'+2}\left[\kappa_a^{(\rm eq-eq)}+\kappa_a^{(\rm dyn-dyn)}+\kappa_a^{(\rm eq-dyn)}\right].
\eea
The coefficients $\kappa_a^{(\rm eq-eq)}$, $\kappa_a^{(\rm dyn-dyn)}$, and $\kappa_a^{(\rm eq-dyn)}$ are defined in Appendix \ref{sec:app:alternative_eom} and their properties described in Appendix \ref{sec:properties_of_coefficient}. This solution (eq. [\ref{eq:inhomog_soln}]), which is analogous to that of the linear amplitude equation (eq. [\ref{eq:linear_amp_eqn}]), has contributions from the three distinct forms of driving: equilibrium tide-equilibrium tide coupling $\kappa_a^{(\rm eq-eq)}$, dynamical tide-dynamical tide coupling $\kappa_a^{(\rm dyn-dyn)}$, and equilibrium tide-dynamical tide coupling $\kappa_a^{(\rm eq-dyn)}$. We first consider the properties of the solution for modes with frequency $\omega_a\la\omega_0$ (high-order g-modes to low-order p-modes) and then for modes with frequency $\omega_a \gg \omega_0$ (high-order p-modes).

\subsection{$\omega_a \la \omega_0$}

\begin{figure}
\epsscale{1.1}
\plotone{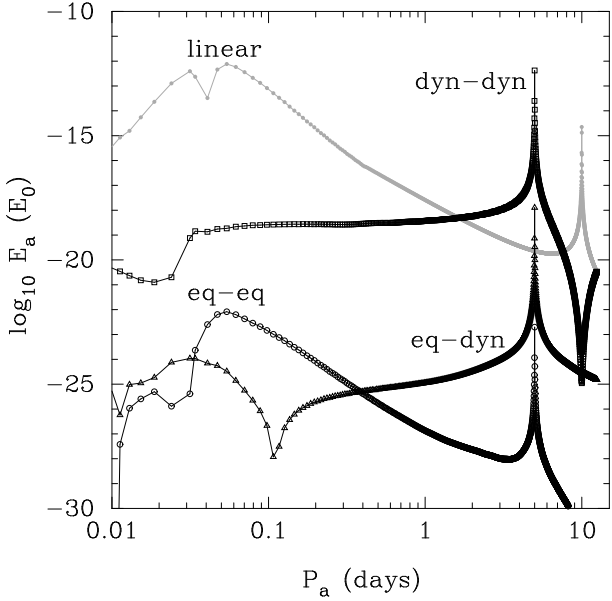}
\caption{Orbit-averaged nonlinear energy $E_{a, nl}$ as a function of mode period $P_a$ accounting only for the inhomogeneous driving terms in the nonlinear amplitude equation (\ref{eq:ra_amp_eqn}). The parameters are $\ell=\ell'=m=m'=k-1=k'-1=2$, $M'=M=M_\odot$, $e=0.1$, and $P_{\rm orb}=10\trm{ days}$. The three contributions to the driving---$\kappa_a^{(\rm eq-eq)}$, $\kappa_a^{(\rm dyn-dyn)}$, and $\kappa_a^{(\rm eq-dyn)}$---are each shown as separate curves. For comparison, the linear energy $E_{a, \rm lin}$ is also shown (eq. [\ref{eq:Ealin}]; cf. Fig. \ref{fig:ElinP}).}
\label{fig:EinhomogP}
\end{figure}

Figure \ref{fig:EinhomogP} shows the orbit-averaged nonlinear energy\footnote{The total energy $E_{a}=|q_a|^2=|q_{a, \rm lin}+r_a|^2$ contains a cross term involving the product of $q_{a, \rm lin}$ and $r_a$. However, since we assume in Figure \ref{fig:EinhomogP} that $r_a$ and $q_{a, \rm lin}$ oscillate at different harmonics of the orbit, the orbit-averaged cross term is zero. In cases where $q_{a, \rm lin}$ and $r_a$ oscillate at the same frequency (i.e., for particular values of $k, m$ etc.), the cross term is constant and the time-averaged total energy is not simply $|q_{a, \rm lin}|^2+|r_a|^2$ but can be larger or smaller depending on the magnitude and phase of $q_{a, \rm lin}$ and $r_a$.} $E_{a, \rm nl}\equiv|r_a|^2$ as a function of mode period for each form of inhomogeneous driving, as compared to the linear solution $E_{a, \rm lin}$ (eq. [\ref{eq:Ealin}] and Fig. \ref{fig:ElinP}). We see that for $\kappa_a^{(\rm eq-eq)}$ coupling and $\kappa_a^{(\rm eq-dyn)}$ coupling,  $E_{a, \rm nl} \ll E_{a, \rm lin}$  over nearly the entire range of modes shown (low-order p-modes to $n\approx10^3$ g-modes). This suggests that for these modes, the energy dissipation from $\kappa_a^{(\rm eq-eq)}$ coupling and $\kappa_a^{(\rm eq-dyn)}$ coupling is insignificant compared to that due to the linear tide. The only exception is the nonlinear resonance at $P_a\simeq P_{\rm orb}/(k+k')$ due to $\kappa_a^{(\rm eq-dyn)}$ coupling. However, since the energy in this resonance is below that of the dynamical tide (the linear resonance at $P_a\simeq P_{\rm orb}/k$), it cannot be a significant additional source of dissipation. 

For $\kappa_a^{(\rm dyn-dyn)}$ coupling, by contrast, Figure \ref{fig:EinhomogP} shows that in the case of solar binaries there are many modes for which $E_{a, \rm nl}\gg E_{a, \rm lin}$. Furthermore, the energy in the nonlinear resonance is $\sim100$ times larger than the energy in the linear resonance.\footnote{Resonances that are coincidentally better than average will shift the relative magnitude of the $E_{a, \rm nl}$ and $E_{a, \rm lin}$ peaks.  In Figure \ref{fig:EinhomogP} the linearly and nonlinearly resonant modes both have resonances that are typical.} This suggests that $\kappa_a^{(\rm dyn-dyn)}$ coupling may be an important source of nonlinear dissipation for the dynamical tide, possibly even more important than parametric coupling to daughter pairs.

It is important to note that $\kappa_a^{(\rm dyn-dyn)}$ coupling cannot lead to enhanced dissipation relative to the mechanical power carried by the dynamical tide (i.e., relative to the upper bound calculated by \citealt{Goodman:98}). This is because $\kappa_a^{(\rm dyn-dyn)}$ coupling always occurs below the radiative-convective interface (Appendix \ref{sec:properties_of_coefficient}) and thus below the region where the dynamical tide is launched. This also implies that if the dynamical tide breaks in the core, then the nonlinearly resonant modes cannot actually reach the large energies shown in Figure \ref{fig:EinhomogP}; the other forms of three-wave interactions (LNC and NNC) must become important well before the modes reach these energies.

 Since $E_{a, \rm nl}\propto \varepsilon^4$ while $E_{a, \rm lin}\propto \varepsilon^2$, we find that in the case of hot Jupiter systems, $E_{a, \rm nl}\ll E_{a, \rm lin}$ for all modes. We therefore do not expect inhomogeneous driving to be an important source of dissipation in planetary systems. This difference between planetary and stellar companions highlights why it is probably very misleading to extrapolate tidal dissipation parameters inferred from observations of one set of systems to a very different set of systems.

\subsection{$\omega_a \gg \omega_0$}

In \S~\ref{sec:inhomog_coef} we found that  $\kappa_a^{(\rm eq-dyn)}$  and $\kappa_a^{(\rm dyn-dyn)}$ can be significant for high-order p-modes (see the right panel of Fig. \ref{fig:linlin_coef}). Since the frequency of these modes is above the acoustic cutoff  of the solar atmosphere ($\approx 60\omega_0$), they do not reflect at the solar surface and cannot therefore form standing waves. 
Equation (\ref{eq:inhomog_soln}), which assumes the driven modes are standing waves, does not therefore describe the driving of these high-order p-modes. Instead, one must treat the driving in the local limit; presumably the p-modes cannot reach amplitudes anywhere near those of equation (\ref{eq:inhomog_soln}) since their group travel time across the driving region is much smaller than the driving period. We defer the calculation of p-mode driving to future work.

\section{Local Nonlinear Interactions}

\label{sec:global_vs_local}

To this point we have assumed that the perturbations are all global standing waves. However, a perturbation can only be treated as a global standing wave if its growth and damping times are longer than its group travel time across the star. 
In this section we show that in nonlinear theory this
is not necessarily the case. The nonlinear growth rate of daughters in the core of the star
due to driving by the dynamical tide can be faster than the rate at which the daughters cross the interaction region. We specialize to the case of dynamical tide driving since equilibrium tide driving, when present, is in the the standing wave limit (at least for solar type stars).  

 The stability calculations, which are the focus of this paper, are independent of the uncertainty of traveling versus standing waves that is present in full nonlinear theory. This is because the daughter growth rates are infinitesimal at threshold, and thus the growth times are always longer than the group travel time very near threshold. However, the breakdown of the standing wave approximation does have implications for the saturation of the nonlinear perturbations since it implies that the excited waves can grow and damp locally. This, in turn, can affect the rate at which energy and angular momentum are taken out of the orbit. 

Consider an inward propagating traveling wave $a$ launched at the radiative-convective boundary $r_c$ by the linear tide. As it approaches the core, this parent wave excites
daughter waves through three-wave interactions. Due to the sharp
increase in $d\kappa/d\ln r$ in the core (see Figs. \ref{fig:kappa}
and \ref{fig:xir_xih}), the local nonlinear growth rate
$\Gamma_{bb}(r)$ of a self-coupled daughter wave $b$ increases rapidly as the parent
approaches $r\simeq0$ . If $\Gamma_{bb}(r) t_{r, b}(r)\ga 1$, where
\bea 
\label{eq:tgroup_local}
t_{r,b}(r) \approx\frac{r}{v_{r,b}(r)}\approx \frac{\Lambda_b N(r)}{\omega_b^2} 
\eea
 is the time it takes a daughter with group
velocity $v_{r,b}\approx \omega_b/k_{r,b}$ to traverse a region of size $r$ and $\Lambda_b^2=\ell_b(\ell_b+1)$,
then the standing wave approximation breaks down and the daughter must be treated as a traveling wave.  
In Appendix \ref{sec:app:local_growth_rate} we show that the local
nonlinear growth rate of a self-coupled daughter $b$ due to three-wave
interactions with a parent $a$ is 
\bea 
\label{eq:Gammabb}
\Gamma_{bb}(r)&\simeq&
2\omega_b T |q k_{r} \xi_{r}|_a  
\non&\simeq& 
16\omega_b\left(\frac{T}{0.2}\right)\left(\frac{r}{R}\right)^{-2}\left(\frac{E_a}{E_0}\right)^{1/2}
\Lambda_a P_{a,10} 
\eea
where $q_a$ is the parent amplitude, $T$ is the angular integral
given by equation (\ref{eq:Tang}), and the expression in the second
line is appropriate for $r\la 0.05R$ (see eq. [\ref{eq:xirWKB}]). The
nonlinear interactions are strongest near the parent's inner turning
point $r_{\rm min}$ where $N(r)=\omega_a\simeq2\omega_b$; the radial
group travel time of the daughter across this region is $t_{r,b}(r_{\rm
  min}) \approx 2\Lambda_b/\omega_b$ and therefore at $r_{\rm min}$
\bea 
\left.\Gamma_{bb}t_{r,b}\right|_{r_{\rm min}} \approx 4 \Lambda_b T |q k_{r}
\xi_{r}|_a. 
\label{eq:local}
\eea 
In a solar model $dN/dr\simeq
98\omega_0$ at $r\la0.05R$ and thus $r_{\rm min}/R \approx
1.3\times10^{-4} P_{a,10}^{-1}$. For the  $\ell_a=2$ 
parent most resonant with the linear tide ($q_a=q_{a, \rm dyn}$; eq. [\ref{eq:qa_dyn}]), 
we find that for a synchronous solar binary
\bea 
\Gamma_{bb} t_{r,b}|_{r_{\rm min}} 
\approx 2\times10^3\Lambda_b e P_{10}^{1/6}
\eea
and the standing wave approximation is invalid unless
$\Lambda_b e\la 10^{-3}$. For a slowly rotating star orbited by a planet we find
\bea
\label{eq:local_planet}
\Gamma_{bb} t_{r,b}|_{r_{\rm min}} 
\approx 0.3\Lambda_b \left(\frac{M'/M}{10^{-3}}\right) P_{10}^{1/6}.
\eea
and the standing wave approximation is valid for $M'/M \la 10^{-3}/\Lambda_b$. 
The dashed line in Figure \ref{fig:massperiod} shows this condition for the case $\ell_b=2$.  

\citet{Goodman:98} computed $|q k_r \xi_r|_a$ as a local measure of
the nonlinearity of waves excited by the linear tide (see also
\citealt{Ogilvie:07}). When $|q k_r \xi_r|_a\ga 1$, the parent wave
overturns the stratification during part of its cycle. \citet{Barker:10} (see also \citealt{Barker:11}) carried out numerical simulations of gravity waves approaching the core of a solar type star. They found that when the overturn criteria of \citet{Goodman:98} is satisfied, the incoming parent wave is fully absorbed in the core and therefore deposits all of its energy there; if it is not satisfied, the parent reflects at its inner turning point and travels back out to the stellar surface virtually undamped.  Goodman \& Dickson's overturn criteria is therefore equivalent to a traveling wave condition for the \emph{parent}.  Here we have
shown that when the parent's amplitude is a factor of $4\Lambda_b T$
smaller than this overturning amplitude, the \emph{daughters} it excites are in the traveling wave limit.

 For $\ell_b\la2$ our daughter traveling wave condition is very similar to Goodman \& Dickson's parent 
traveling wave condition. For $\ell_b \gg 1$, equation (\ref{eq:local}) suggests that the daughter
standing wave approximation may fail even when $|qk_r \xi_r|_a \ll 1$,
i.e, even when the parent is far from overturning the stratification. 
We also showed in \S~\ref{sec:dyntide_collective} that there are collectively unstable daughters at
mass ratios well below the overturn criteria (Figure \ref{fig:massperiod}) with growth rates that are $\approx N/2\gg1$ times larger than that of three wave systems (that of eq. [\ref{eq:Gammabb}]). The cases $\Lambda_b\gg1$ and $N\gg2$ both therefore suggest that the daughter traveling wave limit may
extend to significantly lower planetary masses than indicated by the dashed
line in Figure \ref{fig:massperiod}. However, it is not clear whether this means that the parent is also in the traveling wave limit, and thus fully absorbed, for these lower masses.   Assessing whether this is 
the case will require a better understanding of the daughter driving and saturation 
in the core of solar-type stars. 

We note that both cases ($\Lambda_b\gg1$ and $N\gg2$) involve the excitation of a potentially large number of high $\ell$ resonant daughters (and since $\omega\propto \ell/n$, very high $n$ daughters). Capturing these interactions numerically therefore requires very high spatial resolution. This may explain why \citet{Barker:10} do not seem to observe either of these effects in their numerical simulations.

\section{Summary and Conclusions}
\label{sec:conclusions}

This paper is an initial investigation into the importance of
nonlinear fluid dynamics for the tidal evolution of close binary
systems. We derive a formalism for including nonlinear interactions in
tidal theory and describe the physical effects that arise from these
nonlinearities.  While the ultimate goal is to understand how
nonlinear effects may alter the rate of circularization and
synchronization in binaries, such calculations are deferred to a
future paper. In this paper, we instead focus on determining the
conditions under which the standard linear theory approximation is
(in)valid. We present detailed results for the tidal forcing of a
sun-like star by a stellar or planetary companion, but the formalism
we derive is more general and is applicable to nonlinear tides in
stars, planets, or compact objects.

Most previous studies of tides have made the linear approximation,
accounting for only linear order perturbations to the background
state.  Our formalism includes the leading order nonlinear
corrections. These corrections have two seemingly different physical
origins: (1) \emph{Internal nonlinear interactions} which couple three
waves to each other.  These interactions enable waves that are
linearly excited by the tide to transfer energy to waves that are not
linearly excited.  This nonlinear coupling includes both resonant and
non-resonant nonlinear interactions (\S\S~\ref{sec:nonlinear_tide},
\ref{sec:stability_analysis}, \& \ref{sec:inhomog_driving}).  If the
linearly excited waves systematically lose energy to other waves via
nonlinear coupling, then the nonlinear interactions can substantially
modify the orbital/rotational evolution relative to that predicted by
linear theory. (2) \emph{External nonlinear interactions} which
resonantly couple two internal waves directly to the tide through
their gravitational multipole moments (\S~\ref{sec:simple_nltide}). These interactions allow waves that are not
linearly resonant with the tide to be directly driven by the tidal
potential. The damping of these nonlinearly driven waves can also act
as a source of enhanced dissipation relative to linear theory.

We have presented two related approaches to solving for the nonlinear
tidal response of a star.  In the first, we take the background state
of the star to be spherical and unperturbed by the companion and we
expand the Lagrangian displacement and velocity associated with the
tide (both linear and nonlinear) as a sum over eigenmodes, with each
mode weighed by its amplitudes $q_a$ (\S~\ref{sec:method1}).  This
extends one of the standard methods for studying linear tides to
include nonlinear interactions.  In the second method, we instead take
the background state of the star to include the linear tidal solution
and we expand the nonlinear correction to the tidal solution as a sum
over eigenmodes, with each mode weighed by a different mode amplitude
$r_a$ (\S \ref{sec:method2}).  These two methods are formally
equivalent but we find that each is useful for understanding different
aspects of tides in stars.  The second method of solving explicitly
for the nonlinear correction to the tidal response of a star is
particularly useful for understanding the nonlinear stability of
standard linear tidal theory (\S~\ref{sec:stability_analysis}).

Since we focus on slowly rotating stars, the stellar modes included in
our treatment are p-modes and g-modes; the latter are particularly
important because they can have periods comparable to that of the
binary system.  The orthogonality and completeness of the stellar
eigenmodes enables us to convert the nonlinear partial differential
fluid equations into a coupled network of ordinary differential
equations describing the evolution of each mode amplitude.  We now
summarize the physics contained in these equations, focusing for
concreteness on the method of expanding the full tidal solution
(linear and nonlinear) as a sum over eigenmodes.  In this case, the
resulting amplitude equations have linear terms and nonlinear
terms. The linear terms are standard (e.g., \citealt{Press:77}) and
include driving by the tidal force and damping by radiative
diffusion.\footnote{ While dissipation due to eddy viscosity in
  convection zones is important for the long wavelength equilibrium
  tide, we ignore it in this paper, focusing instead on thermal
  diffusion damping in the radiative zone, which is more important for
  short wavelength, low frequency g-modes in the radiative core.} This is the basic physics included in most
previous studies of tides, with one addition: by solving the mode
amplitude equations, our formalism allows the modes to be dynamic in
their interaction with the orbit, rather than assuming harmonic
response at the forcing frequency.

The nonlinear terms in the equations of motion are parameterized by the
coefficients $\kappa_{abc}$ and $U_{ab}$ (see eqs. [\ref{eq:kappa}]
and [\ref{eq:Uab}] in \S~\ref{sec:eom}); these describe the internal
(three wave) and external (nonlinear tidal driving) nonlinear
interactions, respectively.  Accurately calculating these coupling
coefficients is nontrivial -- these technical details are described in
Appendix \ref{sec:app:coef_in_amp_eqn}, where we pay considerable
effort to ensuring that the coupling coefficients can be accurately
calculated numerically.

Three wave coupling has been considered extensively in the literature,
both in stellar seismology (e.g., \citealt{Dziembowski:82,Wu:01}) and
more specifically in the context of tidally excited oscillations in
binary systems (e.g., \citealt{Kumar:96}).  Nonlinear tidal driving
has not, to our knowledge, been studied in any detail before 
(\citealt{Papaloizou:81} considered
nonlinear tidal driving but neglected three-wave coupling which 
we found cancels strongly with nonlinear tidal driving).  Both
types of interactions can lead to a variety of physical effects
including the non-resonant excitation of higher frequency modes and
the resonant excitation of lower frequency modes (parametric
instability).We have focused our analysis on determining (1) the conditions under
which the linear tidal flow is unstable to the parametric instability
(\S\S~\ref{sec:dyntide} \& \ref{sec:eqtide}) and (2) the efficiency of
nonresonant excitation of g-modes and p-modes by nonlinear coupling to
the linear tidal flow (\S~\ref{sec:inhomog_driving}).

In the course of computing nonlinear coupling coefficients and growth
rates for solar-type stars, we found the {\em a priori} surprising
result that the external nonlinear driving of g-modes ($U_{ab}$) is
almost completely canceled by a portion of the three wave coupling to
the linear tide ($\kappa_{abc}$).  The implication is that these two
effects are intimately related. The nature of this relationship can be
most easily appreciated if we explicitly solve for the nonlinear
correction to the linear tidal solution (method 2 in \S~
\ref{sec:second_order_eom}) and consider the linear flow as a
superposition of an ``equilibrium tide" and a ``dynamical tide"; the
former is the nearly hydrostatic part of the linear tidal response and
the latter is the wave-like, resonant part of the response. The three
wave coupling coefficient that describes the nonlinear driving of a
pair of `daughter' waves then has a contribution from the equilibrium
tide $\kappa_{ab}^{(\rm eq)}$ (eq. [\ref{eq:kappa_eq}]) and the
dynamical tide $\kappa_{ab}^{(\rm dyn)}$ (eq. [\ref{eq:kappa_dyn}]).
Since the equilibrium tide contains the vast majority of the tidal
energy, one might expect it to be more prone to nonlinear instability
than the dynamical tide (e.g., \citealt{Press:75,Kumar:98a}).  In
fact, we find the opposite: the daughters internal driving via three
wave coupling to the equilibrium tide nearly cancels with their
external nonlinear driving by the tidal potential ($\propto
U_{ab}$). As a result, the effective coupling of daughters to the
equilibrium tide is much weaker than their coupling to the dynamical
tide, i.e., $|\kappa_{ab}^{(\rm eq)}| \ll |\kappa_{ab}^{(\rm dyn)}|$
(see \S~\ref{sec:driving_rate} and Appendices \ref{sec:app:alternative_eom} and \ref{sec:properties_of_coefficient}).  Physically, this is
because internal nonlinear driving by the equilibrium tide and
nonlinear tidal driving are fundamentally part of the same process;
together they describe the nonlinear excitation of daughter modes by
the nearly hydrostatic response of the star to its companion.  Another
reason for the weak nonlinear driving by the equilibrium tide in
solar-type stars is that very little of the energy of the equilibrium
tide resides in the core of the star where the low frequency g-modes
that the equilibrium tide attempts to drive have most of their energy.
At a technical level, the subtlety of calculating the equilibrium tide
driving correctly highlights the importance of the detailed
calculation presented in this paper.  For example, an order of
magnitude estimate of the nonlinear driving by the shear of the
equilibrium tide ($\kappa_{ab}^{(\rm eq)} \sim 1$) overestimates the
true driving by almost two orders of magnitude
(\S~\ref{sec:eq_tide_driving_rate} and Appendix \ref{sec:Jeq}; Figs. \ref{fig:Jplus2kap} and \ref{fig:Jplus2kap_dn}).

The equilibrium tide driving of daughter modes, when present, is
relatively evenly spread throughout the radiative zone (left panel of
Fig. \ref{fig:Jplus2kap_dn}); by contrast, the driving by the
dynamical tide is highly concentrated in the core of the star
(Fig. \ref{fig:kappa}).  Hence the two mechanisms for seeding fluid
instability are physically quite different.  Quantitatively, we find
that the equilibrium tide is only unstable to nonlinear driving of resonant
g-modes in solar binaries if the orbital period is $\lesssim 2-5$ days
(\S~\ref{sec:eqtide}).  This implies that instability of the
equilibrium tide in slowly rotating solar-type stars cannot help
explain the observed circularization of solar binaries out to orbital
periods of $\simeq 10-15$ days. An interesting remaining possibility
is that the equilibrium tide in {\em rotating} solar-type stars may
efficiently couple nonlinearly to inertial waves.  In particular, inertial
waves in solar type stars have most of their energy in the outer
convection zone where much of the energy in the equilibrium tide also
resides.

Recently, the elliptical instability has been invoked as a source of nonlinear driving in close binary systems (\citealt{Lebars:10, Cebron:11}; see also \citealt{Press:75, Seguin:76}). The elliptical instability is clearly related to the equilibrium tide instability considered here; both instabilities involve the destabilization of internal modes of oscillation by the long wavelength, hydrostatic, tidal perturbation. They are not equivalent, however, as the elliptical instability drives inertial modes whereas we focus on g-modes.  
 
For solar type stars, we find that the dynamical tide driving rates of
short wavelength g-modes ($\propto \kappa_{ab}^{(\rm dyn)}$) are
orders of magnitude larger than the equilibrium tide driving rates.
In order to determine when linear theory is valid, we computed the
parametric instability growth rates over a range of orbital periods
and companion masses (\S~\ref{sec:dyntide}).  We find that for orbital
periods from days to weeks, the dynamical tide part of the linear
tidal flow is unstable, even for companion masses much smaller than a
Jupiter mass (see Fig. \ref{fig:massperiod}).  The linear tidal
solution in the star is invalid even for a 10 Earth mass planet if the
orbital period is $\lesssim 5$ days! The degree to which the true
tidal flow differs from the linear one depends on the saturation of
these nonlinear instabilities, which will be investigated in a future
paper.  This question is of particular current interest given the
large number of low mass planets orbiting solar-type stars being
discovered by transit surveys such as the \emph{Kepler} telescope
\citep{Borucki:11}.  The orbital properties of these systems are
likely shaped in part by the nonlinear instabilities discussed in this
paper. 

In our study of the parametric instability of the dynamical tide we
found that it is subject to a ``collective'' version of the parametric instability. In the
literature, parametric growth rates have traditionally been derived by
considering how a single pair ($N\le 2$) of coupled short wavelength
daughters interact with the time-dependent background of a parent mode
(e.g., \citealt{Dziembowski:82}). We have extended this analysis to
allow for $N\gg2$ daughters, each of which is coupled to the parent
and the other daughters.  We find that for solar type stars, large
sets of daughters are coherently driven by the dynamical tide
($N\approx 10^2-10^3$ for periods $P\la10\trm{ days}$ and companions
$M'/M\ga few\times10^{-4}$). When collectively unstable, the daughters
all grow as a single coherent unit, i.e., maintaining phase coherence.
This coherence is self-consistently generated by the nonlinear
interactions even if it is not present in the initial
amplitudes/phases of the modes
(Fig. \ref{fig:simple_3wave_collective}).  In the collective
parametric instability, the growth rates of the daughters are
significantly larger than in standard three wave coupling, by a factor
of $\approx N/2$ (Fig. \ref{fig:growth_collective}).  As a result,
the collective instability may have important implications for
the nonlinear damping of the dynamical tide. More generally, it will also be important to revisit previous astrophysical applications of parametric instability to understand the consequences of the restrictive focus on three-mode coupling, which does not necessarily capture the parametric instability threshold or fastest growth rate.

In their study of the dynamical tide in solar-type binaries, \citet{Goodman:98}
 noted that short-wavelength gravity waves traveling
inwards from the radiative-convective boundary (where they are excited)
reach sufficiently large amplitudes that the waves ``break'' at the
center of the star.  \citet{Barker:10} and \citet{Barker:11} confirmed
this result numerically.  For hot Jupiter companions with masses $\lesssim 2 M_{\rm
  Jupiter}$, however, they found that the incoming gravity waves do
not break in the core of the star (see also \citealt{Ogilvie:07}). 

In an effort to understand what happens in the regime where the incoming wave does not break in the core, \citet{Barker:11b} 
performed a stability analysis of a non-linear standing internal gravity wave in the central regions of a cylindrical ``star". Although their methods differ from ours, they obtain (non-collective) daughter driving rates similar to ours (see \S~\ref{sec:dyn_tide_driving_rate}). Like us, \citet{Barker:11b} do not attempt to solve for the saturation of the unstable daughters and thus do not explicitly solve for the tidal $Q$ of the star. However, they argue that the driving rate of the fastest growing daughter mode places a lower bound of $Q\ga 10^7$. We are not convinced that this result is correct for two reasons. First, both our study and theirs find that the parent excites many daughter pairs, many of which have (non-collective) growth rates just slightly below that of the fastest growing mode. These excited daughters will each  drain energy from the parent. Since there are so many such rapidly growing daughters, it seems possible that the parent will lose more energy to all of these modes than it does to the single fastest growing mode. Secondly, \citet{Barker:11b} do not seem to observe the collective instability, and thus might be underestimating the rate at which daughters drain energy from the parent.\footnote{In the absence of collective driving, the driving rate $\Gamma_{bb}\propto \omega/n$ (see \S~\ref{sec:dyn_tide_driving_rate}). Collective driving increases the driving rate of each daughter by a factor of approximately $N\approx n$ and therefore the collective driving rate $\Gamma_{bb}\propto \omega$. \citet{Barker:11b} find $\Gamma_{bb}\propto \omega/n$ even above the collectively instability threshold.}  One possible explanation for why \citet{Barker:11b} do not see the collective instability is that their diffusivity, which they chose in order to obtain numerically converged solutions, is set so high that it artificially suppresses collective driving.  In \S~\ref{sec:dyntide} we show that the companion (planet) mass above which the collective instability sets in lies well above the threshold for the standard three-mode instability (but still well below the wave breaking threshold; see Fig. \ref{fig:massperiod}). Thus, if their diffusivity is set too high compared to the true value in a solar-type star, they might be in the regime where standard driving is unstable but collective driving is artificially stabilized.

In addition to the driving of resonant high $\ell$ daughter waves, there is efficient non-resonant three mode coupling (nonlinear inhomogeneous driving) between the dynamical tide, the equilibrium tide,  and a wide
range of stellar p-modes and g-modes (\S~\ref{sec:inhomog_driving} and Figs.~\ref{fig:EinhomogP} \& \ref{fig:linlin_coef}).  How this coupling saturates is, however,
unclear.  Even if these alternative nonlinear couplings turn out to be
more important than wave breaking in damping the dynamical tide, they
are unlikely to modify the orbital evolution due to the dynamical tide
calculated by \citet{Goodman:98}; this is because the linear energy
input rate into the dynamical tide is independent of its damping rate
in the limit of efficient damping.  Nonetheless, identifying the
correct damping mechanism for the dynamical tide in solar binaries is
important because it determines where in the star the tidal energy is
dissipated.

In some cases, we find that the parametric growth times due to driving
by the dynamical tide can be so short that the excited daughter waves
do not have time to travel through the core of the star (at the group
velocity) before their amplitudes are expected to become highly
nonlinear (\S~\ref{sec:global_vs_local}).  This is particularly true for the ``collective''
parametric instability (Fig. 10).  In this case, the standing wave
approximation employed here may become inaccurate and a traveling wave
point of view may be more appropriate. While this does not alter our
conclusions about the stability of the linear solution, it will likely
affect how the nonlinearities saturate and thus the rate of tidal
dissipation.  

To conclude, we note that the general formalism developed in this
paper can be extended to the tidal forcing of other types of stars.
For solar-type stars, the resonant g-mode amplitudes are large at the
center of the star, where the equilibrium tide amplitude is small. In
other stars, resonant wave amplitudes may be large at the stellar
surface, where the equilibrium tide is also large. Equilibrium tide
driving may thus play a more prominent role in those stars.

\acknowledgments The authors would thank Peter Goldreich, Pawan Kumar, and Yanqin Wu
for useful conversations during the development of this work and the referee for detailed and helpful comments that improved the manuscript.   This work was supported by NSF AST-0908873 and NASA  NNX09AF98G.   P. A. is an Alfred P. Sloan Fellow and received support from the Fund for Excellence in Science and Technology from the University of Virginia. E. Q. was
supported in part by the Miller Institute for Basic Research in
Science, University of California Berkeley, and the David and Lucile
Packard Foundation.

\begin{appendix}

\section{Coefficients in the amplitude equation}
\label{sec:app:coef_in_amp_eqn}

In this Appendix we derive expressions for the coefficients of the two forms of
amplitude equation (eqs. [\ref{eq:modeampeqn}] and [\ref{eq:ra_amp_eqn}]). In \S~\ref{sec:app:xi_lin_calc} we describe our calculation of the linear tidal displacement $\vec{\xi}_{\rm lin}$ needed to calculate some of the coefficients.  Expressions for the
linear-order coefficients $\gamma_a$ and $U_a$ are given in
\S~\ref{sec:app:linear_coef}; for the nonlinear tide $U_{ab}$ in
\S~\ref{sec:app:Uab}; and for the three-mode coupling coefficient
$\kappa_{abc}$ in \S~\ref{sec:app:kappa}. Analytic approximations to
$\kappa_{abc}$ are given in \S~\ref{sec:app:kappa_aprox} and the correction to 
$\kappa_{abc}$ when one of the modes is $\vec{\xi}_{\rm lin}$ rather than an eigenmode
 is given in \S~\ref{sec:app:kapU}.  
 
 Following \citet{Schenk:02}, these
coefficients are found by expanding the displacement $\vec{\xi}$ as a
sum of linear, adiabatic eigenmodes (eq. [\ref{eq:xiexpansion}]) and
contracting each mode $\vec{\xi}_a$ with the linear and nonlinear
internal (pressure, buoyancy, perturbed gravity) forces $\vec{f}_1$
and $\vec{f}_2$ and external (tidal) forces $\vec{a}_{\rm tide}$
(eq. [\ref{eq:atide}]) according to 
\bea 
\langle \vec{\xi}_a,\vec{F}\rangle &\equiv& \frac{1}{E_0}\int d^3x \rho \vec{\xi}_a^\ast
\cdot \vec{F}, 
\eea 
where $\vec{F}$ is one of the forces $\vec{f}_1$,
$\vec{f}_2$, or $\rho \vec{a}_{\rm tide}$
and we choose a normalization in which each mode has the same energy 
$E_0=GM^2/R$. The normalization integral for mode $a$ is then (eq. [\ref{eq:Enorm}])
\bea
\label{eq:normintegral}
E_0 & = & 2 \times \int d^3x \left(\frac{1}{2} \omega_a^2 \rho \left| \vec{\xi}_a \right|^2
+ {\rm \ potential\ energy} \right)
\\ &=& 2 \times 2 \times \frac{1}{2} \omega_a^2 \int d^3x \rho \left|
\vec{\xi}_a
\right|^2
\non &\simeq& 2 \omega_a^2 \Lambda_a^2 \int dr r^2 \rho(r) \xi_{a,h}^2(r),
\nonumber
\eea
where one factor of two arises since both a mode and its 
physically indistinct complex conjugate mode are included in the energy
(see eq. [\ref{eq:xiexpansion}]) and a second factor of two arises from the fact that the potential energy equals the kinetic energy. In the last equality we specialized to the case of low
frequency g-modes for which the horizontal displacement $\xi_h$ is much larger than the
vertical displacement $\xi_r$. This approach to solving the nonlinear partial differential equation (\ref{eq:xieqn}) is sometimes referred to as the Galerkin method, a particular implementation of the method of weighted residuals in which the weighting functions are chosen to be the eigenfunctions of the linear system (see, e.g., \citealt{Finlayson:72}). The expression for the linear force $\vec{f}_1[\vec{\xi}]$ is standard (see, e.g., \citealt{LyndenBell67, Schenk:02}). The linear eigenfunctions $\vec{\xi}_a$ and frequencies $\omega_a$ are
found by solving the eigenvalue problem
$\vec{f}_1[\vec{\xi}_a]=-\rho\omega_a^2\vec{\xi}_a$ using the Aarhus
adiabatic oscillation package \citep{Dalsgaard:08}. We use a $5\trm{ Gyr}$ old solar model taken from the EZ code \citep{Paxton:04}. The expression for the
leading order nonlinear interaction forces
$\vec{f}_2[\vec{\xi},\vec{\xi}]$ and the nonlinear tide is derived in
\citet{Schenk:02} and given by the nonlinear terms in their equations (4.8) and (4.9).

\begin{figure}
\epsscale{1.1}
\plottwo{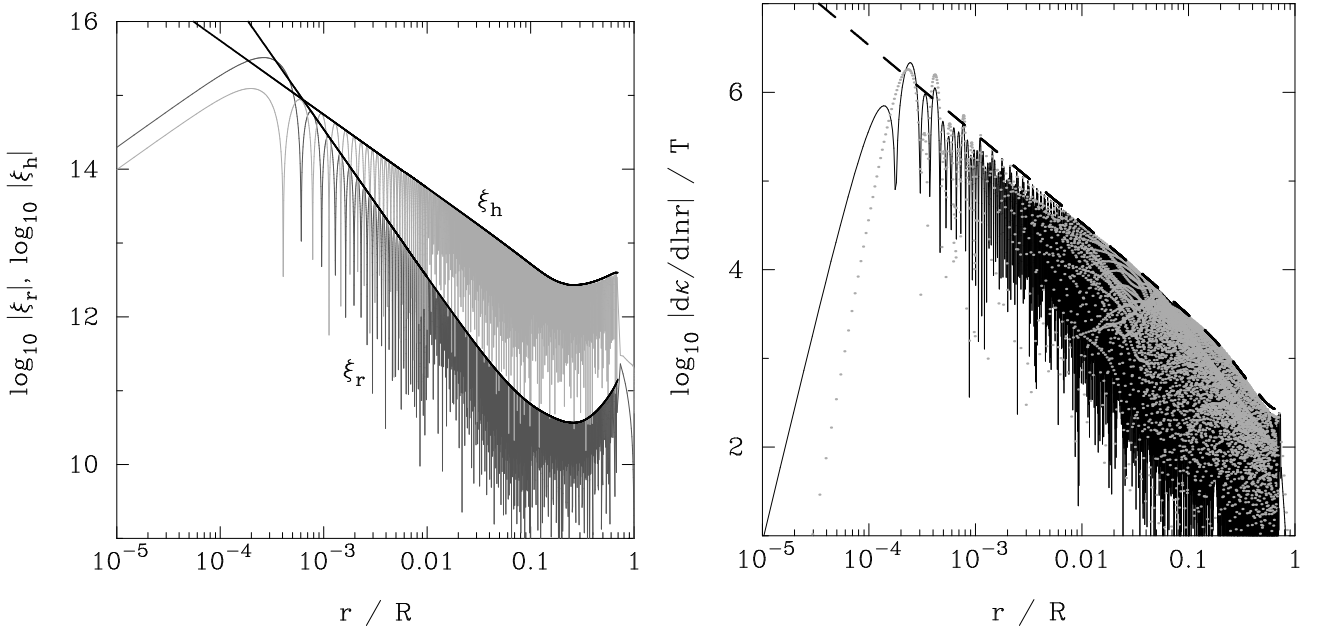}{fig15b.ps}
\caption{{\em Left panel}: Radial eigenfunction $\xi_r$ ({\em dark gray line})
and horizontal eigenfunction $\xi_h$ ({\em light gray line}) of the mode
$(\ell_a, n_a)=(2, 486)$ for a solar model. This mode is linearly resonant with
the tide for $P_{\rm orb}=10\trm{ days}$ (see Figure \ref{fig:ElinP}). The solid
black lines show the WKB approximation to the mode (eqs. [\ref{eq:xirWKB},
\ref{eq:xihWKB}] without the sinusoidal factors). {\em Right panel}: Radial
profile of the three-mode coupling coefficient $d\kappa_{abc}/d\ln r$ ({\em
solid black line}; divided by the angular integral $T$) of a parent with $(\ell_a, n_a)=(2, 486)$ and a self-coupled
daughter $b=c$ with $(\ell_b, n_b)=(2, 971)$. The full $\kappa_{abc}$ expression
(\ref{eq:e411}-\ref{eq:e419}) is dominated by the second term (\ref{eq:e412}),
plotted here as grey disconnected points. The dashed line is the WKB
approximation to the amplitude of $d\kappa_{abc}/d\ln r$ (eq.
[\ref{eq:kappa_aprox_WKB}]).}
\label{fig:xir_xih}
\vspace{0.1cm}
\end{figure}

In Figure \ref{fig:xir_xih} we show the radial profile of the g-mode
eigenfunction that is linearly resonant with the tide for $P_{\rm
  orb}=10\trm{ days}$ for a solar model. Also shown is the WKB
approximation to this high-order mode (see, e.g., \citealt{Unno:89}),
\bea
\label{eq:xirWKB}
\xi_r(r) &=& A(r) \sin \varphi(r),\\
\label{eq:xihWKB}
\xi_h(r) &=&A(r)\frac{k_r r}{\Lambda^2} \cos \varphi(r),
\eea
where  $\Lambda\equiv\ell(\ell+1)$, $k_{r}=\Lambda N/\omega r$ is the radial wavenumber of the mode, 
$\phi(r) \simeq \int_0^r k_r dr +\pi/4$ is its phase, 
\bea
\label{eq:normWKB}
A^2(r)=\left(\frac{E_0\Delta P}{2\pi^2}\right)\frac{1}{\rho r^3 N},
\eea
and $\Delta P=2\pi^2/\int_0^{r_c} N d\ln r \simeq 1.27/\omega_0$. For
$r\la0.05R$, $N/r\simeq 98\omega_0/R$ and $\rho\simeq 150\trm{ g cm}^{-3} \simeq 26\ M/R^3$, and to a good approximation 
$A(r)\simeq 3.5\times10^{8}{\rm cm} (r/R)^{-2} \simeq (R/200) (r/R)^{-2}$. This WKB expression is valid in the propagating region (i.e., where $\omega < N$) and, as we show below, is useful for estimating the magnitude of some of the coefficients in the amplitude equation.

\subsection{Linear tidal displacement $\vec{\xi}_{\rm lin}$}
\label{sec:app:xi_lin_calc}

In order to calculate some of the coupling coefficients (e.g., $\kappa_{bc}^{(\rm eq)}$ and $\kappa_{bc}^{(\rm dyn)}$ defined in eqs. [\ref{eq:kappa_eq}] and [\ref{eq:kappa_dyn}]) we must calculate the equilibrium tide and dynamical tide displacements $\vec{\xi}_{\rm eq}$ and $\vec{\xi}_{\rm dyn}$. We do so by numerically solving the linearized momentum, mass, and energy equations governing the adiabatic response to a perturbing potential $U$
\bea
\label{eq:inhomog_eqn1}
\rho \ddot{\vec{\xi}}_{\rm lin}&=&-\grad \delta p + \vec{g}\delta \rho  - \rho \grad(\delta \phi+U)\\
\label{eq:inhomog_eqn2}
\delta \rho &=&- \grad\cdot(\rho \vec{\xi}_{\rm lin}) \\
\label{eq:inhomog_eqn3}
\delta \rho &=& \frac{\delta p}{c^2}+\frac{\rho N^2}{g} \xi_{r, {\rm lin}}.
\eea
Here $\vec{\xi}_{\rm lin}=\vec{\xi}_{\rm eq}$+$\vec{\xi}_{\rm dyn}$ is the Lagrangian displacement vector of the linear tide, $\delta$ indicates an Eulerian perturbation, and the other symbols have their usual meaning. We can separate the radial, angular, and time dependence of the fluid variables in the usual way. Defining the potential $\psi = \delta p/\rho + \delta \phi+U$, the radial momentum, continuity, and Poisson equation give
\bea
\frac{d\psi}{dr}&=&\frac{N^2}{g}\left(\psi - \delta \phi-U\right)-(N^2-\omega^2)\xi_{r, \rm lin}\\
\frac{d\xi_{r, \rm lin}}{dr}&=&\left(\frac{g}{c^2}-\frac{2}{r}\right)\xi_{r, \rm lin} +\frac{\Lambda^2 \psi}{\omega^2 r^2}-\frac{\psi-\delta\phi-U}{c^2}\\
\label{eq:poisson}
\frac{d\delta g}{dr}&=&-\frac{2}{r}\delta g + \frac{\Lambda^2}{r^2}\delta \phi + 4\pi G \rho\left(\frac{\psi-\delta \phi-U}{c^2}+\frac{N^2}{g}\xi_{r, \rm lin}\right),
\eea
where $\delta g=d\delta \phi/dr$. Given $\psi$, the horizontal displacement is $\xi_{h, \rm lin}=\psi / r\omega^2$. At the center we impose the regularity conditions $\xi_{r, \rm lin}=\ell \xi_{h, \rm lin}$ and $\delta \phi \propto r^\ell$. At the surface, the condition $d\delta \phi/dr=-(\ell+1)\delta \phi/r$ picks out the solution that decreases outward and we require the fluid to be hydrostatic by imposing $\delta p = \rho g \xi_{r, \rm lin}$. 

We \emph{define} the equilibrium tide displacement $\vec{\xi}_{\rm eq}$ as the solution to equations (\ref{eq:inhomog_eqn1}-\ref{eq:inhomog_eqn3}) in the limit $\omega\rightarrow 0$. We find\footnote{These expressions are valid as long as $N^2\neq0$; they therefore apply in radiative regions ($N^2>0$) and convective regions ($N^2<0$). Given our definition of $\vec{\xi}_{\rm eq}$, we have $\vec{\xi}_{\rm dyn}\neq0$ in a convective region (see \citealt{Goodman:98, Terquem:98}).}
\bea
\label{eq:xir_eq}
\xi_{r, \rm eq}&=&-\frac{\delta \phi_{\rm eq}+U}{g},\\
\xi_{h, \rm eq}&=&\frac{1}{r\Lambda^2}\frac{d}{dr}\left(r^2 \xi_{r, \rm eq}\right),
\eea
$\delta p_{\rm eq}=\rho g\xi_{r, \rm eq}$, and $\delta\rho_{\rm eq}=-\xi_{r, \rm eq} d\rho/dr$; we compute $\delta \phi_{\rm eq}$ by numerically solving Poisson's equation (\ref{eq:poisson}). Upon solving for $\vec{\xi}_{\rm lin}$ and $\vec{\xi}_{\rm eq}$ we find $\vec{\xi}_{\rm dyn}=\vec{\xi}_{\rm lin}-\vec{\xi}_{\rm eq}$.

\subsection{Linear damping rate $\gamma_a$ and linear tide coefficient $U_a$}
\label{sec:app:linear_coef}

\begin{figure}
\epsscale{1.1}
\plottwo{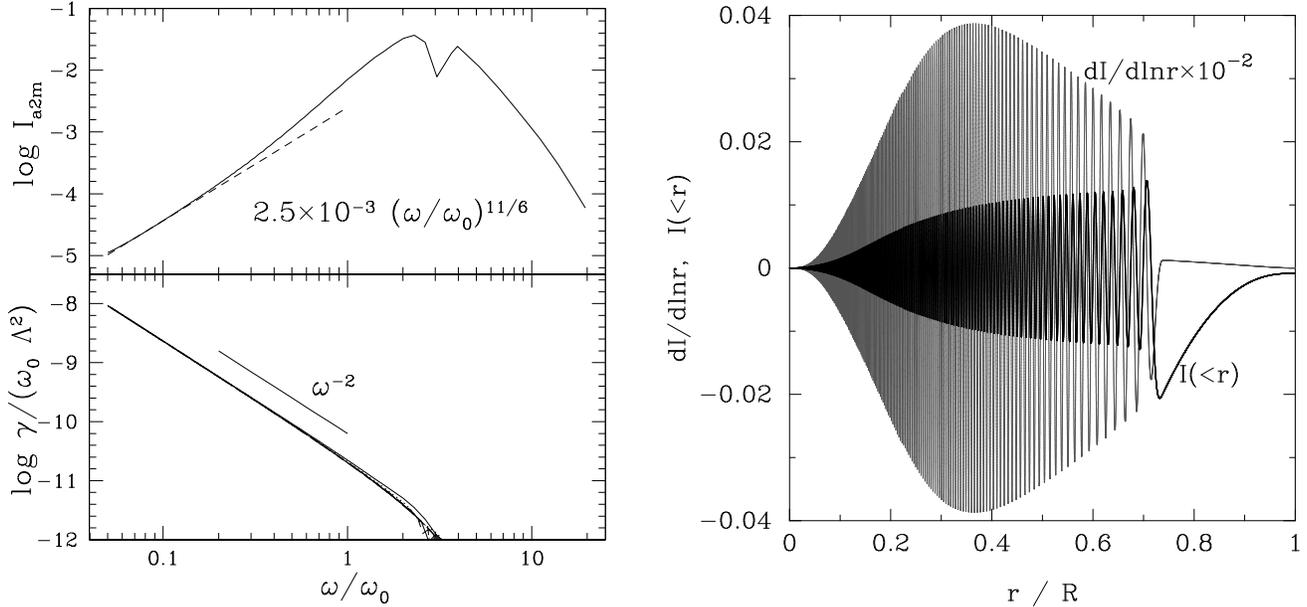}{fig16b.ps}
\caption{
{\em Left panel}: 
The top plot shows the $\ell=2$ linear
  overlap integral $I_{a2m}$ as a function of mode frequency in units
  of $\omega_0\equiv(GM/R^3)^{1/2}$ for a solar model. The solid line
  is the numerical result, while the dashed line is the analytic
  approximation (eq. [\ref{eq:linear_overlap_aprox}]). The bottom plot
  shows the damping rate due to radiative diffusion. Lines
  representing $\ell=2-8$ modes are shown; they very nearly lie on top
  of each other after the $\Lambda^2$ scaling has been removed (see
  eq. [\ref{eq:linear_gamma_aprox}]). 
  {\em Right panel}: Radial profile of $I_{a\ell m}$ for a solar model. The parameters are $\ell=2$ and $(\ell_a, n_a)=(2, 485)$, which corresponds to a mode period of $P_a\simeq 5\trm{ days}$. }
\label{fig:overlap_gamma}
\end{figure}

The temperature fluctuations that accompany the density perturbations of the excited g-modes are smoothed out by radiative diffusion. This is the dominant linear damping mechanism for high-order g-modes. We calculate its rate $\gamma_a$ by computing the quasi-adiabatic work integral (see, e.g., \citealt{Unno:89, Terquem:98, Goodman:98}). The results, shown in the bottom left panel of Figure \ref{fig:overlap_gamma}, are well-fit by the formula
\bea
\label{eq:linear_gamma_aprox}
\gamma_a(\ell, \omega)\simeq2 \times 10^{-11}\Lambda_a^2 \left(\frac{\omega_0}{\omega_a}\right)^2 \omega_0 \propto k_{r,a}^2.
\eea
Since the radiative diffusion is proportional to the second derivative of the temperature fluctuation, the damping rate of short-wavelength perturbations scales as $k_r^2$.

The overlap integral 
\bea
\label{eq:Ialm_appA}
I_{a\ell m}\equiv \frac{1}{MR^\ell}\int d^3x \rho \vec{\xi}_a^\ast \cdot \grad \left(r^\ell Y_{\ell m}\right)=\frac{1}{MR^\ell}\int dr r^{\ell+2} \delta \rho_a 
= -\frac{2\ell+1}{4\pi}\frac{\delta \phi(R)}{GM/R},
\eea
upon which the linear tide coefficient $U_a$ depends, is shown in the top left panel of Figure \ref{fig:overlap_gamma} for $\ell=2$. The second and third expressions in equation (\ref{eq:Ialm_appA}) can be derived by integrating by parts (see \citealt{Zahn:70}).  The radial profile of $I_{a\ell m}$ for a high-order mode is shown in the right panel of Figure \ref{fig:overlap_gamma}. Because the linear overlap  $I_{a\ell m}$ involves an integral over a single high-order mode, it is highly oscillatory from $r=0$ out to the radiative-convective interface at $r\simeq0.7 R_\odot$; the oscillations cancel almost perfectly throughout this region. The main contribution to $I_{a\ell m}$ comes from a region near the radiative-convective interface; since the mode becomes evanescent in this region, its wavelength becomes very long, allowing it to finally couple well with the large scale tidal potential \citep{Zahn:75}. 

For mode periods $P_a\ga 1 \trm{ day}$, \bea
\label{eq:linear_overlap_aprox}
I_{a2m}\simeq 2.5\times10^{-3}\left(\frac{\omega_a}{\omega_0}\right)^{11/6}.
\eea
\citet{Goodman:98} also found an $\omega^{11/6}$ scaling by matching approximate solutions of the inhomogeneous linear fluid equation across the radiative-convective boundary. As a check of the numerical accuracy of our calculation, we verified that our overlap integrals satisfy the sum rule 
\bea
\sum_a \omega_a^2 I_{a2m}^2 = 5G\int dx \rho x^4,
\eea
to high accuracy, where $x=r/R$ (see \citealt{Reisenegger:94a}; note that he uses a different eigenfunction normalization).

\subsection{Nonlinear tide coefficient $U_{ab}$}
\label{sec:app:Uab}

We now calculate the nonlinear  overlap integral $J_{ab\ell m}$ (eq. [\ref{eq:Jablm}]), upon which the nonlinear tide coefficient $U_{ab}$ depends. Using the covariant basis with vectors $\epsilon _i = h_i e_i$ where $h_r = 1, h_\theta=r, h_\phi=r \sin\theta$, the components of the Lagrangian displacement vector are
\beq
\vec{\xi}_a = \left[\xi_a^r,  \xi_a^\theta,  \xi_a^\phi \right]=\left[ a_r Y_a, \frac{a_h}{r}\frac{\partial Y_a}{\partial \theta},\frac{a_h}{r\sin^2\theta} \frac{\partial Y_a}{\partial \phi}\right],
\eeq
where $a_r=a_r(r)$ and $a_h=a_h(r)$, and $Y_a\equiv Y_{\ell_a, m_a}(\theta, \phi)$ is the spherical harmonic function. The integrand of $J_{ab\ell m}$ then consists of the following terms:
\bea 
\xi_a^r \xi_b^r (r^\ell Y_t)_{;rr} &=& \ell(\ell-1) r^{\ell-2} a_r b_r Y_a Y_b Y_{t}  \non
\xi_a^r \xi_b^\theta (r^\ell Y_t)_{;r \theta} &=& (\ell-1)r^{\ell-2}a_r b_h Y_a \frac{\partial Y_b}{\partial \theta}\frac{\partial Y_t}{\partial \theta}\non
\xi_a^r \xi_b^\phi(r^\ell Y_t)_{;r \phi} &=& (\ell-1)r^{\ell-2}a_r b_h  \frac{1}{\sin^2\theta}Y_a\frac{\partial Y_b}{\partial \phi}\frac{\partial Y_t}{\partial \phi}\non
\xi_a^\theta \xi_b^\theta(r^\ell Y_t)_{;\theta \theta} &=& r^{\ell-2}a_h b_h \frac{\partial Y_a}{\partial \theta}\frac{\partial Y_b}{\partial \theta}\left[ \frac{\partial^2 Y_t}{\partial \theta^2}+\ell Y_t \right]\non
\xi_a^\theta \xi_b^\phi(r^\ell Y_t)_{;\theta \phi} &=& r^{\ell-2}a_h b_h \frac{1}{\sin^2\theta} \frac{\partial Y_a}{\partial \theta}\frac{\partial Y_b}{\partial \phi}\left[\frac{\partial^2 Y_t}{\partial \theta\partial \phi}-\frac{\cos\theta}{\sin\theta} \frac{\partial Y_t}{\partial \phi}\right]\non
\xi_a^\phi \xi_b^\phi(r^\ell Y_t)_{;\phi \phi} &=& r^{\ell-2}a_h b_h \frac{1}{\sin^4\theta} \frac{\partial Y_a}{\partial \phi}\frac{\partial Y_b}{\partial \phi}\left[\frac{\partial^2 Y_t}{\partial \phi^2}+\sin\theta\cos\theta \frac{\partial Y_t}{\partial \theta} +\ell\sin^2\theta Y_t \right],
\eea
where $\ell$ and $Y_t$ refer to the harmonic of the tidal potential and the $\theta r, \phi r$, and $\phi \theta$ terms not shown are analogous to their symmetric counterpart. The angular integrations can be done analytically. Following the notation of Wu \& Goldreich (2001; see also \citealt{Kumar:96}), we define  
\bea
\label{eq:Tang}
T &\equiv& \int d\Omega \, Y_a Y_b Y_c
\\&=&\left[\frac{(2\ell_a+1)(2\ell_b+1)(2\ell_c+1)}{4\pi}\right]^{1/2}\nonumber
\left(\begin{array}{ccc}
\ell_a & \; \ell_b& \; \ell_c \\
m_a & \;  m_b & \;  m_c
\end{array}\right)
\left(\begin{array}{ccc}
\ell_a & \; \ell_b& \;\ell_c \\
0 &\; 0 &\; 0
\end{array}\right)
,\\
\label{eq:Fang}
F_a &\equiv&\int d\Omega \, Y_a \left[\frac{\partial Y_b}{\partial \theta}\frac{\partial Y_c}{\partial \theta} + 
\frac{1}{\sin^2\theta}\frac{\partial Y_b}{\partial \phi}\frac{\partial Y_c}{\partial \phi}\right]= \frac{T}{2}\left(\Lambda_b^2 + \Lambda_c^2 - \Lambda_a^2\right),\\
\label{eq:Gang}
G_a &\equiv& \int d\Omega \left[g^{im} g^{jn} (\nabla_i \nabla_j
Y_a)(\nabla_m Y_b)(\nabla_n Y_c)\right]=\frac{T}{4}\left[\Lambda_a^4 -
\left(\Lambda_b^2-\Lambda_c^2\right)^2\right], \eea where $g^{ij}$ is
the metric tensor on the unit sphere, the matrices on the right hand side of equation (\ref{eq:Tang}) are Wigner 3-$j$ symbols,
and other terms (e.g., $F_b$, $F_c$) are found by
permuting indices; note that these expressions apply to any 3
spherical harmonics denoted by $a, b$, and $c$.  For the particular
case of calculating $J_{ab\ell m}$, ``$c$'' here stands for the tidal
potential labeled by $\ell$ and $m$.  With these definitions we find
\bea 
J_{ab\ell m} &=& \frac{1}{MR^\ell}\int d^3x\, \rho \, \xi_a^{i}
\xi_b^j (r^\ell Y_{\ell m})_{;ij} 
\non &=& 
\frac{1}{MR^\ell} \int dr
\, \rho r^\ell \left[\ell(\ell-1)a_r b_r T + (\ell-1) a_r b_h F_a +
(\ell-1)a_h b_r F_b\right.
\non &&\left.\hspace{2.8cm} + a_h b_h\left(G_t+\ell F_t\right)\right].
\label{eq:Jablm_dr}
\eea
The angular integral $T$ is subject to the selection rules $|\ell_b-\ell_c| \le \ell_a \le \ell_b + \ell_c$ with $\ell_a+\ell_b+\ell_c$ even and $m_a+m_b+m_c =0$. These rules, which ensure conservation of angular momentum during the nonlinear interactions, allow modes with $\ell_a,\ell_b\neq2$ to couple to the $\ell=2$ component of the tide. The linear overlap integral $I_{a\ell m}$, by contrast, involves a product of two spherical harmonics and thus vanishes unless the angular degree of the mode equals that of the tide. 

In the propagating zone of low frequency g-modes,
$\Lambda \xi_h \gg \xi_r$  and 
the last term in (\ref{eq:Jablm_dr}) is dominant. However, as we explain in \S~\ref{sec:driving_rate}, this term cancels with the inhomogeneous piece of the three-wave coupling coefficient $\kappa_{bc\ell m}^{(I)}$.

\subsection{Nonlinear coupling coefficient $\kappa_{abc}$}
\label{sec:app:kappa}

In this appendix we derive the form of the three-wave coupling
coefficient.  We follow the treatment
of Wu \& Goldreich (2001, hereafter WG01; see also \citealt{Wu:98}),
who found that significant cancellations between large terms in the
coupling coefficient can lead to decreased numerical accuracy. They
showed that many of the cancellations can be removed analytically
using integration by parts, resulting in a mathematically equivalent
expression that can be integrated with much higher numerical accuracy.

The final result we derive below differs in four respects from
WG01. First, their equation (A15) should be multiplied by $-3/E_0$ to
obtain our normalization. Second, their equation (A1) neglects a term
involving the gravitational potential of the unperturbed background
(this term is not the perturbed gravitational potential that one can
justifiably ignore when the Cowling approximation is valid). We find
that this term is large, and leads to some cancellation in their
equation (A15). Third, we believe there are errors in terms 5, 7 and 9
of their equation (A15), as discussed below. These errors are also
contained in their equation (A14). Fourth, we include the perturbed
gravitational potential, both in the original form of the 3-wave
Lagrangian and when using the linear equations of motion to simplify
the expressions. Lastly, for clarity, we write out all permutations of
mode indices explicitly.

Our final expression for $\kappa_{abc}$ is given by lines (\ref{eq:e411}-\ref{eq:e418}). While quite complicated in general, we show in \S\S~\ref{sec:app:kappa_aprox} and \ref{sec:app:kappa_aprox_low_order_parent} that the expressions simplify greatly when the daughters are short wavelength modes. In particular, we show that $d\kappa_{abc}/d\ln r$ approximately equals the energy density of the daughters $dE/dr$ times the shear of the parent $k_r \xi_r$. This is true regardless of whether the parent is a short or long wavelength wave (e.g., dynamical tide or equilibrium tide).

For brevity, in this subsection we use the notation that the
displacement vector for a mode $a$ is $\vec{a}$, with radial and
horizontal components $a_r$ and $a_h$, and similarly for modes $b$ and
$c$. The three-wave coupling coefficient is given by equation (4.20)
of \citet{Schenk:02}: 
\bea 
 \kappa_{abc} &=& \frac{1}{2E_0} \int d^3x\
p \bigg[ \left\{ (\Gamma_1-1)^2 + \frac{\partial \Gamma_1}{\partial
\ln \rho} \Big\rfloor_s \right\} \grad \cdot \vec{a} \grad \cdot
\vec{b} \grad \cdot \vec{c}
\non && + (\Gamma_1-1) \left( a^i_{;j} b^j_{;i}
\grad \cdot \vec{c} + b^i_{;j} c^j_{;i} \grad \cdot \vec{a} + c^i_{;j}
a^j_{;i} \grad \cdot \vec{b} \right) 
 + a^i_{;j} b^j_{;k} c^k_{;i} + a^i_{;j} c^j_{;k} b^k_{;i} \bigg] 
 \non && -
\frac{1}{2E_0} \int d^3x \rho \left[ a^i b^j \delta \phi_{c;ij} + b^i
c^j \delta \phi_{a;ij} + c^i a^j \delta \phi_{b;ij} + a^ib^jc^k
\phi_{;ijk} \right].
\label{eq:kappa_original}
\eea
This expression is symmetric with respect to the three modes. The semicolon denotes a covariant derivative.

To evaluate this expression, WG01 first performed the angular integrations. Using WG01's notation, these integrations can be expressed in terms of the angular integrals $T$, $F_a$, $G_a$ (eqs. [\ref{eq:Tang}, \ref{eq:Fang}, \ref{eq:Gang}]), and 
\bea 
S &\equiv& \frac{1}{2} \left( \Lambda_a^2 F_a + \Lambda_b^2 F_b + \Lambda_c^2 F_c   \right)
\label{eq:S}
\\ 
V_a &\equiv& \Lambda_b^2 \Lambda_c^2 T - F_a - S.
\eea
The selection rules on the angular integral $T$ (\S~\ref{sec:app:Uab}) enforce angular momentum conservation during the 3-wave interactions.

After performing the angular integrals, WG01 showed that many terms cancel, and several terms could be integrated by parts. Our expression for the remaining terms, using WG01 equation (A5-A8), is
\bea
\kappa_{abc}  & = &   \frac{1}{2E_0} \int dr \left[
r^2 p \left\{ \Gamma_1(\Gamma_1-2) + \frac{\partial \Gamma_1}{\partial \ln \rho} \Big\rfloor_s \right\}
\grad \cdot \vec{a} \grad \cdot \vec{b} \grad \cdot \vec{c} T
\label{eq:e1l1}
\right. \\ && \left.
+ 2T \rho g a_r b_r c_r 
+ (F_a+S) \rho g a_r b_h c_h
+ (F_b+S) \rho g b_r c_h a_h
+ (F_c+S) \rho g c_r a_h b_h
\nonumber \right. \\ && \left.
- \Lambda_a^2 T \rho g a_h b_r c_r
- \Lambda_b^2 T \rho g b_h c_r a_r
- \Lambda_c^2 T \rho g c_h a_r b_r
- 2S \rho g a_h b_h c_h
\label{eq:e1l2}
\right. \\ && \left.
+ \left( \frac{da_r}{dr} \frac{db_r}{dr} + \frac{2}{r^2} a_r b_r \right) r^2 \Gamma_1 p \grad \cdot \vec{c} T
+ \left( a_r \frac{db_h}{dr} + b_r \frac{da_h}{dr} - \frac{d}{dr} \left( a_h b_h \right) \right)
r \Gamma_1 p \grad \cdot \vec{c} F_c
\nonumber \right. \\ && \left.
- \left( \Lambda_b^2 a_r b_h + \Lambda_a^2 a_h b_r \right) \Gamma_1 p \grad \cdot \vec{c} T
+ a_h b_h \Gamma_1 p \grad \cdot \vec{c} V_c
\nonumber \right. \\ && \left.
+ \left( \frac{db_r}{dr} \frac{dc_r}{dr} + \frac{2}{r^2} b_r c_r \right) r^2 \Gamma_1 p \grad \cdot \vec{a} T
+ \left( b_r \frac{dc_h}{dr} + c_r \frac{db_h}{dr} - \frac{d}{dr} \left( b_h c_h \right) \right)
r \Gamma_1 p \grad \cdot \vec{a} F_a
\nonumber \right. \\ && \left.
- \left( \Lambda_c^2 b_r c_h + \Lambda_b^2 b_h c_r \right) \Gamma_1 p \grad \cdot \vec{a} T
+ b_h c_h \Gamma_1 p \grad \cdot \vec{a} V_a
\nonumber \right. \\ && \left.
+ \left( \frac{dc_r}{dr} \frac{da_r}{dr} + \frac{2}{r^2} c_r a_r \right) r^2 \Gamma_1 p \grad \cdot \vec{b} T
+ \left( c_r \frac{da_h}{dr} + a_r \frac{dc_h}{dr} - \frac{d}{dr} \left( c_h a_h \right) \right)
r \Gamma_1 p \grad \cdot \vec{b} F_b
\nonumber \right. \\ && \left.
- \left( \Lambda_a^2 c_r a_h + \Lambda_c^2 c_h a_r \right) \Gamma_1 p \grad \cdot \vec{b} T
+ c_h a_h \Gamma_1 p \grad \cdot \vec{b} V_b
\label{eq:e1l3}
\right. \\ && \left.
- r^2 \rho a_r b_r c_r \frac{d^2g}{dr^2} T
- r^2 \rho \frac{d}{dr} \left( \frac{g}{r} \right) \left( F_a a_r b_h c_h + F_b a_h b_r c_h + F_c a_h b_h c_r \right)
\label{eq:e1l4}
\right. \\ && \left.
- r^2 \rho \left( 
a_r b_r \frac{d^2 \delta \phi_c}{dr^2}
+ b_r c_r \frac{d^2 \delta \phi_a}{dr^2}
+ c_r a_r \frac{d^2 \delta \phi_b}{dr^2}
\right) T
\nonumber \right. \\ && \left.
- r^2 \rho \left( a_r b_h F_a + b_r a_h F_b \right) \frac{d}{dr} \left( \frac{\delta \phi_c}{r} \right)
- r^2 \rho \left( b_r c_h F_b + c_r b_h F_c \right) \frac{d}{dr} \left( \frac{\delta \phi_a}{r} \right)
\nonumber \right. \\ && \left.
- r^2 \rho \left( c_r a_h F_c + a_r c_h F_a \right) \frac{d}{dr} \left( \frac{\delta \phi_b}{r} \right)
\nonumber \right. \\ && \left.
- r^2 \rho a_h b_h \left( \frac{1}{r} \frac{d\delta \phi_c}{dr} F_c + \frac{\delta \phi_c}{r^2} G_c \right)
- r^2 \rho b_h c_h \left( \frac{1}{r} \frac{d\delta \phi_a}{dr} F_a + \frac{\delta \phi_a}{r^2} G_a \right)
\nonumber \right. \\ && \left.
- r^2 \rho c_h a_h \left( \frac{1}{r} \frac{d\delta \phi_b}{dr} F_b + \frac{\delta \phi_b}{r^2} G_b \right)
\label{eq:e1l5}
\right].
\eea
The additional terms not included in WG01 are the ``$a^i b^j c^k
\phi_{;ijk}$" terms on line \ref{eq:e1l4} arising from the background
gravity, and the ``$a^i b^j \delta \phi_{c;i,j}$ terms on line
\ref{eq:e1l5} arising from perturbed gravity.

To this point, the linear equations of motion have not been used. We
now follow WG01 and integrate by parts, simplifying the resulting
expressions using the equations of motion. While WG01's equation (A9)
still holds, we must include the perturbed gravity as follows: 
\bea
\frac{d\xi_r}{dr} & =& - \frac{2}{r} \xi_r + \frac{\Lambda^2}{r} \xi_h
+ \grad \cdot \vec{\xi}
\label{eq:eom0}
\\
\Gamma_1 p \grad \cdot \vec{\xi} & = & \rho g \xi_r - \omega^2 \rho r \xi_h + \rho \delta \phi
\label{eq:eom1}
\\
\frac{d}{dr} \left( \Gamma_1 p \grad \cdot \vec{\xi} \right)
& = & \frac{\Lambda^2}{r} \rho g \xi_h - \left( \omega^2 + \frac{2g}{r} - \frac{dg}{dr} \right) \rho \xi_r
+ \rho \frac{d\delta \phi}{dr}.
\label{eq:eom2}
\eea
Note that equations (\ref{eq:eom1}) and (\ref{eq:eom2}) are the \emph{homogeneous} equations of motion (i.e., they do not include inhomogeneous terms involving the tidal potential $U$) and are therefore only appropriate if all three modes are solutions of the homogeneous equations. As we show in \S~\ref{sec:app:kapU}, when one of the modes is replaced by a solution of the inhomogeneous equations of motion (i.e., the linear tidal displacement $\vec{\xi}_{\rm lin}$), there are additional terms in the final expression for the coupling coefficient. 

The largest terms on line \ref{eq:e1l3} have the form ``$d/dr(a_h b_h)$". Integrating these three terms by parts, plugging in equation (\ref{eq:eom2}), and canceling terms against the last term ``$-2S\rho g a_h b_h c_h$" on line \ref{eq:e1l2} we find
\bea
&&\hspace{-0.5cm} 
\int dr  \left[ 
- 2S \rho g a_h b_h c_h 
- \frac{d}{dr} \left( a_h b_h \right) r \Gamma_1 p \grad \cdot \vec{c} F_c
- \frac{d}{dr} \left( b_h c_h \right) r \Gamma_1 p \grad \cdot \vec{a} F_a
- \frac{d}{dr} \left( c_h a_h \right) r \Gamma_1 p \grad \cdot \vec{b} F_b
\right]
 \nonumber \\   & =& 
\int dr  \left[
- 2S \rho g a_h b_h c_h
+ a_h b_h \frac{d}{dr} \left( r \Gamma_1 p \grad \cdot \vec{c} \right) F_c
+ b_h c_h \frac{d}{dr} \left( r \Gamma_1 p \grad \cdot \vec{a} \right) F_a
+ c_h a_h \frac{d}{dr} \left( r \Gamma_1 p \grad \cdot \vec{b} \right) F_b
\right]
\nonumber \\  & =&
\int dr  \left[
a_h b_h \left( \Gamma_1 p \grad \cdot \vec{c} -\left\{ \omega_c^2 + \frac{2g}{r} - \frac{dg}{dr} \right\} r \rho c_r
+ r \rho \frac{d\delta \phi_c}{dr} \right) F_c
\nonumber \right. \\ && \left.
+ b_h c_h \left( \Gamma_1 p \grad \cdot \vec{a} -\left\{ \omega_a^2 + \frac{2g}{r} - \frac{dg}{dr} \right\} r \rho a_r 
+ r \rho \frac{d\delta \phi_a}{dr} \right) F_a
\nonumber \right. \\ && \left.
+ c_h a_h \left( \Gamma_1 p \grad \cdot \vec{b} -\left\{ \omega_b^2 + \frac{2g}{r} - \frac{dg}{dr} \right\} r \rho b_r
+ r \rho \frac{d\delta \phi_b}{dr} \right) F_b
\right]
\eea
where we have used equation (\ref{eq:S}) in the cancellation.

Next, line \ref{eq:e1l3} contains terms involving $d\xi_h/dr$. These terms can be integrated by parts and then simplified using equation (\ref{eq:eom2}). The derivatives of the form $d\xi_r/dr$ can then be eliminated from line \ref{eq:e1l3} using equation (\ref{eq:eom0}). The resulting expressions contain derivatives only in the divergences $\grad \cdot \vec{\xi}$. For terms containing just one factor of $\grad \cdot \vec{\xi}$, we use equation (\ref{eq:eom1}). After significant cancellation of terms, and collection of like terms, we find
\bea 
\kappa_{abc}  & = &   \frac{1}{2E_0} \int dr \left[
r^2 p \left\{ \Gamma_1(\Gamma_1+1) + \frac{\partial \Gamma_1}{\partial \ln \rho} \Big\rfloor_s \right\}
\grad \cdot \vec{a} \grad \cdot \vec{b} \grad \cdot \vec{c} T 
\label{eq:e2l1}
\right. \\ && \left.
- r \rho a_h b_h c_h \left( \omega_a^2 G_a + \omega_b^2 G_b + \omega_c^2 G_c \right)
\label{eq:e2l2}
\right. \\ && \left.
+ \rho \left( r \frac{dg}{dr} - g \right) \left\{ a_r b_h c_h F_a + b_r c_h a_h F_b + c_r a_h b_h F_c \right\}
\label{eq:e2l3}
\right. \\ && \left.
+ rp \Gamma_1 \grad \cdot \vec{a} \grad \cdot \vec{c} \left\{ - 4b_r + b_h \Lambda_b^2 \right\} T
+ rp \Gamma_1 \grad \cdot \vec{b} \grad \cdot \vec{a} \left\{ - 4c_r + c_h \Lambda_c^2 \right\} T
\nonumber \right. \\ && \left.
+ rp \Gamma_1 \grad \cdot \vec{c} \grad \cdot \vec{b} \left\{ - 4a_r + a_h \Lambda_a^2 \right\} T
\label{eq:e2l4}
\right. \\ && \left.
+ 6p\Gamma_1 \left\{ a_r b_r \grad \cdot \vec{c} + b_r c_r \grad \cdot \vec{a} + c_r a_r \grad \cdot \vec{b} \right\} T
\label{eq:e2l5}
\right. \\ && \left.
+ 2\rho g a_r b_r c_r T
\label{eq:e2l6}
\right. \\ && \left.
- r \rho a_r b_h c_h \left\{ (\omega_a^2 - 3\omega_b^2 - 3\omega_c^2)F_a - 2(\omega_b^2 F_b + \omega_c^2 F_c)  \right\}
\nonumber \right. \\ && \left.
- r \rho b_r c_h a_h \left\{ (\omega_b^2 - 3\omega_c^2 - 3\omega_a^2)F_b - 2(\omega_c^2 F_c + \omega_a^2 F_a)  \right\}
\nonumber \right. \\ && \left.
- r \rho c_r a_h b_h \left\{ (\omega_c^2 - 3\omega_a^2 - 3\omega_b^2)F_c - 2(\omega_a^2 F_a + \omega_b^2 F_b)  \right\}
\label{eq:e2l7}
\right. \\ && \left.
- \rho \left( 4g + r \frac{dg}{dr} \right) \left\{ \Lambda_a^2 a_h b_r c_r +\Lambda_b^2 b_h c_r a_r + \Lambda_c^2 c_h a_r b_r  \right\} T
\nonumber \right. \\ && \left.
+ r \rho a_h b_r c_r \left\{ \omega_b^2F_b + \omega_c^2 F_c \right\}
+ r \rho b_h c_r a_r \left\{ \omega_c^2F_c + \omega_a^2 F_a \right\}
+ r \rho c_h a_r b_r \left\{ \omega_a^2F_a + \omega_b^2 F_b \right\}
\label{eq:e2l8}
\right. \\ && \left.
- r^2 \rho a_r b_r c_r \frac{d^2g}{dr^2} T
- r^2 \rho \frac{d}{dr} \left( \frac{g}{r} \right) \left( F_a a_r b_h c_h + F_b a_h b_r c_h + F_c a_h b_h c_r \right)
\label{eq:e2l9}
\right. \\ && \left.
+ \rho a_h b_h \left\{ r \frac{d\delta \phi_c}{dr} F_c + \delta \phi_c G_c \right\}
+ \rho b_h c_h \left\{ r \frac{d\delta \phi_a}{dr} F_a + \delta \phi_a G_a \right\}
+ \rho c_h a_h \left\{ r \frac{d\delta \phi_b}{dr} F_b + \delta \phi_b G_b \right\}
\nonumber \right. \\ && \left.
+ \rho a_r b_h\left\{ - r \frac{d\delta \phi_c}{dr} F_c + \delta \phi_c(-3\Lambda_b^2T + F_c) \right\}
+ \rho a_h b_r\left\{ - r \frac{d\delta \phi_c}{dr} F_c + \delta \phi_c(-3\Lambda_a^2T + F_c) \right\}
\nonumber \right. \\ && \left.
+ \rho b_r c_h\left\{ - r \frac{d\delta \phi_a}{dr} F_a + \delta \phi_a(-3\Lambda_c^2T + F_a) \right\}
+ \rho b_h c_r\left\{ - r \frac{d\delta \phi_a}{dr} F_a + \delta \phi_a(-3\Lambda_b^2T + F_a) \right\}
\nonumber \right. \\ && \left.
+ \rho c_r a_h\left\{ - r \frac{d\delta \phi_b}{dr} F_b + \delta \phi_b(-3\Lambda_a^2T + F_b) \right\}
+ \rho c_h a_r\left\{ - r \frac{d\delta \phi_b}{dr} F_b + \delta \phi_b(-3\Lambda_c^2T + F_b) \right\}
\label{eq:e2l10}
\right. \\ && \left.
- r^2 \rho \left(
a_r b_r \frac{d^2 \delta \phi_c}{dr^2}
+ b_r c_r \frac{d^2 \delta \phi_a}{dr^2}
+ c_r a_r \frac{d^2 \delta \phi_b}{dr^2}
\right) T
\nonumber \right. \\ && \left.
- r^2 \rho \left( a_r b_h F_a + b_r a_h F_b \right) \frac{d}{dr} \left( \frac{\delta \phi_c}{r} \right)
- r^2 \rho \left( b_r c_h F_b + c_r b_h F_c \right) \frac{d}{dr} \left( \frac{\delta \phi_a}{r} \right)
\nonumber \right. \\ && \left.
- r^2 \rho \left( c_r a_h F_c + a_r c_h F_a \right) \frac{d}{dr} \left( \frac{\delta \phi_b}{r} \right)
\nonumber \right. \\ && \left.
- r^2 \rho a_h b_h \left( \frac{1}{r} \frac{d\delta \phi_c}{dr} F_c + \frac{\delta \phi_c}{r^2} G_c \right)
- r^2 \rho b_h c_h \left( \frac{1}{r} \frac{d\delta \phi_a}{dr} F_a + \frac{\delta \phi_a}{r^2} G_a \right)
\nonumber \right. \\ && \left.
- r^2 \rho c_h a_h \left( \frac{1}{r} \frac{d\delta \phi_b}{dr} F_b + \frac{\delta \phi_b}{r^2} G_b \right)
\label{eq:e2l11}
\right].
\eea

We find three differences between our expression and WG01's equation (A15). First we find that their term 5 should be $-3\Gamma_1 p \grad \cdot \vec{a} b_r c_r T$, instead of $-(3\Gamma_1+1) p \grad \cdot \vec{a} b_r c_r T$. Next, their term 7 should be $-\rho g a_r b_r c_r T/3$ instead of $2pr^{-1} a_r b_r c_r T$. In their term 9, we find $5\rho g - 2p/r$ should be replaced by $4\rho g$. Lastly, inclusion of the background gravity terms found in line \ref{eq:e2l9}, which are not contained in WG01's equation (A1), cancel line \ref{eq:e2l3}. Hence neglect of the background gravity terms is not justified.

Partial cancellation occurs between the $\delta \phi$ terms arising from integration by parts (line \ref{eq:e2l10}) and the terms found in our equation (\ref{eq:kappa_original}) contained in line \ref{eq:e2l11}. In particular, the large terms of the form $\rho a_h b_h \delta \phi_c G_c$ cancel out.

The coupling coefficient as written on lines \ref{eq:e2l1}-\ref{eq:e2l11} significantly reduces the cancellation error compared to the original form (our eq. [\ref{eq:kappa_original}]).  Yet near the center of solar-type stars we find that there is still considerable cancellation between the terms of type $a_r b_r \grad \cdot \vec{c}$, $a_r b_r c_r$, and $a_h b_r c_r$. These terms rise the fastest toward the center, and come to dominate there. However, after integration they cancel against each other. For high radial order g-modes, these individual terms can be orders of magnitude larger than their sum. We transform these terms as follows: 
\bea
&& \int dr \left[ 6 p \Gamma_1 \left\{ a_r b_r \grad \cdot \vec{c} + b_r c_r \grad \cdot \vec{a} 
+ c_r a_r \grad \cdot \vec{b} \right\}
+ \rho a_r b_r c_r \left( 2g - r^2 \frac{d^2g}{dr^2} \right)
\nonumber \right. \\ && \left.
- \rho \left( 4g + r \frac{dg}{dr} \right) \left\{ \Lambda_a^2 a_h b_r c_r + \Lambda_b^2 b_h c_r a_r
+ \Lambda_c^2 c_h a_r b_r \right\}
 \right]
\label{eq:e3l1}
 \\ & =& 
 \int dr \left[
18 \rho g a_r b_r c_r 
- 6 \rho r \left\{ \omega_a^2 a_h b_r c_r + \omega_b^2 b_h c_r a_r + \omega_c^2 c_h a_r b_r \right\}
+ \rho a_r b_r c_r \left( 2g - r^2 \frac{d^2g}{dr^2} \right)
\nonumber \right. \\ && \left.
- \rho \left( 4g + r \frac{dg}{dr} \right) \bigg\{ 
\left( r \frac{da_r}{dr} + 2a_r - r \grad \cdot \vec{a} \right) b_r c_r
+ \left( r \frac{db_r}{dr} + 2b_r - r \grad \cdot \vec{b} \right) c_r a_r
\nonumber \right. \\ && \left.
+ \left( r \frac{dc_r}{dr} + 2c_r - r \grad \cdot \vec{c} \right) a_r b_r
\bigg\}
+ 6\rho \left( a_r b_r \delta \phi_c + b_r c_r \delta \phi_a + c_r a_r \delta \phi_b \right) T
\right]
\label{eq:e3l2}
\\ & =&
\int dr \left[
\rho a_r b_r c_r \left\{ 20 g - r^2 \frac{d^2g}{dr^2} - 24g - 6r\frac{dg}{dr} \right\}
- 6 \rho r \left\{ \omega_a^2 a_h b_r c_r + \omega_b^2 b_h c_r a_r + \omega_c^2 c_h a_r b_r \right\}
\nonumber \right. \\ && \left.
- \rho \left( 4g + r \frac{dg}{dr} \right) r \frac{d}{dr} \left( a_r b_r c_r \right)
+  \rho \left( 4g + r \frac{dg}{dr} \right) r \left\{ \grad \cdot \vec{a} b_r c_r
+ \grad \cdot \vec{b} c_r a_r + \grad \cdot \vec{c} a_r b_r \right\}
\nonumber \right. \\ && \left.
+ 6\rho \left( a_r b_r \delta \phi_c + b_r c_r \delta \phi_a + c_r a_r \delta \phi_b \right) T
\right]
\label{eq:e3l3}
\\ & =&
\int dr \left[
\rho a_r b_r c_r \frac{d\ln \rho}{d\ln r} \left( 4g + r \frac{dg}{dr} \right) 
- 6 \rho r \left\{ \omega_a^2 a_h b_r c_r + \omega_b^2 b_h c_r a_r + \omega_c^2 c_h a_r b_r \right\}
\nonumber \right. \\ && \left.
+  \rho \left( 4g + r \frac{dg}{dr} \right) r \left\{ \grad \cdot \vec{a} b_r c_r 
+ \grad \cdot \vec{b} c_r a_r + \grad \cdot \vec{c} a_r b_r \right\}
\nonumber \right. \\ && \left.
+ 6\rho \left( a_r b_r \delta \phi_c + b_r c_r \delta \phi_a + c_r a_r \delta \phi_b \right) T
\right].
\label{eq:e3l4}
\eea Going from line \ref{eq:e3l1} to line \ref{eq:e3l2}, we plugged
equation (\ref{eq:eom2}) into the first three terms and equation
(\ref{eq:eom0}) into the last three terms. This expression was
simplified in line \ref{eq:e3l3}. Going from line \ref{eq:e3l3} to
line \ref{eq:e3l4}, we integrated the term $d/dr(a_r b_r c_r)$ by
parts and simplified. The integrand is now not so steeply peaked
toward $r=0$. Further, the last terms of the form $a_h b_r c_r$ are
now proportional to $\omega_a^2$, not the large frequency $g/r$. 

The terms involving $\delta \phi$ can also be further simplified in order to remove terms that cancel against each other near the center. The $\delta \phi_c$ terms can be transformed as follows
\bea
&&\int dr \rho  \left[a_r b_r \left( 6 \delta \phi_c - r^2 \frac{d^2\delta \phi_c}{dr^2} \right)T -a_r b_h\left\{r \frac{d\delta \phi_c}{dr} F_c + \delta \phi_c\left(3\Lambda_b^2 T-F_c\right)+r^2F_a \frac{d}{dr}\left(\frac{\delta \phi_c}{r}\right)\right\}
\right. \nonumber \\&& \left.
-a_h b_r\left\{r \frac{d\delta \phi_c}{dr} F_c + \delta \phi_c\left(3\Lambda_a^2 T-F_c\right)+r^2F_b \frac{d}{dr}\left(\frac{\delta \phi_c}{r}\right)\right\} \right]
\label{eq:dphi11}\\ && =
\int dr \rho  \left[a_r b_r \left( 6 \delta \phi_c - r^2 \frac{d^2\delta \phi_c}{dr^2} \right)-  \left( \Lambda_a^2 a_h b_r + \Lambda_b^2 b_h a_r \right) \left( r \frac{d\delta \phi_c}{dr} + 2 \delta \phi_c \right) \right]T
\label{eq:dphi12}\\&& = -\int dr \rho  \left[a_r b_r \left(2 \delta \phi_c + 4r \frac{d\delta \phi_c}{dr} + r^2 \frac{d^2\delta \phi_c}{dr^2} \right) 
+r \left(\frac{d(a_r b_r)}{dr} - a_r \grad\cdot\vec{b} - b_r \grad \cdot \vec{a} \right) \left( r \frac{d\delta \phi_c}{dr} + 2 \delta \phi_c \right) \right]T
\label{eq:dphi13}
\\ &&= \int dr \rho  \left[\frac{d\ln \rho}{d\ln r} a_r b_r + r\left( a_r \grad\cdot\vec{b} + b_r \grad \cdot \vec{a}\right)\right] \left( r \frac{d\delta \phi_c}{dr} + 2 \delta \phi_c \right) T
\label{eq:dphi14}
\eea
To get the first equality we used (\ref{eq:Fang}), to get the second we used equation (\ref{eq:eom0}), and to get the last we integrated by parts. The $\delta \phi_a$ and $\delta \phi_b$ terms yield analogous expressions.

We thus obtain our final result:
\bea
 \kappa_{abc}  & = &   \frac{1}{2E_0} \int dr \left[
T r^2 p \left\{ \Gamma_1(\Gamma_1+1) + \frac{\partial \Gamma_1}{\partial \ln \rho } \right\}
\grad \cdot \vec{a} \grad \cdot \vec{b} \grad \cdot \vec{c}
\label{eq:e411}
\right. \\ && \left.
+ Trp\Gamma_1 \left(\grad \cdot \vec{a} \grad \cdot \vec{c} \left\{ b_h \Lambda_b^2 -4b_r \right\}
+\grad \cdot \vec{a} \grad \cdot \vec{b} \left\{ c_h \Lambda_c^2 - 4c_r \right\}
\right. \right.\non && \left.\left.
+\grad \cdot \vec{b} \grad \cdot \vec{c}  \left\{ a_h \Lambda_a^2-4a_r  \right\}\right)
\label{eq:e412}
\right. \\ && \left.
+ T \frac{d\rho}{d\ln r} \left( 4g + r\frac{dg}{dr} \right)  a_r b_r c_r
\label{eq:e413}
\right. \\ && \left.
+ T \rho r \left( 4g + r\frac{dg}{dr} \right) \left\{ \grad \cdot \vec{a}  b_r c_r + \grad \cdot \vec{b} c_r a_r
+ \grad \cdot \vec{c} a_r b_r\right\}
\label{eq:e414}
\right. \\ && \left.
-\rho r a_h b_h c_h \left\{ \omega_a^2 G_a + \omega_b^2 G_b + \omega_c^2 G_c \right\}
\label{eq:e415}
\right. \\ && \left.
- \rho r a_r b_h c_h \left\{  (\omega_a^2-3\omega_b^2-3\omega_c^2)F_a - 2(\omega_b^2F_b+\omega_c^2 F_c) \right\}
\right. \nonumber \\ && \left.
- \rho r b_r c_h a_h \left\{  (\omega_b^2-3\omega_c^2-3\omega_a^2)F_b - 2(\omega_c^2F_c+\omega_a^2 F_a) \right\}
\right. \nonumber \\ && \left.
- \rho r c_r a_h b_h \left\{  (\omega_c^2-3\omega_a^2-3\omega_b^2)F_c - 2(\omega_a^2F_a+\omega_b^2 F_b) \right\}
\label{eq:e416}
\right.  \\ && \left.
+ \rho r a_h b_r c_r \left\{ \omega_b^2 F_b + \omega_c^2 F_c - 6\omega_a^2 T \right\}
\right. \nonumber \\ && \left. + \rho r b_h c_r a_r \left\{ \omega_c^2 F_c + \omega_a^2 F_a - 6 \omega_b^2 T \right\}
\right. \nonumber \\ && \left. + \rho r c_h a_r b_r \left\{ \omega_a^2 F_a + \omega_b^2 F_b - 6 \omega_c^2  T \right\}
\label{eq:e417}
\right. \\ && \left.
+\rho\left\{\frac{d\ln \rho}{d\ln r} a_r b_r + r\left( a_r \grad\cdot\vec{b} + b_r \grad \cdot \vec{a}\right) \right\}\left( r \frac{d\delta \phi_c}{dr} + 2 \delta \phi_c \right) T
\right.  \nonumber \\ && \left.
+ \rho\left\{\frac{d\ln \rho}{d\ln r} a_r c_r + r\left( a_r \grad\cdot\vec{c} + c_r \grad \cdot \vec{a}\right) \right\}\left( r \frac{d\delta \phi_b}{dr} + 2 \delta \phi_b \right)T 
\right.  \nonumber  \\ && \left.
+\rho \left\{\frac{d\ln \rho}{d\ln r} b_r c_r + r\left( b_r \grad\cdot\vec{c} + c_r \grad \cdot \vec{b}\right)\right\} \left( r \frac{d\delta \phi_a}{dr} + 2 \delta \phi_a \right) T\right].
\label{eq:e418}
\label{eq:e419}
\eea
Since $d\rho/dr$ can be written in terms of $N$, no numerical derivatives are needed in this expression.

In the right panel of Figure \ref{fig:xir_xih} (see also Fig. \ref{fig:kappa}) we plot $d\kappa_{abc}/d\ln r$ as a function of radius for a parent $a$ resonant with the linear tide and a self-coupled daughter $b=c$ with period $P_b\simeq 2 P_a = 10 \trm{ days}$.

\subsubsection{Analytic estimate of $\kappa_{abc}$ for three high-order g-modes in solar-type stars}
\label{sec:app:kappa_aprox}

We now derive an analytic estimate of $\kappa_{abc}$ for the coupling of three high-order g-modes (see also \citealt{Dziembowski:82}). Over  such a mode's propagation region (i.e., where $\omega < N$), the WKB approximation (eqs. [\ref{eq:xirWKB}, \ref{eq:xihWKB}]) is very accurate (see left panel of Figure \ref{fig:xir_xih}). Over most of this region $\omega \ll N/\Lambda$ and thus $\xi_h \gg \xi_r$. Below the inner turning point in the core ($\omega > N$), the mode can be approximated with Bessel functions. The gravitational perturbation $\delta \phi$ due to the mode is negligible throughout the star (i.e., the Cowling approximation is appropriate).  An inspection of our final $\kappa_{abc}$ expression (lines \ref{eq:e411}-\ref{eq:e419}) reveals that as a result of these properties,  the second group of terms (line \ref{eq:e412}) is the largest of the 8 groups of terms by at least a factor of a few throughout all but the innermost part of the star (see grey points in right panel of Figure \ref{fig:xir_xih}). 

First consider the region where $\omega \ll N/\Lambda$ for all three modes. Using the WKB approximation (eqs. [\ref{eq:xirWKB}, \ref{eq:xihWKB}]) we find 
\bea
\label{eq:kappa_term2_aprox_WKB}
\frac{d\kappa_{abc}}{d\ln r}&\simeq& \left.\frac{d\kappa_{abc}}{d\ln r}\right|_{\rm term-2}
\simeq\frac{T r^2 p}{2E_0 \Gamma_1 H^2} \Lambda_a^2 a_h b_r c_r\\
&\simeq& 
\label{eq:kappa_aprox_WKB}
6 \left(\frac{\Lambda_a}{\sqrt{6}}\right) \left(\frac{P_a}{5 \trm{ days}}\right) \left(\frac{T}{0.2}\right)  \left(\frac{g}{\Gamma_1 H N^2} \right) \left(\frac{N\Delta P}{5}\right)^{3/2}\left(\frac{\rho}{1 \trm{ g cm}^{-3}}\right)^{-1/2}  \left(\frac{r}{R}\right)^{-5/2}. 
\eea
Here $H$ is the pressure scale height and we used equation (\ref{eq:eom1}). The $b_r c_h$ and $b_h c_r$ terms  in equation (\ref{eq:e412}) are negligible since they involve products like $\sin \phi_b \cos \phi_c\simeq \sin(\phi_b-\phi_c)/2$ and thus very nearly vanish (the daughters must be similar by momentum conservation and thus $\phi_b\simeq \phi_c$). The $b_r c_r$ term, by contrast, involves the product $\sin \phi_b \sin \phi_c\simeq \cos(\phi_b-\phi_c)/2$. The factor $g/\Gamma_1 H N^2$ varies from $\simeq 1-4$ between the center and convective zone of a solar-type star. The expression on the second line gives only the component that varies slowly with background quantities and not the high-frequency oscillatory component $\cos \phi_a \sin\phi_b \sin\phi_c$. As shown by the dashed line in the right panel of Figure \ref{fig:xir_xih}, equation (\ref{eq:kappa_aprox_WKB})  is a very good approximation to the amplitude of $d\kappa/d\ln r$ over nearly the entire star. Note that since the local energy of a mode within a shell of thickness $dr$ is $dE\simeq \rho r^2 N^2 \xi_r^2 dr$
we see by (\ref{eq:kappa_term2_aprox_WKB}) that insofar as $b_r\simeq c_r$,
\bea
\label{eq:kappa_krxir}
\frac{d\kappa_{abc}}{d\ln r}\approx 
T \frac{d(E_b/E_0)}{d\ln r} \frac{da_r}{dr} \simeq 
T \frac{d(E_b/E_0)}{d\ln r} k_{r, a} a_r,
\eea
i.e., $d\kappa/d \ln r$ is proportional to the parent's local shear $da_r/dr$; we will use this expression to evaluate the local three-wave coupling in Appendix \ref{sec:app:local_growth_rate}.

Now consider the region in the core between the parent's inner turning point $r_a\simeq \omega_a (dN/dr)^{-1}$ and the daughters' at $r_{b, c} \simeq  r_a/2$. The daughters are again given by the WKB approximation while the parent is evanescent and given by Bessel functions $a_r \simeq A x j_{\ell_a}(x)$ and $a_h\simeq (r\Lambda_a^2)^{-1}d(r^2a_r)/dr$, where $A$ is given by equation (\ref{eq:normWKB}) and $x=k_{r, a} r$.
For $x\ll 1$ and $\ell_a=2$, $j_2\simeq x^2/8$. Plugging in values
corresponding to the core of a solar model ($dN/dr\simeq 98
\omega_0/R$, $r_a\simeq 1.3\times10^{-4}P_{a, 10}^{-1} R$, $\rho\simeq
150\trm{ g cm}^{-3}$) we find
\bea
 \frac{d\kappa_{abc}}{d\ln
r}&\simeq& \left.\frac{d\kappa_{abc}}{d\ln r}\right|_{\rm term-2}
\simeq\frac{T r^2 p}{E_0 \Gamma_1 H^2} a_r b_r c_r
\simeq9\times10^4 \left(\frac{P_a}{5 \trm{ days}}\right)^2 \left(\frac{T}{0.2}\right) \left(\frac{r}{r_a}\right).
\eea
In the standing wave limit (see \S~\ref{sec:global_vs_local}), the global coupling is dominated by the region $r_a/2\la r \la r_a$ and for resonant coupling $P_a\simeq P_{\rm orb} / 2$ we find
\bea
\label{eq:kappa_analytic}
\kappa_{abc} \simeq 5\times 10^4 \left(\frac{T}{0.2}\right)\left(\frac{P_{\rm orb}}{10 \trm{ days}}\right)^2,
\eea
in good agreement with the full numerical calculation.

\subsubsection{Analytic estimate of $\kappa_{abc}$ for the equilibrium tide coupled to two high-order g-modes in solar-type stars}
\label{sec:app:kappa_aprox_low_order_parent}

We now consider the case where the parent $a$ is the equilibrium tide, which corresponds to the case $\kappa_{abc}=\kappa_{bc}^{(H, \rm eq)}$ (see \S\S~\ref{sec:driving_rate} and \ref{sec:app:kapU}). We then have $a_r\approx  (r/R)^{\ell+2} R$ and $\Lambda^2 a_h \approx (\ell+4)a_r$ at $r\sim R$ and one can show that six of the eight terms in $\kappa_{abc}$ are of similar magnitude (the exceptions are terms 1 and 7 [lines \ref{eq:e411} and \ref{eq:e417}], which are much smaller and can be ignored). For example, since $\grad\cdot \vec{a}=0$ and $\grad\cdot \vec{b}\simeq b_r / \Gamma_1 H$ for short wavelength daughter g-modes, term 2 (\ref{eq:e412}) is approximately
\bea
\left.\frac{d\kappa_{abc}}{d\ln r}\right|_{\rm term-2}\approx \frac{Tg}{2\Gamma_1 H N^2}\frac{\rho r^2 N^2 b_r c_r}{E_0}\ell a_r \approx  \frac{Tg\ell}{2\Lambda \Gamma_1 H N^2}\frac{d(E_b/E_0)}{dr} \left(\frac{c_r}{b_r}\right) \Lambda a_r.
\eea
 The ratio $c_r/b_r$ implies that $\kappa_{bc}^{(H, \rm eq)}$ is small unless $b_r\simeq c_r$ (i.e., $|n_b-n_c|\la 1$), since otherwise the integrand oscillates rapidly about zero on the scale of the long wavelength parent. The other five terms are of similar magnitude (although their signs can vary). Since $g\ell /2\Lambda\Gamma_1 H N^2\approx 1$ throughout the radiative zone where the daughters propagate, we find
\bea
\label{eq:dkapHeq_dlnr}
\frac{d\kappa_{bc}^{(H, \rm eq)}}{d\ln r}\approx T \frac{d(E_b/E_0)}{d r} \Lambda a_r \approx T \frac{d(E_b/E_0)}{d\ln r}\Lambda \left(\frac{r}{R}\right)^{\ell +1},
\eea
where the second equality applies at $r\sim R$. Therefore, as we found in the case of a high-order parent (eq. [\ref{eq:kappa_krxir}]), the local coupling of daughters to the equilibrium tide is proportional to the local shear, $d\xi_{r, \rm{ eq}}/dr$, of the equilibrium tide. Furthermore,  because the daughter energy density is nearly independent of daughter frequency and angular degree, $\kappa_{bc}^{(H, \rm eq)}$ is a weak function of both.

\subsection{Linear tide coupling coefficient}
\label{sec:app:kapU}

In deriving our expression for $\kappa_{abc}$, we used the linear homogeneous equations of motion (eqs. [\ref{eq:eom1}] and [\ref{eq:eom2}]) to simplify the divergence terms in line \ref{eq:e1l3}. Our final expression (lines \ref{eq:e411}-\ref{eq:e419}) is therefore only appropriate when all three waves are solutions of the homogeneous equations. In order to determine the nonlinear coupling between the linear tide $\vec{\xi}_{\rm lin}$ and a pair of daughters $(\vec{b}, \vec{c})$, we must instead simplify the corresponding divergence terms in line \ref{eq:e1l3} using the inhomogeneous equations of motion; i.e., replace $\delta \phi\rightarrow\delta \phi + \overline{U}$  in equations  (\ref{eq:eom1}) and (\ref{eq:eom2}), where the normalized tidal potential $\overline{U}(\vec{x},t)=-(GM/R)(r/R)^\ell Y_{\ell m}(\theta,\phi) \exp(-i\omega t)$.  This results in a modified coupling coefficient $\kappa_{bc}^{(U)}=\kappa_{bc}^{(H)} + \kappa_{bc}^{(I)}$: the homogeneous part, $\kappa_{bc}^{(H)}$, is given by lines \ref{eq:e411}-\ref{eq:e419} with $\vec{a}$ replaced by $\overline{\vec{\xi}}_{\rm lin}$, where  $\overline{\vec{\xi}}_{\rm lin}$ is the solution to equations (\ref{eq:inhomog_eqn1}-\ref{eq:inhomog_eqn3}) with $U\rightarrow \overline{U}$; the inhomogeneous part 
\bea
\label{eq:kappa_bclm}
 \kappa_{bc}^{(I)} &=& -\frac{1}{2MR^{\ell}}\int dr \rho r^{\ell} \left[6Tb_r c_r -b_r c_h\left\{3\Lambda_c^2 T+(\ell-1)F\right\}-b_h c_r \left\{3\Lambda_b^2 T+(\ell-1)F\right\}
 \right.\non&&\left.\hspace{2.5cm}+b_h c_h\left(G+\ell F\right)\right],
\eea 
where $F$ and $G$ are $F_a$ and $G_a$ with $(\ell_a, m_a)\rightarrow (\ell, m)$. 

Comparing $\kappa^{(I)}_{bc}$ and $J_{bc\ell m}$ (eq. [\ref{eq:Jablm_dr}]), we see that the leading order terms $b_h c_h$ for g-modes cancel in the $J_{bc\ell m}+ 2\kappa^{(I)}_{bc}$ portion of the equilibrium tide coupling coefficient $\kappa_{bc}^{(\rm eq)}$ (eq. [\ref{eq:kappa_eq}]). Using equation (\ref{eq:eom0}) we find
\bea
J_{bc\ell m}+ 2\kappa^{(I)}_{bc}&=&
-\frac{T(\ell+2)}{MR^\ell}\int dr \rho r^\ell \left[-(\ell+1)b_r c_r - r\frac{\partial}{\partial r}\left(b_r c_r\right) + r b_r \grad \cdot\vec{c}+rc_r \grad\cdot\vec{b}\right]
\non &=&-\frac{T(\ell+2)}{MR^\ell}\int dr \rho r^\ell \left[\frac{\partial \ln \rho}{\partial \ln r}b_r c_r + r b_r \grad \cdot\vec{c} + rc_r \grad\cdot\vec{b}\right]
\label{eq:Jplus2kapI}
\\&\simeq&-\frac{T(\ell+2)}{MR^\ell}\int dr \rho r^\ell \left[\frac{\partial \ln \rho}{\partial \ln r} +\frac{2 r}{\Gamma_1 H}\right]b_r c_r,
\eea
where in the second line we used integration by parts and in the third line we made the WKB approximation $\grad\cdot \vec{\xi}\simeq \xi_r / \Gamma_1 H$ (this approximation is well-satisfied throughout the star, including the convective zone). Within the radiative zone, the fractional difference between  $J_{bc\ell m}$ and $2\kappa^{(I)}_{bc}$ is therefore $\sim b_r c_r / b_h c_h \sim (\omega / N)^2\sim10^{-5} P_{10}^{-2}$ and we see that there is a large cancellation between these terms. Furthermore, because the $b_r c_r$ terms dominate the integrand and the radial displacement $\xi_r$ is a weak function of period and $\ell$ (see eq. [\ref{eq:xirWKB}]), $J_{bc\ell m}+2\kappa^{(I)}_{bc}$ is a weak function of period and $\ell$. Finally, using $N^2=-g[d\ln \rho/dr+1/\Gamma_1 H]$ one can show that the integrand of $J_{bc\ell m}+2\kappa^{(I)}_{bc}$ is similar in magnitude to $\kappa_{bc}^{(H, \rm eq)}$ (see eq. [\ref{eq:dkapHeq_dlnr}]).

\subsection{Coefficients of the `Method 2' amplitude equation}
\label{sec:app:alternative_eom}

Here we derive expressions for the coefficients in the method 2 form of the 
amplitude equation  (\S~\ref{sec:method2}, eq. [\ref{eq:ra_amp_eqn}]). We first expand the coefficients in terms of the tidal harmonics: 
\bea
\label{eq:Ua_nl_full}
V_{a}(t)&=& \left(\frac{M'}{M}\right)^2 \sum_{k\ell m}\sum_{k'\ell' m'}
W_{\ell m}W_{\ell' m'}\left[J_a^{(\rm eq)}+J_a^{(\rm dyn)}\right]  X_{k}^{\ell m} X_{k'}^{\ell' m'}\left( \frac{R}{a} \right)^{\ell+\ell'+2}  e^{-i(k+k')\Omega t},\hspace{0.0cm}\\
\label{eq:kap_a_full}
K_a(t) &=& \left(\frac{M'}{M}\right)^2 \sum_{k\ell m}\sum_{k'\ell' m'}
W_{\ell m}W_{\ell' m'}\left[\tilde{\kappa}_{a}^{(\rm eq-eq)}+\tilde\kappa_{a}^{(\rm dyn-dyn)}+2\tilde{\kappa}_{a}^{(\rm eq-dyn)}\right]  X_{k}^{\ell m} X_{k'}^{\ell' m'}\left( \frac{R}{a} \right)^{\ell+\ell'+2}  e^{-i(k+k')\Omega t},\hspace{0.0cm}\\
\label{eq:kap_ab_full}
K_{ab}(t) &=&\frac{M'}{M} \sum_{k\ell m}
W_{\ell m}\left[\kappa_{ab}^{(I)}+\kappa_{ab}^{(H,{\rm eq})}+\kappa_{ab}^{(H,{\rm dyn})}\right] X_{k}^{\ell m}\left( \frac{R}{a} \right)^{\ell+1}  e^{-ik\Omega t},
\eea
The two other method 2 coefficients, $U_{ab}(t)$ and $\kappa_{abc}$, are described in \S\S~\ref{sec:app:Uab} and \ref{sec:app:kappa}, respectively.  By writing the  linear displacement as a sum of the equilibrium tide and dynamical tide displacements $\vec{\xi}_{\rm lin}=\vec{\xi}_{\rm eq}+\vec{\xi}_{\rm dyn}$, we have separated coefficients into corresponding components, as indicated by the superscripts ``eq" and ``dyn". 

 The various coupling coefficients in the definition of $K_{ab}$ are defined in \S~\ref{sec:app:kapU}. They describe the parametric driving of daughter pairs $(a,b)$ considered in \S~\ref{sec:stability_analysis}; we showed there that the equilibrium tide part of $2K_{ab}$ cancels strongly with $U_{ab}$ and therefore the equilibrium tide part of the sum $U_{ab}+2K_{ab}$ is much smaller than its individual terms. 
 
The coefficients $V_a$ and $K_a$ describe the nonlinear inhomogeneous driving of a mode $a$. The term $V_a$ corresponds to nonlinear tidal driving by the force $(\vec{\xi}_{\rm lin}\cdot\grad)\grad U$; one of its components is given by $J_a^{(\rm dyn)} = J_{ab}(\vec{b}\rightarrow\overline{\vec{\xi}}_{\rm dyn})$ where the arrow indicates that the corresponding eigenmode in the expression should be replaced by $\overline{\vec{\xi}}_{\rm dyn}$, the dynamical tide part of the solution of (\ref{eq:inhomog_eqn1}-\ref{eq:inhomog_eqn3}) with $U$ replaced by the normalized tidal potential $\overline{U}(\vec{x},t)=-(GM/R)(r/R)^\ell Y_{\ell m}(\theta,\phi) \exp(-i\omega t)$. The other component $J_a^{(\rm eq)}$ is defined similarly but with $\overline{\vec{\xi}}_{\rm eq}$ instead of $\overline{\vec{\xi}}_{\rm dyn}$. The term $K_a$ describes three-``mode" coupling between mode $a$ and the linear tide and is directly analogous to $\kappa_{abc}$ but with modes $b$ and $c$ at the linear tide values.  The components of $K_a$ are given by
\bea
\tilde{\kappa}_a^{(\rm eq-eq)}&=&2\kappa_{ab}^{(I)} (\vec{b}\rightarrow\overline{\vec{\xi}}_{\rm eq})+\kappa_{abc} (\vec{b},\vec{c}\rightarrow\overline{\vec{\xi}}_{\rm eq})\\
\tilde{\kappa}_a^{(\rm dyn-dyn)}&=&2\kappa_{ab}^{(I)} (\vec{b}\rightarrow\overline{\vec{\xi}}_{\rm dyn})+\kappa_{abc} (\vec{b},\vec{c}\rightarrow\overline{\vec{\xi}}_{\rm dyn})\\
\tilde{\kappa}_a^{(\rm eq-dyn)}&=&\kappa_{abc}(\vec{b}\rightarrow\overline{\vec{\xi}}_{\rm eq},\vec{c}\rightarrow\overline{\vec{\xi}}_{\rm dyn}).
\eea
 The terms $\tilde{\kappa}_a^{(\rm eq-eq)}$ and $\tilde{\kappa}_a^{(\rm dyn-dyn)}$ describe driving of mode $a$ by equilibrium tide-equilibrium tide and dynamical tide-dynamical tide coupling, respectively, while the term $\tilde{\kappa}_a^{(\rm eq-dyn)}$ describes driving by the cross interaction between the equilibrium tide and dynamical tide.  By the same arguments used to derive equation (\ref{eq:Jplus2kapI}) we have
\bea
 J_a^{({\rm dyn})}+2\kappa_{ab}^{(I)} (\vec{b}\rightarrow\overline{\vec{\xi}}_{\rm dyn})&=&
 -\frac{T(\ell+2)}{MR^\ell}\int dr \rho r^\ell\left[\frac{\partial \ln \rho}{\partial \ln r} a_r \overline{\xi}_{r,{\rm dyn}}+r a_r \grad\cdot \overline{\vec{\xi}}_{\rm dyn} +r\overline{\xi}_{r,{\rm dyn}}\grad\cdot\vec{a}\right].
 \non
\eea
The equivalent term involving the equilibrium tide can be simplified further because $\grad\cdot\overline{\vec{\xi}}_{\rm eq}=0$:
\bea
 J_a^{({\rm eq})}+2\kappa_{ab}^{(I)} (\vec{b}\rightarrow\overline{\vec{\xi}}_{\rm eq})&=&
 -\frac{T(\ell+2)}{MR^\ell}\int dr \rho r^\ell\left[\frac{\partial \ln \rho}{\partial \ln r} a_r \overline{\xi}_{r,{\rm eq}} +r\overline{\xi}_{r,{\rm eq}}\grad\cdot\vec{a}\right]
\non &=&\frac{T(\ell+2)}{MR^\ell}\int dr r^{\ell+1}\delta \rho_a\overline{\xi}_{r,{\rm eq}}.
 \label{eq:Ja_plus_2kap_eq}
\eea
As in equation (\ref{eq:Jplus2kapI}), we find that there is significant cancellation between  $J_a^{({\rm dyn})}$ and $2\kappa_{ab}^{(I)} (\vec{b}\rightarrow\overline{\vec{\xi}}_{\rm dyn})$ and likewise for the equivalent equilibrium tide expression. It is therefore convenient to define the three types of nonlinear inhomogeneous driving coefficients as follows:
\bea
\label{eq:kap_eqeq}
\kappa_a^{(\rm eq-eq)}&\equiv&  J_a^{({\rm eq})}+\tilde{\kappa}_a^{(\rm eq-eq)}\\
\label{eq:kap_dyndyn}
\kappa_a^{(\rm dyn-dyn)}&\equiv&  J_a^{({\rm dyn})}+\tilde{\kappa}_a^{(\rm dyn-dyn)}\\
\label{eq:kap_eqdyn}
\kappa_a^{(\rm eq-dyn)}&\equiv&  2\tilde{\kappa}_a^{({\rm eq-dyn})}.
\eea
The properties of these three coefficients are described in \S~\ref{sec:inhomog_coef}.

\section{Properties of the nonlinear coupling coefficients}
\label{sec:properties_of_coefficient}

\begin{deluxetable}{ll}
\tablewidth{0pc}
\tablecaption{COUPLING COEFFICIENTS}
\tablehead{
\colhead{Symbol} & \colhead{Coupling of}} 
\startdata
$\kappa_{abc}$  & Three eigenmodes  (\ref{eq:kappa})\nl
$\kappa_{ab}^{(\rm eq)}$  &Equilibrium tide and two eigenmodes  (\ref{eq:kappa_eq})  \nl
$\kappa_{ab}^{(\rm dyn)}$  & Dynamical tide and two eigenmodes  (\ref{eq:kappa_dyn}) \nl
$\kappa_{a}^{(\rm eq-eq)}$  &   Equilbrium tide to itself and an eigenmode (\ref{eq:kap_eqeq})\nl
$\kappa_{a}^{(\rm dyn-dyn)}$   & Dynamical tide to itself and an eigenmode (\ref{eq:kap_dyndyn}) \nl
$\kappa_{a}^{(\rm eq-dyn)}$  & Equil. tide, dyn. tide, and an eigenmode  (\ref{eq:kap_eqdyn}) \nl
\enddata
\tablecomments{Numbers in parentheses refer to the equation where the coefficient is defined. All $\kappa$'s not equal to $\kappa_{abc}$ are simply versions of $\kappa_{abc}$ with $c$ and/or $b$ equal to components of the linear tide.} 
\label{tab:coup_coef}
\end{deluxetable}

In this appendix we describe the properties of the nonlinear coupling coefficients \emph{for a solar model}. 
 For reference, we list the various coupling coefficients in Table \ref{tab:coup_coef}. We first discuss the two coupling coefficients corresponding to parametric driving: the dynamical tide coefficient $\kappa_{bc}^{(\rm dyn)}$ (\S~\ref{sec:kappa_dyn}) and the equilibrium tide coefficient $\kappa_{bc}^{(\rm eq)}$ (\S~\ref{sec:Jeq}).  We show that nonlinear driving by the dynamical tide $\Gamma_{bc}^{(\rm dyn)}\propto \kappa_{bc}^{(\rm dyn)}$ and equilibrium tide $\Gamma_{bc}^{(\rm eq)}\propto \kappa_{bc}^{(\rm eq)}$ are, in general, highly localized, with most of the coupling occurring at a specific location in the
star: near the core in the case of $\Gamma_{bc}^{(\rm dyn)}$ and at the radiative-convective interface in the case of  $\Gamma_{bc}^{(\rm eq)}$. The only exception is equilibrium tide driving of self-coupled daughters, which tends to be more global and, as a result, can potentially lead to driving throughout the star. 

In \S~\ref{sec:inhomog_coef} we describe the coupling coefficients corresponding to nonlinear inhomogeneous driving $V_a$ and $K_a$. Just as the linear inhomogeneous term $U_a$ can be expressed in terms of a dimensionless coefficient $I_{a\ell m}$ times a tidal factor $\sim\varepsilon$ (see eq. [\ref{eq:Ua_via_Ia}]), the nonlinear inhomogeneous term $V_a+K_a$ can be expressed in terms of dimensionless coupling coefficients times a tidal factor $\sim\varepsilon^2$ . By expressing the linear tide as the sum of the $\omega=0$ equilibrium tide and the dynamical tide, $\vec{\xi}_{\rm lin}=\vec{\xi}_{\rm eq}+\vec{\xi}_{\rm dyn}$, these coefficients can be broken into three types of interactions: equilibrium tide-equilibrium tide coupling $\kappa_a^{(\rm eq-eq)}$,  dynamical tide-dynamical tide coupling $\kappa_a^{(\rm dyn-dyn)}$, and the cross coupling between the equilibrium tide and dynamical tide $\kappa_a^{(\rm eq-dyn)}$. We define these coefficients in \S~\ref{sec:app:alternative_eom} (eqs. [\ref{eq:kap_eqeq}], [\ref{eq:kap_dyndyn}], and [\ref{eq:kap_eqdyn}]). In general, we find that for a solar binary $\varepsilon \kappa_a^{(\rm eq-eq)}$ is much smaller than $I_{a\ell m}$. However, both $\varepsilon\kappa_a^{(\rm dyn-dyn)}$ and $\varepsilon\kappa_a^{(\rm eq-dyn)}$ can be much larger than $I_{a\ell m}$ for certain modes (including p-modes). As we discuss in \S~\ref{sec:inhomog_driving}, this implies that these modes are nonlinearly driven to energies that are much larger than their linear values. 

In order to calculate these various coefficients we need to know the linear tidal response, as described by the equilibrium and dynamical tides $\vec{\xi}_{\rm eq}$ and $\vec{\xi}_{\rm dyn}$. We describe how we calculate these in \S~\ref{sec:app:xi_lin_calc}.

\subsection{Dynamical tide coupling coefficient $\kappa_{bc}^{(\rm dyn)}$}
\label{sec:kappa_dyn}

\begin{figure*}
\epsscale{1.1}
\plottwo{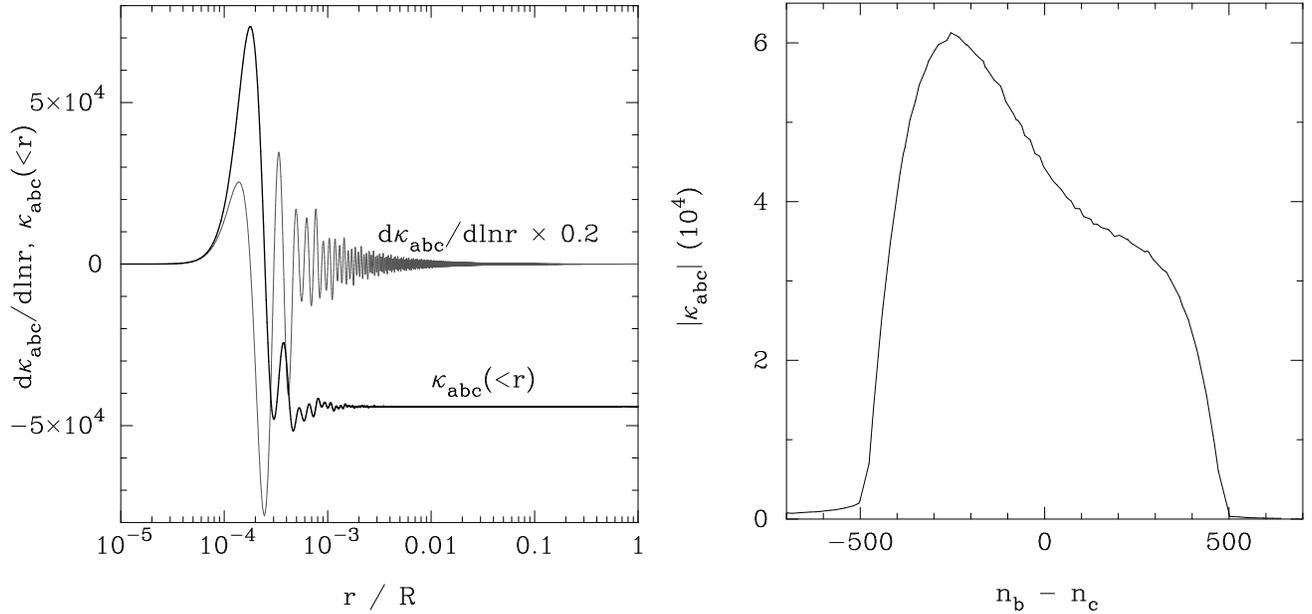}{fig2b.ps}
\caption{\emph{Left panel}: Radial profile of the three-mode coupling coefficient $\kappa_{abc}$ for a solar model. The parent mode is $(\ell_a, m_a,n_a)=(2,0, 486)$ and the daughter is a self-coupled mode ($b=c$) with $(\ell_b, m_b,n_b)=(2, 0,971)$. The periods of these modes are $P_a \simeq P_b/2\simeq5 \trm{ days}$. \emph{Right panel}:  $|\kappa_{abc}|$ as a function of $\Delta n = n_b-n_c$. The parent mode $a$ and daughter mode $b$ are the same as in the left panel. The other daughter mode, whose radial order $n_c$ we vary, has $(\ell_c, m_c)=(2, 0)$.}
\label{fig:kappa}
\end{figure*}

The spatial structure of the dynamical tide coupling coefficient $\kappa_{bc}^{(\rm dyn)}$ is approximately equal to that of the three mode coupling coefficient $\kappa_{abc}$ when mode $a$ is the parent eigenmode most linearly resonant with the tide. This is because the dynamical tide is typically dominated by this single mode (e.g., the mode with $n=486$ in Fig. \ref{fig:ElinP}). The approximate magnitude of $\kappa_{bc}^{(\rm dyn)}$ can be found by multiplying $\kappa_{abc}$ by $\omega_a I_{a\ell m}/(\Delta_a-i\gamma_a)$ (see eq. [\ref{eq:qa_lin}]). Since it is more common to express three-wave coupling coefficients in terms of three eigenmodes, we show results for $\kappa_{abc}$ rather than $\kappa_{bc}^{(\rm dyn)}$.

In the left panel of Figure \ref{fig:kappa} we show the integrand of $\kappa_{abc}$ for a linearly resonant parent coupled to a self-coupled daughter $b=c$.   The coupling coefficient $\kappa_{abc}$ involves an integral over an odd number (three) of high-order modes. The 
integrand peaks in the core since the radial and horizontal
displacements $(\xi_r,\xi_h)$ of high order modes increase with decreasing
$r$. For a standing wave, nearly all the contribution to $\kappa_{abc}$
comes from a small region between the daughter's inner turning point
where $\omega_b \simeq N$ and the parent's inner turning point where
$\omega_a \simeq N$; since $\omega_b<\omega_a$ and $dN/dr > 0$, the
parent's turning point lies above the daughter's. In this region,
the parent is evanescent and the displacements $\xi_{a,r},\xi_{a,h}
\propto r^{\ell_a-1}$ do not oscillate. As long as the
factor $\cos\left[ \left( k_b -k_c \right)r \right]$ is roughly constant in this region, the modes add
coherently and can exchange energy through nonlinear interactions.

In the right panel of Figure \ref{fig:kappa} we show $|\kappa_{abc}|$ as a function of $\Delta n = n_b-n_c$ for $\ell=2$ modes. The radial order of the parent mode $a$ is $n_a=486$. For $|\Delta n|\la n_a$, we find that $|\kappa_{abc}|$ is approximately constant and close to its maximum value (to within a factor of a few); otherwise it is much smaller than the maximum value. This is because in the propagation zone for each
wave, the integrand is proportional to a factor $\cos(k_ar)\cos(k_br)
\cos(k_cr)$, where $k_a$ is the radial wavenumber of mode $a$, etc. This
product of three cosines can be combined into a sum of terms of the
form $\cos\left[ \left( k_a \pm k_b \pm k_c \right)r \right]$. The coupling coefficient is thus maximized when $k_a \pm k_b \pm k_c\simeq 0$, the usual condition for momentum conservation. For high-order modes this is roughly equivalent to $|n_b - n_c| \la n_a$ (see also \citealt{Wu:01}). Hence although there is no rigorous selection rule for the 
radial direction, as there is for $\ell$ and $m$, there is an approximate 
selection rule. When momentum conservation is satisfied in the appropriate region, the magnitude of $\kappa_{abc}$ is near its maximum value. When it is not satisfied, the coupling coefficient is much smaller than this maximum possible value.

The approximate selection rule $|n_b - n_c| \la n_a$ implies that there can be a very large number daughter pairs with similar values of $\kappa_{abc}$. As a result, each daughter couples well to not only one other daughter but to $N\approx n_a$ other daughters. As we show in \S~\ref{sec:dyntide_collective}, this implies that collective driving can be very important for the dynamical tide.

\subsection{Equilibrium tide coupling coefficient $\kappa_{bc}^{(\rm eq)}$}
\label{sec:Jeq}

\begin{figure*}
\epsscale{1.1}
\plottwo{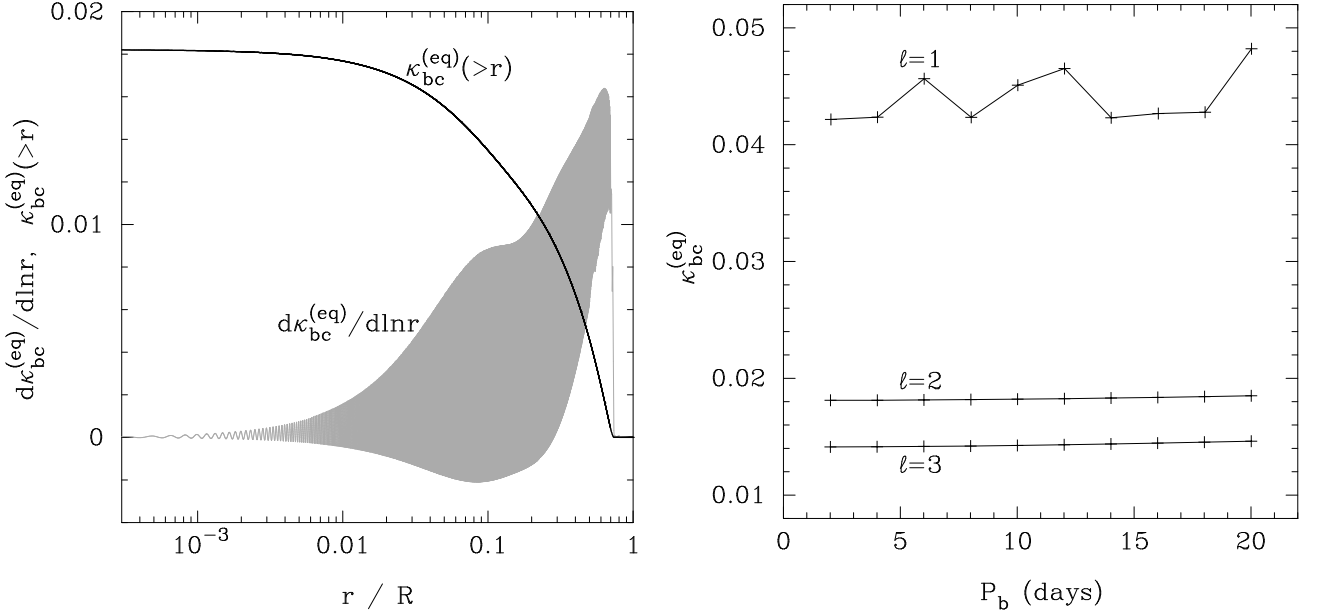}{fig3b.ps}
\caption{\emph{Left panel}: Radial profile of the equilibrium tide coupling coefficient 
$\kappa_{bc}^{(\rm eq)}$ for a solar model. The parameters of the equilibrium tide are $(\ell,m)=(2,0)$ and the daughter is a self-coupled mode ($b=c$) with
  $(\ell_b, m_b,n_b)=(2, 0,971)$, which corresponds to a period of
  $P_b\simeq 10\trm{ days}$.  The individual oscillations in the
  integrand $dJ/d\ln r$ are too closely spaced to be seen well at the
  resolution of the figure. \emph{Right panel}:  $\kappa_{bc}^{(\rm eq)}$ as a function of period $P_b$  for self-coupled daughters $b=c$ with $m=m_b=0$ and $\ell_b=1, 2, 3$.}
\label{fig:Jplus2kap}
\end{figure*}

\begin{figure*}
\epsscale{1.1}
\plottwo{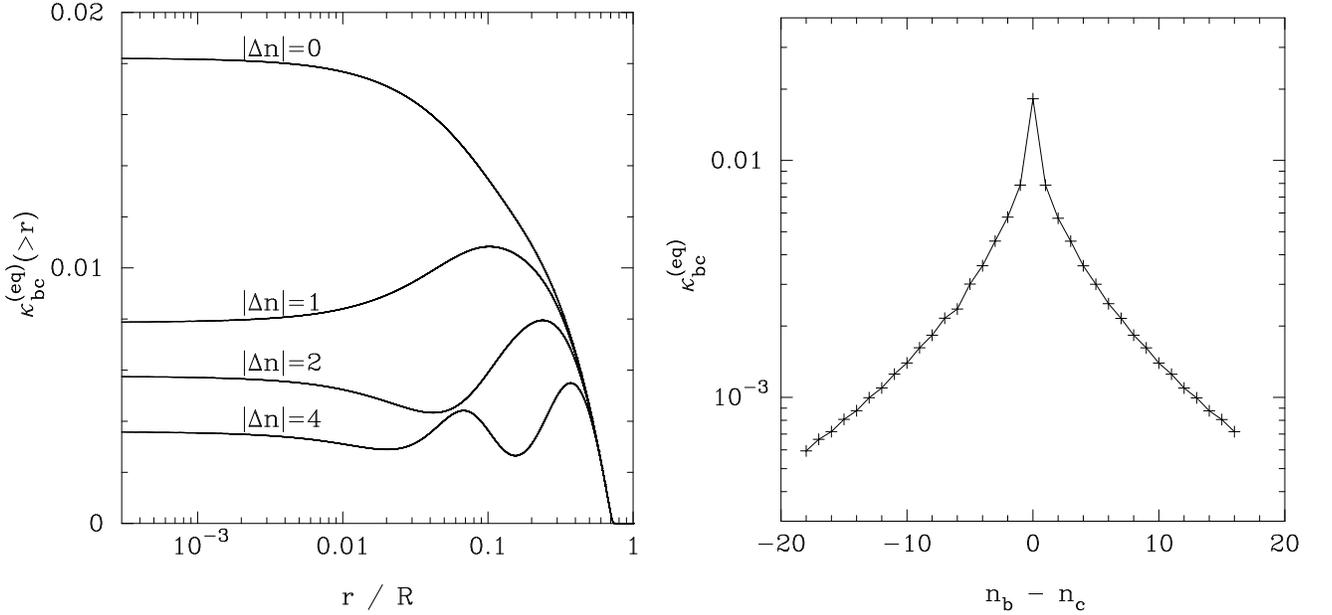}{fig4b.ps}
\caption{\emph{Left panel}: Radial profile of the equilibrium tide coupling coefficient 
$\kappa_{bc}^{(\rm eq)}$ for the $(\ell,m)=(2,0)$ equilibrium tide coupled to different daughter pairs $(b,c)$. The four curves are for $(\ell_b, m_b,n_b)=(2, 0,971)$ and $(\ell_c,m_c,n_c)=(2,0, [971,972,973,975])$. 
\emph{Right panel}: $\kappa_{bc}^{({\rm eq})}$ as a function of $\Delta n = n_b-n_c$ for $\ell_b=\ell_c=2$, $m=m_b=m_c=0$, and $P_b\simeq P_c\simeq 10\trm{ days}$.}
\label{fig:Jplus2kap_dn}
\end{figure*}

We show the  radial dependence of $\kappa_{bc}^{(\rm eq)}$ for the equilibrium tide coupled to a self-coupled daughter mode $b=c$ in Figure \ref{fig:Jplus2kap} and for daughter modes $b\neq c$ in Figure \ref{fig:Jplus2kap_dn}. Unlike $\kappa_{abc}$, the integrand of the equilibrium tide coupling coefficient $\kappa_{bc}^{(\rm eq)}$ contains a product of an even number of modes. More precisely, since the displacement due to the equilibrium tide increases monotonically with increasing radius and there is a factor of $\cos\left[(k_b \pm k_c) r \right]$ in the integrand, momentum conservation requires that locally $|k_b|\simeq|k_c|$. The coupling is therefore maximized for self-coupled modes $b=c$; as the number of nodes differs by $|\Delta n|=|n_b-n_c|=0,1,2\ldots$,
the equilibrium tide coupling decreases rapidly due to cancellation between the modes. As a result, unlike with the dynamical tide, collective driving does not appear to be important for the equilibrium tide.  

We also find that for self-coupling, $\kappa_{bc}^{(\rm eq)}$ is not strongly concentrated in radius. However, as $|\Delta n|$ increases the coupling becomes increasingly concentrated in the region just below the radiative-convective interface. As shown in the right panel of  Figure \ref{fig:Jplus2kap}, $\kappa_{bc}^{({\rm eq})}$ is also a weak function of daughter period and $\ell$.

\subsection{Inhomogeneous coupling coefficients}
\label{sec:inhomog_coef}

\begin{figure*}
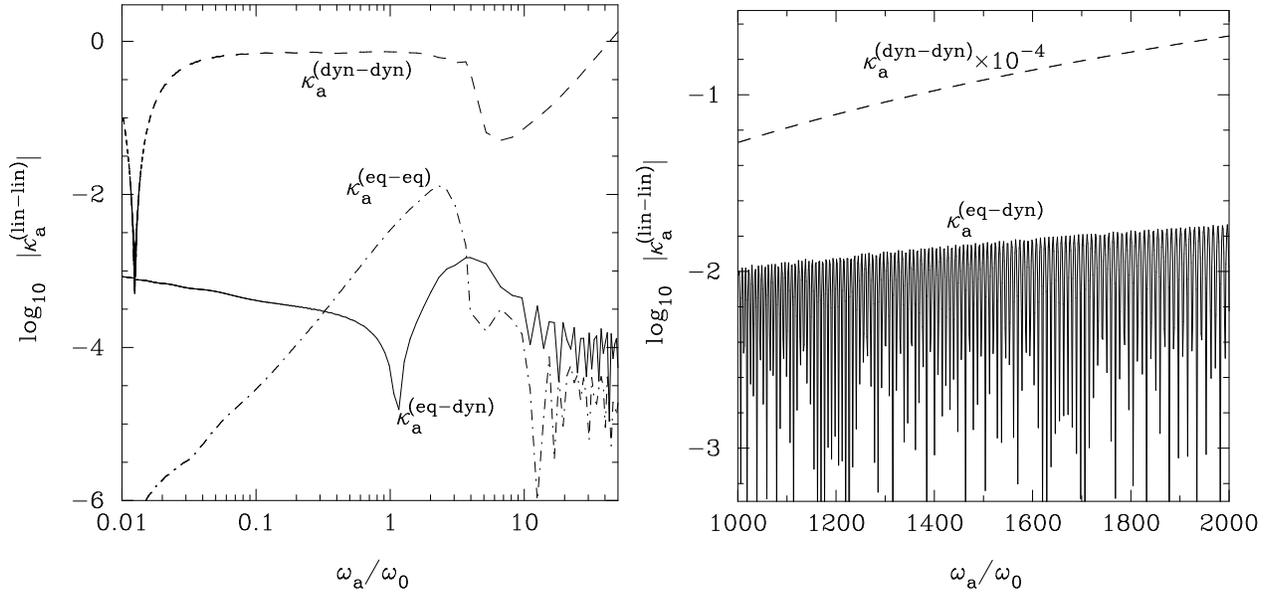

\begin{center}$
\begin{array}{cc}
\includegraphics[width=3.2in]{fig5a.ps} &
\includegraphics[width=3.26in]{fig5b.ps}
\end{array}$
\end{center}
\caption{The nonlinear inhomogeneous coupling coefficients $\kappa_a^{(\rm eq-eq)}$, $\kappa_a^{(\rm dyn-dyn)}$, and $\kappa_a^{(\rm eq-dyn)}$ as a function of mode frequency $\omega_a$ in units of $\omega_0=(GM/R^3)^{1/2}$ for a solar model. The daughters are all $(\ell_a, m_a)=(2, 0)$ modes coupled to the $(\ell, m)=(2, 0)$ dynamical tide at a $P=10\trm{ day}$ driving period. Those in the left panel range from high-order g-modes to low order p-modes while those in the right panel are high-order p-modes ($700 \la n_a \la 1400$). In the right panel we divide $\kappa_a^{(\rm dyn-dyn)}$ by $10^4$ in order to show it on the same scale as $\kappa_a^{(\rm eq-dyn)}$.}
\label{fig:linlin_coef}
\end{figure*}

In Figure \ref{fig:linlin_coef} we show the magnitudes of the three nonlinear inhomogeneous coupling coefficients  $\kappa_a^{(\rm eq-eq)}$, $\kappa_a^{(\rm dyn-dyn)}$, and $\kappa_a^{(\rm eq-dyn)}$. These represent driving of a mode $a$ due to equilibrium tide-equilibrium tide coupling,  dynamical tide-dynamical tide coupling, and equilibrium tide-dynamical tide coupling, respectively. The modes in the left panel range from high-order g-modes with $n\approx 10^3$ to low-order p-modes with $n\la30$. Since the acoustic cutoff of the solar atmosphere is $\approx 60\omega_0$, these p-modes are trapped. The right panel shows results for high-order p-modes that are all well above the acoustic cutoff and therefore do not reflect at the solar surface.

The integrand of the coupling coefficient $\kappa_a^{(\rm eq-eq)}$ is similar to that of the linear overlap $I_{a\ell m}$ (compare eqs. [\ref{eq:Ialm_appA}] and [\ref{eq:Ja_plus_2kap_eq}]). They therefore have very similar dependences on $\omega_a$ (cf. Fig. \ref{fig:overlap_gamma}) and for $\ell=2$ we find $\kappa_a^{(\rm eq-eq)}\sim\varepsilon I_{a\ell m}$. Driving by $\kappa_a^{(\rm eq-eq)}$, which draws all of its energy from just the equilibrium tide, therefore appears to be insignificant compared to linear driving.

The coupling coefficient $\kappa_a^{(\rm dyn-dyn)}$ is by far the largest of the three coefficients, typically by at least 2-3 orders of magnitude. From scaling arguments we see that its magnitude is roughly equal to $\kappa_{abb}(I_b\omega_b / \Delta_b)^2$, where mode $b$ here refers to the dynamical tide mode and $\kappa_{abb}\approx 5\times10^4 P_{b, 10}^2$ (eq. [\ref{eq:kappa_analytic}]).  We find that $\kappa_a^{(\rm dyn-dyn)}$ is large as long as the frequency of mode $a$ (the mode being driven) is larger than the frequency of the dynamical tide. This is because there is then a region below mode $a$'s inner turning point where $a$ is evanescent and the integrand does not oscillate about zero. This behavior is identical to that of $\kappa_{bc}^{(\rm dyn)}$ (\S~\ref{sec:kappa_dyn}) except that there the dynamical tide serves as the high frequency mode in the triplet whereas here mode $a$ serves as the high frequency mode. For the case of a solar binary, $\varepsilon \kappa_a^{(\rm dyn-dyn)}$  is larger than $I_{a\ell m}$ for g-modes with periods $P_a \ga 1\trm{ day}$. This suggests that inhomogeneous driving by the dynamical tide may be important for these modes. We discuss this possibility in \S~\ref{sec:inhomog_driving}.

\begin{figure}
\epsscale{0.55}
\plotone{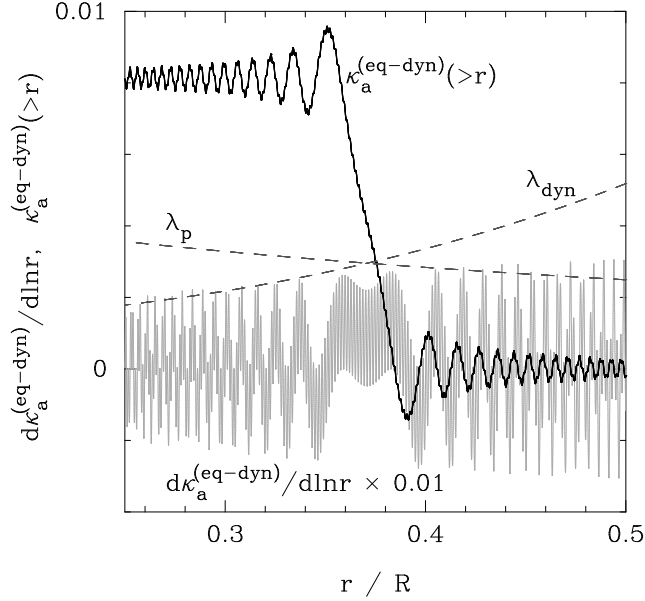}
\caption{Radial profile of $\kappa_a^{(\rm eq-dyn)}$ for the high order p-mode $(\ell_a=2,  n_a=1200, m_a=0)$ coupled to the $(\ell,m)=(2,0)$ equilibrium tide and the $(\ell, m)=(2, 0)$ dynamical tide at a $P=10\trm{ day}$ driving period. The dashed lines show the variation of the radial wavelength of the p-mode $\lambda_{\rm p}$ and the dynamical tide $\lambda_{\rm dyn}$ in units of $R$.}
\label{fig:kap_linlin_profile}
\end{figure}

We find that $\kappa_a^{(\rm dyn-dyn)}$ is especially large for high-order p-modes and increases with increasing $\omega_a$ (right panel of Figure \ref{fig:linlin_coef}). Physically, this is because high-order p-modes penetrate deeply into the solar core, where the amplitude of the dynamical tide peaks. The p-modes turning point occurs where $\omega_a$ equals the Lamb frequency $S_{\ell_a}(r)=\Lambda_a c_s/r$. The sound speed is nearly constant in the core ($c_s\simeq 5\times10^7 \trm{ cm s}^{-1}$) and high-order $\ell_a=2$ p-modes do not reflect until they reach a depth $r/R\approx0.003(\omega_a/10^3\omega_0)^{-1}$. In \S~\ref{sec:inhomog_driving} we briefly discuss whether these p-modes, which cannot form standing waves since they are above the acoustic cutoff, can be driven to significant amplitudes. 

The coupling coefficient $\kappa_a^{(\rm eq-dyn)}$ is significantly larger than $\kappa_a^{(\rm eq-eq)}$  for high-order g- and p-modes. For high-order g-modes, both coefficients are dominated by  coupling within the convection zone as all three waves in the triplet are long wavelength in the convection zone and can thus couple well there. However, unlike $\kappa_a^{(\rm eq-eq)}$, we find that $\kappa_a^{(\rm eq-dyn)}$ is weakly dependent on $\omega_a$ for g-modes. This is because the integrand of the latter depends on $(\xi_{r, a}, \xi_{h, a})$, both of which vary weakly with $\omega_a$ within the convection zone, where the mode is evanescent. By contrast, $\kappa_a^{(\rm eq-eq)}$ depends on $\delta \rho_a$ (see eq. [\ref{eq:Ja_plus_2kap_eq}]) which is a strong function of $\omega_a$ within the convection zone. 

For high-order p-modes, the coupling $\kappa_a^{(\rm eq-dyn)}$ does not occur within the convection zone since the p-modes are oscillatory there. Instead, the coupling occurs in the radiative interior where the p-mode wavelength $\lambda_{\rm p} \approx 2\pi c_s/\omega_a$ is comparable to the dynamical tide wavelength $\lambda_{\rm dyn}\approx 2\pi \omega r/\Lambda N$. Since $\lambda_{\rm p}$ decreases monotonically with $r$ and $\lambda_{\rm dyn}$ increases monotonically with $r$, there is a region where the two wavelenths cross for p-modes with $\omega_a$ in a certain range. In Figure \ref{fig:kap_linlin_profile} we show the radial profile of  $\kappa_a^{(\rm eq-dyn)}$ over such a region for an $n_a=1200$ p-mode. The integrand of $\kappa_a^{(\rm eq-dyn)}$ oscillates symmetrically about zero in regions where the wavelengths are very different. However, at $r\simeq0.37R$ the two wavelengths cross (dashed lines) and the two modes couple coherently over a region of size $\approx0.05R$. We find that over the frequency range $10^2P_5\la \omega_a/\omega_0\la 10^4 P_5$ there is a radius in the radiative interior of a solar-type star where $\lambda_{\rm dyn}=\lambda_{\rm p}$; a small fraction of this frequency range is shown in the right panel of Figure \ref{fig:linlin_coef}.

\section{Interaction energy and orbital evolution}
\label{sec:app:Interaction_energy}

Here we derive the acceleration of the secondary due to the oscillation modes excited in the primary assuming the modes are all standing waves. This is most easily accomplished by plugging the tidal potential (eq. [\ref{eq:U}]) into the interaction Hamiltonian (eq. [\ref{eq:Hint}]), and using the integrals defined in equations (\ref{eq:Ialm}) and (\ref{eq:Jablm}). We find
\bea
H_{\rm int} & = & 
- \sum_{a\ell m} \frac{GMM'W_{\ell m}R^\ell}{D^{\ell+1}}
\left[I_{a\ell m} q_a^\ast 
+ \frac{1}{2} \sum_{b} 
J_{ab\ell m} q_a q_b\right] e^{-im\Phi},
\label{eq:exHint}
\eea
In the center of mass frame, this interaction energy gives the following radial and angular accelerations:
\bea
a_D & = & - \frac{1}{\mu} \frac{\partial H_{\rm int}}{\partial D}
=- \frac{G(M+M')}{R^2} \sum_{a\ell m} (\ell+1)
W_{\ell m}
\left( \frac{R}{D} \right)^{\ell+2}
\left(I_{a\ell m} q_a^\ast
+ \frac{1}{2}\sum_b J_{ab\ell m} q_a q_b \right) e^{-im\Phi},
\label{eq:aD}
\\
a_\Phi & = & - \frac{1}{\mu D} \frac{\partial H_{\rm int}}{\partial \Phi}
=- \frac{G(M+M')}{R^2} \sum_{a\ell m} i m
W_{\ell m}
\left( \frac{R}{D} \right)^{\ell+2}
\left( I_{a\ell m} q_a^\ast 
+ \frac{1}{2} \sum_b J_{ab\ell m} q_a q_b \right) e^{-im\Phi},
\label{eq:aPhi}
\eea
where $\mu = MM'/(M+M')$. The expressions in equations (\ref{eq:aD}) and (\ref{eq:aPhi}) may also be derived from the perturbed potential of the primary evaluated at the position of the secondary. In doing so, the nonlinear expression for $\delta \rho$ from  \citet{Schenk:02} must be used to compute the quadrupole moment.

For orbital frequency  $\Omega=[G(M+M')/a^3]^{1/2}$, the radial and angular accelerations due to the tide imply the following changes in the  orbit \citep{Murray:00}:
\bea
\dot{a} & = & \frac{2}{\Omega \sqrt{1-e^2}}
\left[ a_D e \sin\Phi +  a_\Phi \left( 1+e\cos\Phi \right) \right]
\label{eq:adot}\\
\dot{e} & = & \frac{\sqrt{1-e^2}}{\Omega a}
\left[a_D \sin\Phi + a_\Phi \left( \cos\Phi + \cos E \right) \right],
\label{eq:edot}
\eea
where $\cos E=(e+\cos\Phi)/(1+e\cos\Phi)$. Writing the mode amplitude in amplitude-phase form $q_a=A_ae^{-i\varphi_a}$, where $A_a$ and $\varphi_a$ are real, the decay and circularization rates (eqs. [\ref{eq:adot}] and [\ref{eq:edot}]) can be written
\bea
\label{eq:adot_edot}
\left( \begin{array}{c}
\dot{a}/a \\
\dot{e}/e 
\end{array} \right)
& =&  \Omega \sum_{a\ell m k} W_{\ell m} X^{\ell m}_k \left( \frac{R}{a} \right)^\ell  \left( \begin{array}{c}
f_1 \\
f_2
\end{array} \right)
 \Bigg[I_{a\ell m} A_a  \sin \left(\varphi_a - k\Omega t \right) 
 \non &&
 - \frac{1}{2}\sum_b J_{ab\ell m} A_a A_b \sin \left(\varphi_a+\varphi_b + k\Omega t\right) \Bigg],
\eea
where $f_1 = 2k$ and $f_2 = \left(e^{-2}-1\right)\left[k-m/\sqrt{1-e^2}\right]$. At resonances the arguments in the sines can be time independent with a constant lag angle that depends on the damping rates and detunings (see, e.g., eq. [\ref{eq:qa_lin}] and Appendix \ref{sec:app:stability_3mode}). 

\section{Nonlinear equilibrium for a simple three mode system}
\label{sec:app:stability_3mode}

In this appendix we solve for the nonlinear equilibrium (i.e., saturation) of a simple 3 mode network in which a single parent $a$, driven by the linear tide, is coupled to a pair daughters $(b,c)$. We assume that the daughters are distinct, not self-coupled, and not driven by the linear or nonlinear tide. The amplitude equations for this system are thus
\bea
\dot{q}_a +(i \omega_a + \gamma_a)q_a &=&i\omega_a
\left[U_a(t)+ 2\kappa q_b^\ast q_c^\ast \right], \\
\dot{q}_b +(i \omega_b + \gamma_b)q_b &=&2 i\omega_b \kappa q_a^\ast q_c^\ast,\\
\dot{q}_c +(i \omega_c + \gamma_c)q_c &=&2 i\omega_c \kappa q_a^\ast q_b^\ast,
\eea
where $\kappa = \kappa_{abc}$ and the factors of two come from assuming that the modes are distinct from one another. Focusing on a particular harmonic $\omega$ such that $U_a(t)=U_a e^{-i\omega t}$, let $q_\alpha=Q_\alpha e^{-i (\omega_\alpha-\Delta_\alpha) t}$, where $\alpha=\{a,b,c\}$ and $Q_\alpha$ is a constant complex amplitude. If $\Delta_a=\omega_a-\omega$ and $\Delta_b+\Delta_c=\omega+\omega_b+\omega_c$ the time dependences drop out yielding the equilibrium solution 
\bea
\label{eq:parent}
A_a e^{i\delta_a} \left(i \Delta_a + \gamma_a\right) &=&
i\omega_a\left[U_a+
2\kappa A_b A_c e^{-i(\delta_b+\delta_c)} \right], \\
\label{eq:daughter1}
A_b e^{i\delta_b} \left(i \Delta_b + \gamma_b\right)  &=&
2 i\omega_b \kappa A_a A_c  e^{-i(\delta_a+\delta_c)}, \\
\label{eq:daughter2}
A_c e^{i\delta_c}\left(i \Delta_c + \gamma_c\right)&=&
2 i\omega_c \kappa A_a A_b  e^{-i(\delta_a+\delta_b)}.
\eea
where we have written $Q_\alpha=A_\alpha e^{i\delta_\alpha}$ with $A_\alpha$ and $\delta_\alpha$ real.\footnote{In \S~\ref{sec:stability_analysis} we derived the stability criteria for this type of system (eq. [\ref{eq:stability_3wave}]). Here $|\Gamma|=2\kappa A_a\sqrt{\omega_b \omega_c}$ and the linear tidal flow is unstable if the parent's linear energy $E_{a, {\rm lin}}> E_{\rm th}$, where $E_{\rm th}$ is given by equation (\ref{eq:Eth}).}

To solve for the parent's nonlinear equilibrium energy, divide the two daughter equations (\ref{eq:daughter1}) and (\ref{eq:daughter2}) to get
\beq
\frac{A_b^2}{A_c^2} =\frac{\omega_b}{\omega_c}\left(\frac{\Delta_b \Delta_c + \gamma_b \gamma_c - i(\Delta_b \gamma_c -\Delta_c \gamma_b)}{\Delta_b^2 + \gamma_b^2}\right).
\eeq
Since the left hand side is real, the imaginary part on the right hand side must vanish, $\Delta_b /\Delta_c = \gamma_b /\gamma_c$, and we find
\beq
\label{eq:dratio}
\frac{A_b^2}{A_c^2} = \frac{\omega_b \gamma_c}{\omega_c \gamma_b}.
\eeq
Multiplying the two daughter equations we obtain
\bea
A_a^2 &=&  -\frac{\gamma_b \gamma_c}{4\kappa^2 \omega_b \omega_c} 
\left[(1-\mu^2) \cos 2\delta - 2 \mu \sin 2 \delta + i (2 \mu \cos 2 \delta + (1-\mu^2)\sin 2\delta)\right] ,
\eea
where $\mu=(\Delta_b+\Delta_c) / (\gamma_b+\gamma_c)$ and $\delta=\delta_a+\delta_b+\delta_c$.
Since the left hand side is real, the imaginary part must vanish so that
\beq
\label{eq:deltaeq}
\tan 2\delta = - \frac{2\mu}{1-\mu^2}.
\eeq
There is a sign ambiguity which is resolved by noting that $A_a^2 > 0$; if, for example, $0 < \mu < 1$, choose the quadrant such that $\sin2\delta = 2\mu/(1+\mu^2) > 0$ and $\cos2\delta=-(1-\mu^2)/(1+\mu^2) < 0$. We thus find the parent's equilibrium energy
\beq
\label{eq:Ea_eq}
E_{a, {\rm eq}}=A_a^2 = \frac{\gamma_b \gamma_c}{4\kappa^2 \omega_b \omega_c} 
 \left(1+\mu^2 \right), 
\eeq
which we see equals its threshold energy $E_{\rm th}$ (eq. [\ref{eq:Eth}]). 

To solve for the daughters' nonlinear equilbrium energy, rearrange the parent's equation (\ref{eq:parent}) and use equations (\ref{eq:dratio}) and (\ref{eq:Ea_eq}) to get
\bea
A_b^2 &=& 
\frac{\gamma_c\sqrt{1+\mu^2}}{4\kappa^2\omega_a \omega_c} 
\Bigg[\Delta_a \cos \delta + \gamma_a \sin\delta -  \omega_a \frac{U_a}{A_a} \cos(\delta - \delta_a)
+ i \left(\Delta_a \sin \delta - \gamma_a \cos\delta   - \omega_a \frac{U_a}{A_a}\sin(\delta-\delta_a)\right)\Bigg].
\eea
Since $A_b^2$ is real, the imaginary part on the right hand side must vanish
\beq
\omega_a \frac{U_a}{A_a} \sin(\delta - \delta_a) = \Delta_a \sin\delta -\gamma_a \cos\delta,
\eeq
which, with equations (\ref{eq:deltaeq}) and (\ref{eq:Ea_eq}) and the definition of $\delta$, determines $\delta_a, \delta_b,$ and $\delta_c$. The daughters' equilibrium solution is thus given by
\bea
\label{eq:Eb_eq}
E_{b, {\rm eq}}&=&A_b^2 =
\frac{\gamma_c\sqrt{1+\mu^2}}{4\kappa^2\omega_a \omega_c} 
\Bigg[\Delta_a \cos \delta + \gamma_a \sin\delta -  \omega_a \frac{U_a}{A_a} \cos(\delta - \delta_a)\Bigg]
\eea
and $E_{c, {\rm eq}}=E_{b, {\rm eq}} \omega_c\gamma_b /  \omega_b\gamma_c$.
In the limit $E_{a, {\rm lin}} \gg E_{a, {\rm eq}}$, as is appropriate for solar binaries and hot Jupiter systems (see \S~\ref{sec:dyntide}), 
\bea
\label{eq:Eb_eq_limit}
E_{b, {\rm eq}}&=& \sqrt{\frac{\gamma_c \omega_b}{\gamma_b \omega_c}} \left|\frac{U_a}{2\kappa}\right|.
\eea
In the limit $E_{a, {\rm lin}}\rightarrow E_{\rm th}=E_{a, {\rm eq}}$ the term in brackets in (\ref{eq:Eb_eq}) vanishes and $A_b, A_c \rightarrow 0$.

\section{Local nonlinear growth rate}
\label{sec:app:local_growth_rate}

In this Appendix we calculate the local nonlinear growth rate due to three mode coupling and nonlinear tidal driving. The local rate per unit volume at which the nonlinear interactions do work is given by the fluid velocity dotted into the nonlinear force: $\dot{e}(\vec{x})=2\dot{\vec{\xi}}\cdot \left(\vec{f}_2[\vec{\xi},\vec{\xi}] - \rho (\vec{\xi} \cdot \grad)\grad U\right)$. For a self-coupled daughter wave $b$ 
\bea
\dot{e}_b (\vec{x}) &=& 2|\dot{q}_b| \vec{\xi}_b\cdot \left(|q_a||q_b|\vec{f}_2[\vec{\xi}_a,\vec{\xi}_b] - \rho |q_b| (\vec{\xi}_b \cdot \grad)\grad U\right)
\non&=& \frac{\omega|q_b|^2}{r^2} \left\{|q_a|\frac{d^3\kappa_{abb}}{dr d\Omega}
+ \frac{d^3U_{bb}}{drd\Omega}\right\}E_0,
\eea
where we assume that the interactions drive the daughter at half the tidal frequency $\omega/2$. In the WKB approximation, the local energy density of an internal gravity wave is 
\bea
e(r)=\frac{|q|^2}{r^2} \int \frac{d^3E}{drd\Omega} d\Omega \simeq \rho N^2 |q|^2 \xi_r^2 E_0.
\eea
The local nonlinear growth rate of the daughter over a region $\Delta r$ much smaller than $r$ (and the scale over which the background quantities vary) but much larger than its radial wavelength $2\pi/k_{r, b}$ is therefore
\bea
\Gamma_{b}(r)&\simeq& \frac{1}{\Delta r} \int_r^{r+\Delta r} \frac{\dot{e}_b(r)}{e_b(r)}\, dr
\non &=& \frac{\omega}{\Delta r}  \int_r^{r+\Delta r} \frac{E_0}{dE/dr} \left[|q_a| \frac{d\kappa_{abb}}{dr}
+ \frac{dU_{bb}}{dr}\right]\, dr.
\eea 
The product of the phases of the traveling waves (e.g., $\exp[i(\phi_a+2\phi_b)]$) is constant by momentum conservation and we have integrated over angles since the horizontal group velocity is faster than the vertical group velocity by a factor of $k_r/k_h=N/\omega\gg1$. 

By equation  (\ref{eq:kappa_krxir}), the local growth rate of daughters due to three-wave coupling with a parent $a$ is thus
\bea
\Gamma_{b, \kappa}&\equiv& \frac{\omega|q_a|}{\Delta r}  \int_r^{r+\Delta r} \frac{E_0}{dE/dr} \frac{d\kappa_{abb}}{dr}\,dr \simeq \omega T |q k_r \xi_r|_a \simeq \omega T|q\Lambda k_h\xi_h|_a.
\eea
We thus see that for the coupling of high-order g-modes, three-wave interactions act as a Kelvin-Helmholtz-like shear instability that is driven by the parent's horizontal motion $\xi_h$; the growth rate is given by the oscillation frequency times the strength of the horizontal shear $q\Lambda k_h \xi_h$. Whereas overturning of the vertical stratification requires $|q| k_r \xi_r \ga 1$, a g-mode is unstable to three-wave coupling even for $|q| k_r \xi_r \ll 1$, although at small enough amplitude the growth rate of the daughters becomes smaller than their linear damping rate $\gamma_b$ (see, e.g., \citealt{Drazin:77}; \citealt{Sonmor:97}).

\end{appendix}

\bibliographystyle{apj} 
\bibliography{ref}

\end{document}